%% file: ms.tex
\documentclass{lmcs}

\usepackage{imakeidx}
\makeindex
\indexsetup{noclearpage}

\usepackage{hyperref}
\usepackage{amssymb}
\usepackage{stmaryrd}

\theoremstyle{plain}
\usepackage{laws}
\usepackage{listings}

\makeatletter
\def\@setmcodes#1#2#3{{\count0=#1 \count1=#3
  \loop \global\mathcode\count0=\count1 \ifnum \count0<#2
  \advance\count0 by1 \advance\count1 by1 \repeat}}

\DeclareSymbolFont{italic}{OT1}{\rmdefault}{m}{it}
\let\mathit\undefined
\DeclareSymbolFontAlphabet{\mathit}{italic}
\edef\@tempa{\hexnumber@\symitalic}
\@setmcodes{`A}{`Z}{"7\@tempa41}
\@setmcodes{`a}{`z}{"7\@tempa61}
\makeatother

\newif\ifmagictop
\magictopfalse

\newif\iffullpaper
\fullpapertrue

\ifmagictop
\newcommand{\refsto}{\mathrel{\sqsubseteq}}
\newcommand{\refines}{\mathrel{\sqsupseteq}}
\newcommand{\nondet}{\mathbin{\sqcap}}
\newcommand{\Nondet}{\textstyle\mathop{\bigsqcap}}
\newcommand{\sconj}{\mathbin{\sqcup}}
\newcommand{\Sconj}{\textstyle\mathop{\bigsqcup}}

\newcommand{\sconjname}{join}

\newcommand{\InfFP}{\mu}

\else
\newcommand{\refsto}{\mathrel{\succcurlyeq}}
\newcommand{\refines}{\mathrel{\preccurlyeq}}
\newcommand{\nondet}{\mathbin{\vee}}
\newcommand{\Nondet}{\textstyle\mathbin{\bigvee}}
\newcommand{\sconj}{\mathbin{\wedge}}
\newcommand{\Sconj}{\textstyle\mathop{\bigwedge}}

\newcommand{\sconjname}{meet}

\newcommand{\InfFP}{\nu}

\fi

\newcommand{\constant}{invariant}

\newcommand{\pstepd}{\pi}
\newcommand{\estepd}{\epsilon}

\newcommand{\cpstepd}{\boldsymbol{\pi}}
\newcommand{\cestepd}{\boldsymbol{\epsilon}}
\newcommand{\cstepd}{\boldsymbol{\alpha}}
\newcommand{\cpstep}[1]{\mathop{\pi} #1}
\newcommand{\cestep}[1]{\mathop{\epsilon} #1}
\newcommand{\cstep}[1]{\mathop{\alpha} #1}
\newcommand{\cgdd}{\boldsymbol{\tau}}
\newcommand{\cgd}[1]{\mathop{\tau} #1}
\newcommand{\Fin}[1]{#1^{\star}}
\newcommand{\Inf}[1]{#1^\infty}
\newcommand{\Om}[1]{#1^\omega}
\newcommand{\universalrel}{\kw{univ}}
\newcommand{\emptyrel}{\emptyset}
\newcommand{\Opt}[1]{\mathop{\kw{opt}} #1}
\newcommand{\trans}[3]{#1 \xrightarrow{#2} #3}

\newcommand{\kw}[1]{\mathsf{#1}}
\newcommand{\Nil}{\cgdd}
\newcommand{\Abort}{\lightning}
\newcommand{\Chaos}{\kw{chaos}}
\newcommand{\Do}{\mathop{\kw{do}}}
\newcommand{\Else}{\mathbin{\kw{else}}}
\newcommand{\Fi}{\mathop{\kw{fi}}}
\newcommand{\Idle}{\kw{idle}}
\newcommand{\If}{\mathop{\kw{if}}}
\newcommand{\Magic}{\kw{magic}}
\newcommand{\Od}{\mathop{\kw{od}}}
\newcommand{\Skip}{\kw{skip}}
\newcommand{\Term}{\kw{term}}
\newcommand{\Then}{\mathbin{\kw{then}}}
\newcommand{\While}{\mathop{\kw{while}}}

\newcommand{\true}{\textsf{true}}
\newcommand{\false}{\textsf{false}}
\newcommand{\spot}{\mathrel{.}}
\newcommand{\union}{\mathbin{\cup}}
\newcommand{\Union}{\textstyle\mathop{\bigcup}}
\newcommand{\inter}{\mathbin{\cap}}
\newcommand{\Inter}{\textstyle\mathop{\bigcap}}
\newcommand{\Pre}[1]{\{#1\}}
\newcommand{\Post}[1]{\Spec{}{}{#1}}
\newcommand{\Rely}{\mathop{\kw{rely}}}
\newcommand{\rely}[1]{\Rely #1}
\newcommand{\Guarantee}{\mathop{\kw{guar}}}
\newcommand{\guar}[1]{\Guarantee #1}
\makeatletter
\newcommand{\Frame}[2]{\ifx\@empty#1\else#1\!:\!\fi#2}
\def\Spec{\@ifnextchar*{\@Spec}{\@@Spec}}
\def\@Spec*#1#2#3{\ifx\@empty#1\else#1\colon\fi
   [{#2}\ifx\@empty#2\else,~\fi#3]}
\def\@@Spec#1#2#3{\ifx\@empty#1\else
   \begin{array}{@{}l@{}}#1\colon\end{array}\!\!\fi%
   \left[{\begin{array}{@{}l@{}}#2\end{array}}\ifx\@empty#2\else~,~~\fi
   \begin{array}{@{}l@{}}#3\end{array}\right]}
\def\PSpec{\@ifnextchar*{\@PSpec}{\@@PSpec}}
\def\@PSpec*#1#2#3{\ifx\@empty#1\else#1\colon\fi
   \lceil{#2}\ifx\@empty#2\else,~\fi#3\rceil}
\def\@@PSpec#1#2#3{\ifx\@empty#1\else
   \begin{array}{@{}l@{}}#1\colon\end{array}\!\!\fi%
   \left\lceil{\begin{array}{@{}l@{}}#2\end{array}}\ifx\@empty#2\else~,~~\fi
   \begin{array}{@{}l@{}}#3\end{array}\right\rceil}
\makeatother

\renewcommand{\implies}{\mathbin{\Rightarrow}}
\newcommand{\defs}{\mathrel{\triangleq}}
\newcommand{\together}{\mathbin{\Cap}}
\newcommand{\sync}{\mathbin{\otimes}}
\newcommand{\Seq}{\mathbin{;}}
\makeatletter
\def\strut@op#1{\mathop{\mathstrut{#1}}\nolimits}
\newcommand{\power}{\strut@op{\mathbb{P}}}
\newcommand{\finset}{\strut@op{\mathbb{F}}}
\makeatother
\newcommand{\fun}{\mathrel{\rightarrow}}
\newcommand{\Boolean}{{\mathbb B}}

\newcommand{\pnegate}[1]{\overline{#1}}

\newcommand{\dres}{\mathbin{\vartriangleleft}}
\newcommand{\rres}{\mathbin{\vartriangleright}}
\newcommand{\Finrel}[1]{#1^*}
\newcommand{\upto}{\mathbin{\ldotp\ldotp}}
\newcommand{\id}[1]{\kw{id}_{#1}}
\newcommand{\limg}{(\!|}
\newcommand{\rimg}{|\!)}
\makeatletter
\def\comp@sym{\raise 0.6ex\hbox{\small\oalign{\hfil%
        $\scriptscriptstyle\mathrm{o}$\hfil%
        \cr\hfil$\scriptscriptstyle\mathrm{9}$\hfil}}}
\newcommand{\semi}{\mathrel{\comp@sym}}
\makeatother
\newcommand{\dom}[1]{\mathop{\kw{dom}}#1}
\newcommand{\atomic}[1]{\langle#1\rangle} 
\newcommand{\atomicrel}[1]{\left\langle #1 \right\rangle}
\newcommand{\Ominus}{\mathop{\ominus}}
\newcommand{\Oplus}{\mathbin{\oplus}}
\newcommand{\UnarySem}[2]{\mathop{\widehat{#1}}#2}
\newcommand{\BinarySem}[3]{#1\mathbin{\widehat{#2}}#3} \newcommand{\Test}[1]{[\![#1]\!]}
\newcommand{\Eval}[2]{#1_{#2}}
\newcommand{\EqEval}[2]{eq\,#1\,#2}
\newcommand{\GeEval}[2]{ge\,#1\,#2}
\newcommand{\GtEval}[2]{gt\,#1\,#2}
\newcommand{\Update}[2]{update\,#1\,#2}
\newcommand{\TypeOf}[2]{type\_of(#1,#2)}
\newcommand{\Abs}[1]{abs(#1)}
\newcommand{\Subrange}[2]{\{#1 \upto #2\}}

\newcommand{\atom}[1]{\mathsf{#1}}
\newcommand{\ata}{\atom{a}}

\makeatletter
\newdimen\zedtab \zedtab=2em%
\def\t#1{\afterassignment\@t\count@=#1}
\def\@t{\hskip\count@\zedtab}
\DeclareMathSymbol{\@@cat}{\mathbin}{AMSa}{"61}
\def \cat{\mathbin{\raise 0.8ex\hbox{$\mathchar\@@cat$}}}
\makeatother

\newcommand{\TS}[1]{[\![#1]\!]}
\newcommand{\Transrelationeq}{\succeq}
\newcommand{\WFrelation}[2]{#1 \supset #2}
\newcommand{\WFrelationeq}[2]{#1 \supseteq #2}
\newcommand{\Dec}[1]{dec_{\supset}#1}
\newcommand{\Deceq}[1]{dec_{\supseteq}#1}
\newcommand{\posvals}[1]{\widehat{#1}}

\makeatletter
\newcommand{\ChainRel}[1]{\crcr 
  #1~ &
  \@ifnextchar*{\@ChainRelCommment}{}}
\newcommand{\Why}[1]{\mbox{{\color{blue}\hspace*{1em}#1}}}
\def\@ChainRelCommment*[#1]{\Why{#1}
  \crcr & 
  }
\newcommand{\StartRef}[1]{\hspace*{-1.5em}(\ref{#1}) \refsto
  \@ifnextchar[{\@StartRefCommment}{}}
\def\@StartRefCommment[#1]{\mbox{#1}
  \crcr 
  }
\makeatother
\newcommand{\Implies}{\ChainRel{\Rightarrow}}
\newcommand{\IFF}{\ChainRel{\Leftrightarrow}}

\newcommand{\ImpliedBy}{\ChainRel{\Leftarrow}}
\newcommand{\Refsto}{\ChainRel{\refsto}}
\newcommand{\Refines}{\ChainRel{\refines}}
\newcommand{\Equals}{\ChainRel{=}}
\newcommand{\Subseteq}{\ChainRel{\subseteq}}

\newcommand{\tr}{tr}
\newcommand{\tracestates}[1]{#1^S}
\newcommand{\tracetype}[1]{#1^X}
\newcommand{\tracetrans}[1]{#1^T}
\newcommand{\tracekind}[1]{#1^K}
\newcommand{\xtrans}[3]{#1 \xrightarrow{#2} #3}

\newcommand{\ptrans}[2]{#1 \xrightarrow{\pi} #2}
\newcommand{\etrans}[2]{#1 \xrightarrow{\epsilon} #2}
\newcommand{\terminating}{\checkmark}
\newcommand{\aborting}{\dag}
\newcommand{\incomplete}{\bot}

\newcommand{\draftonly}[1]{}

\newcommand{\BB}{\hspace{-3pt}}
\makeatletter
\def\Set{\@ifnextchar*{\@Set}{\@@Set}}
\def\@Set*#1{{\color{purple}\llcorner\begin{array}{l}#1\end{array}\lrcorner}}
\def\@@Set#1{{\color{purple}\llcorner#1\lrcorner}}
\def\SetA{\@ifnextchar*{\@SetA}{\@@SetA}}
\def\@SetA*#1{{\color{purple}\left\lfloor\BB\left\floor\begin{array}{l}#1\end{array}\right\rfloor\BB\right\floor}}
\def\@@SetA#1{{\color{purple}\lfloor\BB\lfloor#1\rfloor\BB\rfloor}}
\def\Rel{\@ifnextchar*{\@Rel}{\@@Rel}}
\def\@Rel*#1{{\color{purple}\left\ulcorner\begin{array}{l}#1\end{array}\urcorner}}
\def\@@Rel#1{{\color{purple}\ulcorner#1\urcorner}}
\def\RelA{\@ifnextchar*{\@RelA}{\@@RelA}}
\def\@RelA*#1{{\color{purple}\left\lceil\BB\left\lceil\begin{array}{l}#1\end{array}\right\rceil\BB\right\rceil}}
\def\@@RelA#1{{\color{purple}\lceil\BB\lceil#1\rceil\BB\rceil}}
\newcommand{\SPre}[1]{\Pre{\Set{#1}}}
\newcommand{\RSpec}[3]{\Spec{#1}{}{\Rel{#3}}}
\newcommand{\Rrely}[1]{\rely{{\Rel{#1}}}}
\newcommand{\Rguar}[1]{\guar{{\Rel{#1}}}}
\newcommand{\RPost}[1]{\Post{\Rel{#1}}}
\newcommand{\Ratomicrel}[1]{\atomicrel{\Rel{#1}}}
\newcommand{\Comprehension}[3]{\{\ifx\@empty#3\else#3 \fi \ifx\@empty#1\else\mid #1\fi \mathrel{.} #2 \}}
\newcommand{\EvalSet}[2]{\Comprehension{}{\Eval{#1}{\sigma} = #2}{\sigma}}
\makeatother

\newcommand\numberthis{\addtocounter{equation}{1}\tag{\theequation}}

\newenvironment{RelatedWork}{\paragraph{Related work}\color{blue}}{}

\usepackage[disable,colorinlistoftodos,textwidth=2cm]{todonotes}
\newcommand{\backgroundintensity}{50}

\newcommand\notein[3]
{\todo[inline,linecolor=red,backgroundcolor=#2!\backgroundintensity]
  {#1 says: #3}
}
\newcommand{\ihin}[1]{\notein{IH}{yellow}{#1}}

\newcommand\todobig[4]{\todo[inline,backgroundcolor=#2!\backgroundintensity,caption={#1 says: #3}]{ 
\begin{minipage}{\textwidth-4pt}#1 says: #3.\par #4\end{minipage}}}
\newcommand{\ihbig}[2]{\todobig{IH}{yellow}{#1}{#2}}

\newcommand{\TODO}[1]{\draftonly{{\color{red}#1}}}

\newcounter{hours}
\newcounter{minutes}
\newcommand{\printtime}{%
  \ifthenelse{\value{hours}<10}{0}{}\thehours:%
  \ifthenelse{\value{minutes}<10}{0}{}\theminutes}
\newbox{\MyDate}
\savebox{\MyDate}{\draftonly{ (\today\ \printtime)}}

\begin{document}

\newif\ifarxiv
\arxivtrue

\ifarxiv
\else
\let\origthepage=\thepage
\makeatletter
\renewcommand{\thepage}{\@arabic\c@page-R}
\makeatother
\input{refcomments}
\clearpage
\let\thepage=\origthepage
\setcounter{page}{1}
\setcounter{section}{0}
\fi

\title[Deriving Laws for Developing Concurrent Programs\usebox{\MyDate}]{Deriving Laws for Developing Concurrent Programs
in a Rely-Guarantee Style\iffullpaper\else: \\ Condensed version without proofs\fi\rsuper*}
\titlecomment{{\lsuper*}This research was supported by
Australian Research Council (ARC)
Discovery Grant
DP190102142.}
\author[I. J. Hayes]{Ian J. Hayes}
\address{%
   The University of Queensland,
   School of Electrical Engineering and Computer Science,
   Brisbane,
   4072,
   Australia}
\email{Ian.Hayes@uq.edu.au}
\author[L. A. Meinicke]{Larissa A. Meinicke}
\address{%
   The University of Queensland,
   School of Electrical Engineering and Computer Science,
   Brisbane,
   4072,
   Australia}
\email{L.Meinicke@uq.edu.au}
\author[P. A. Meiring\usebox{\MyDate}]{Patrick A. Meiring}
\address{%
   The University of Queensland,
   School of Electrical Engineering and Computer Science,
   Brisbane,
   4072,
   Australia}
\email{patrick.meiring@gmail.com}

\begin{abstract}
This paper presents a theory for the refinement of shared-memory concurrent algorithms from specifications.
We augment pre and post condition specifications with 
Jones' rely and guarantee conditions,
all of which are encoded as commands within a wide-spectrum language.
Program components are specified using either partial or total correctness versions of postcondition specifications.
Operations on shared data structures and atomic machine operations (e.g.\ compare-and-swap)
are specified using an atomic specification command.
All the above constructs are defined in terms of a simple core language,
based on a small set of primitive commands and a handful of operators.
A comprehensive set of laws for refining such specifications to code is derived in the theory.
The approach supports fine-grained concurrency, 
avoiding atomicity assumptions on expression evaluation and assignment commands. 
The theory has been formalised in Isabelle/HOL,
and the refinement laws and supporting lemmas have been proven in Isabelle/HOL.

\end{abstract}


\maketitle

\newpage\tableofcontents\newpage

\section{Introduction}\labelsect{introduction}

Our overall goal is to develop a theory for deriving verified shared-memory concurrent programs
from abstract specifications.
A set of threads running in parallel can exhibit a high degree of non-determinism 
due to the myriad possible interleavings of their fine-grained accesses to shared variables.
The set of all threads running in parallel with a thread is referred to as its \emph{environment}
and the term \emph{interference} refers to the changes made to 
the shared variables of a thread by its environment.

\paragraph{The rely/guarantee approach.}

Reasoning operationally about threads that execute under interference is
fraught with the dangers of missing possible interleavings.
A systematic approach to concurrency is required to manage interference.
The approach taken here is based on the rely/guarantee technique of Jones~\cite{Jones81d,Jones83a,Jones83b},
which provides a compositional approach to handling concurrency.

To illustrate the rely/guarantee approach,
we give a Jones-style specification \cite{Jones81d,Jones83a} of an operation to remove an element $i$ from a set (\ref{ex-rem-from-set}).
The interesting aspect of the example
is that in removing $i$ from the set,
interference from the environment may also remove elements from the set,
possibly including $i$.
The set can be represented as a bit-map stored in an array of words.
Removing an element from the set then corresponds to removing an element from one of the words.
Here we focus on the interesting part from the point of view of handling interference,
of removing the element $i$ from a word $w$, 
where accesses to $w$ are atomic.
Words are assumed to contain $N$ bits
and hence the maximum number of elements in a set represented by a single word is $N$.
The variable $i$ is local and hence not subject to interference.
The rely condition is an assumption 
that the environment may neither add elements to $w$ nor change $i$ 
(i.e.\ the rely condition is, $w \supseteq w' \land i' = i$,
where $w$ refers to the initial value of $w$ and $w'$ to its final value and likewise for $i$).
The remove operation guarantees that each program step never adds elements to $w$, 
never removes elements other than $i$, and does not change $i$.
That rules out an (unlikely) implementation that adds additional elements to the set and 
then removes them as well as $i$.
Because $w$ only decreases and $i$ is not modified, 
the precondition, $w \subseteq \Subrange{0}{N-1} \land i \in \Subrange{0}{N-1}$, is an invariant.
The postcondition requires that $i$ is not in $w$ in the final state, (i.e.\ $i' \notin w'$).
\begin{align}
 \begin{array}{l}
   \kw{pre}~w \subseteq \Subrange{0}{N-1} \land i \in \Subrange{0}{N-1} \\
   \kw{rely}~w \supseteq w' \land i' = i \\
   \kw{guar}~w \supseteq w' \land w - w' \subseteq \{i\} \land i' = i \\
   \kw{post}~i' \not\in w' 
  \end{array} \label{ex-rem-from-set}
\end{align}
Note how the requirement to remove $i$ and only $i$ from $w$ is split between the post condition and
the guarantee.
Compare that with the postcondition of, $w' = w - \{ i \}$, of a (sequential) operation
to remove $i$ in the context of no interference.
The sequential postcondition is not appropriate in the context of 
concurrent interference that may remove elements from $w$
because that interference may falsify the sequential postcondition,
while the postcondition $i' \not\in w'$ is stable under the rely condition.

\begin{figure}[ht]
\begin{center}
\input{rely-guar}
\caption{An execution trace of a thread consisting of a sequence of states $\sigma_1$--$\sigma_7$ 
with either program ($\pi$) or environment ($\epsilon$) transitions between successive states.
If the execution trace is from a thread satisfying a rely/guarantee specification, then
if the initial state $\sigma_0$ satisfies the precondition of the specification, $p$,
and all environment transitions satisfy the rely relation $r$,
then all program transitions must satisfy the guarantee relation $g$,
and the postcondition relation $q$ must be satisfied between the initial ($\sigma_0$) and final ($\sigma_7$) states.}
\labelfig{rely-guar}
\end{center}
\end{figure}

The semantic model represents the behaviour of a thread as a set of Aczel traces \cite{Aczel83,DeRoever01}
of the form given in \reffig{rely-guar}.
Aczel traces distinguish atomic steps (or transitions) made by a thread itself, called \emph{program or $\pstepd$ steps} here,
from atomic steps made by its environment, called \emph{environment or $\estepd$ steps} here.
In the rely/guarantee approach, the interference on a thread $c$ is assumed to satisfy a rely condition $r$,
where $r$ is a reflexive, transitive binary relation between program states
that all (atomic) environment steps of $c$ are assumed to satisfy.
Because $c$ itself is part of the environment of the other threads,
$c$ is required to satisfy a guarantee $g$, which is a reflexive relation between states
that all program steps of $c$ must satisfy. 
The guarantee condition of a thread must imply the rely conditions of all threads in its environment.

\paragraph{Concurrent refinement calculus}

The sequential refinement calculus \cite{Morgan94,BackWright98} 
makes use of a wide-spectrum language, 
which extends an executable imperative programming language with specification constructs
that encode preconditions and postconditions as the commands, $\Pre{p}$ and $\Post{q}$, respectively.
A postcondition specification command, $\Post{q}$, where $q$ is a binary relation on programs states,
represents a commitment that the program will terminate and satisfy $q$ between its initial and final states overall.
An assertion command, $\Pre{p}$, where $p$ is a set of states,
represents an assumption that the initial state is in $p$;
it allows any behaviour whatsoever for initial states not in $p$,
and hence from initial states not satisfying $p$, there is no obligation for the program to satisfy its postcondition or terminate.
If the initial state is not in $p$, we say the assertion command $\Pre{p}$ aborts,
i.e.\ it behaves as Dijkstra's abort command \cite{Dijkstra75,Dijkstra76}, denoted by $\Abort$ here.

We extend this approach by encoding Jones' 
rely condition $r$ as the command, $\rely{r}$, and 
his guarantee condition $g$ as the command, $\guar{g}$, 
where $r$ and $g$ are binary relations on program states.
A guarantee command, $\guar{g}$, represents a commitment 
that every (atomic) program step satisfies the relation $g$ between its before and after program states.
A rely command, $\rely{r}$, represents an assumption that all (atomic) environment steps satisfy $r$.
If its environment performs a step not satisfying $r$,
the command, $\rely{r}$, aborts and hence any behaviour whatsoever is allowed from that point on,
in particular, there is no longer an obligation for the program to terminate or to satisfy its postcondition specification overall 
or satisfy its guarantee from that point on.

In order to combine these commands to form a rely/guarantee specification similar to (\ref{ex-rem-from-set}),
we make use of a weak conjunction operator ($\together$) novel to our approach \cite{AFfGRGRACP,FM2016atomicSteps}.
A behaviour of a weak conjunction of two commands, $c \together d$,
must be both a behaviour of $c$ and a behaviour of $d$ up until the point that either $c$ or $d$ aborts,
at which point $c \together d$ aborts. 
If both $c$ and $d$ have no aborting behaviours, then every behaviour of $c \together d$ 
must be a behaviour of both $c$ and $d$, that is, their strong conjunction $c \sconj d$. 
We illustrate the difference between weak and strong conjunction with an example of combining 
two pre-post specifications,%
\footnote{Such operators has been investigated for sequential programs 
\cite{Ward:SpecConj,Groves2002}.}
where sequential composition ($\Seq$) has highest precedence.
\begin{eqnarray}
  \Pre{p_1} \Seq \Post{q_1} \together \Pre{p_2} \Seq \Post{q_2} & = & \Pre{p_1 \inter p_2} \Seq \Post{q_1 \inter q_2} \labelprop{weak-conj-eg} \\
  \Pre{p_1} \Seq \Post{q_1} \sconj \Pre{p_2} \Seq \Post{q_2} & = & \Pre{p_1 \union p_2} \Seq \Post{(p_1 \implies q_1) \inter (p_2 \implies q_2)} \labelprop{strong-conj-eg}  
\end{eqnarray}
The weak conjunction \refprop{weak-conj-eg} aborts if either component aborts, 
as represented by the precondition of $p_1 \inter p_2$ on the right,
and must satisfy both postconditions $q_1$ and $q_2$ otherwise.
The strong conjunction  \refprop{strong-conj-eg} aborts if both can abort, 
as represented by the precondition of $p_1 \union p_2$ on the right,
and from initial states that satisfy $p_1$ it must satisfy postcondition $q_1$ and 
from initial states that satisfy $p_2$ it must satisfy postcondition $q_2$.

\paragraph{Characteristic predicates}

In Hoare logic, preconditions and postconditions are predicates 
that are interpreted with respect to a program state $\sigma$
that gives the values of the program variables,
e.g.\ $\sigma\,x$ is the value of the program variable $x$ in state $\sigma$.
The semantics of a predicate $P$ characterising a set of states is given by $\Set{P}$,
where we use ``${\color{purple}\llcorner}$'' and ``${\color{purple}\lrcorner}$'' as lightweight semantic brackets
and colour the predicate purple to distinguish it,
for example,%
\footnote{The syntax for set comprehension matches that of Isabelle/HOL.}
\begin{eqnarray*}
  \Set{x > 0} & = & \Comprehension{}{\sigma\,x > 0}{\sigma} .
\end{eqnarray*}
Similarly, the semantics of a predicate $R$ characterising a binary relation between states is given by $\Rel{R}$,
where we use ``${\color{purple}\ulcorner}$'' and ``${\color{purple}\urcorner}$'' as lightweight semantic brackets,
and references to a variable $x$ in $R$ stand for its value 
in the before state, $\sigma\,x$, 
and primed occurrences $x'$ stand for the value of $x$ in the after state, $\sigma'\,x$,
as in VDM~\cite{Jones80a}, Z~\cite{Hayes93,Woodcock96} and $\textrm{TLA}^{+}$~\cite{Lamport03},
for example, 
\[
  \Rel{x \geq x'} = \Comprehension{}{\sigma\,x \geq \sigma'\,x}{(\sigma,\sigma')} .
\]
The theory developed in the body of this paper uses the semantic models of sets and relations directly 
so that preconditions use sets of states, 
and relies, guarantees and postconditions use binary relations on states,
rather than their characteristic predicates.
This approach has the advantage of making the theory independent of the particular concrete syntax
used to express characteristic predicates.
We hope you will excuse us not giving an explicit definition of the interpretation of the predicates
used in the examples; the interpretation is straightforward and the particular notation used for predicates
is not of concern for expressing the theory and laws presented in the body of the paper.

\paragraph{Combining commands}

Weak conjunction ($\together$) and sequential composition ($\Seq$) can be used to combine commands into a specification, 
for example, the Jones-style specification (\ref{ex-rem-from-set}) 
is represented by the following command.
\begin{align}
  \begin{array}{ll}
   & \Rrely{w \supseteq w' \land i' = i} \\
   \together
   & \Rguar{w \supseteq w' \land w - w' \subseteq \{i\} \land i' = i} \\
   \together & 
   \SPre{w \subseteq \Subrange{0}{N-1} \land i \in \Subrange{0}{N-1}} 
   \Seq \RPost{i' \not\in w'} 
  \end{array} \label{ex-rem-from-set-spec}
\end{align}
The weak conjunction requires 
that both the guarantee and the postcondition are satisfied by an implementation
unless either the precondition does not hold initially or
the rely condition fails to hold for an environment step at some point,
in which case the whole specification aborts from that point.
The precondition and rely have no effect if the precondition holds initially and the rely condition holds for all environment steps
and hence in this case the behaviour must satisfy both guarantee for every program step 
and postcondition between the initial and final states overall.

The advantage of representing relies and guarantees as separate commands
is that one can develop laws for each construct in isolation as well as in combination with other constructs.
For example, Jones noted that strengthening a guarantee is a refinement.
In our theory, 
strengthening a guarantee corresponds to the refinement of $\guar{g_1}$ to $\guar{g_2}$,
if relation $g_1$ contains in $g_2$, i.e.\ $g_1 \supseteq g_2$.
Note that this law is expressed just in terms of the guarantee command, 
unlike the equivalent law using four-tuples of pre/rely/guar/post conditions, 
which must refer to extraneous (unchanged) pre/rely/post conditions
and needs to be proven in term of their semantics.%
\footnote{The single law given by Jones allows preconditions and rely conditions to be weakened
and guarantees and postconditions to be strengthened. 
We prefer to treat these as four separate laws because commonly only one of these conditions is modified.}

The core theory consists of a lattice of commands with a small set of primitive commands and operators.
Other commands, including $\Pre{p}$, $\Post{q}$, $\guar{g}$, $\rely{r}$ and programming language constructs,
are defined in terms of these primitives.
The theory is built up in stages: 
each stage introduces a new concept or command in the wide-spectrum language
along with a supporting theory of lemmas and laws.
Significant contributions of this paper are the following.
\begin{itemize}
\item
A comprehensive theory for handling postcondition specification commands in the context of interference 
(Sections~\ref{S-specifications} and \ref{S-stability}).
Two forms of specification command are provided: 
one for partial correctness, $\PSpec{}{}{q}$, and the other for total correctness, $\Spec{}{}{q}$.
\item
An atomic specification command, $\atomicrel{p,q}$,
suitable for defining atomic machine operations,
such as compare-and-swap,
and for specifying atomic operations on concurrent data structures (\refsect{atomic-spec}).
\item
A theory of expressions that makes minimal atomicity assumptions making it suitable for 
use in developing fine-grained concurrent algorithms (\refsect{expressions}).
A practical constraint on expressions is that only a single variable in the expression is subject to interference
and that variable is referenced just once in the expression (\refsect{expr-single-reference}).
For example, for an integer variable $x$,
$x+x$ may evaluate to an odd value under interference that modifies $x$ between references,
but $2*x$ has a single reference to $x$ and hence always evaluates to an even value.
\item
Due to the more general treatment of expression evaluation under interference,
we have been able to develop more general laws than those in the existing literature 
\cite{CoJo07,ColemanVSTTE08,Dingel02,Jones81d,Jones83a,Jones83b,PrensaNieto03,Sanan21,SchellhornTEPR14,stolen1991method,XuRoeverHe97} for introducing
assignments (\refsect{assignments}), 
conditionals (\refsect{conditional}) 
and loops (\refsect{loop}).
\item
Recursively defined commands are supported (\refsect{recursion}) and 
are used to define the $\While$ loop (\refsect{loop}).
\item
The algebraic theories have been formalised within Isabelle/HOL \cite{IsabelleHOL}, 
with the language primitives being defined axiomatically and other constructs defined in terms of the primitives.
The laws and lemmas presented in this paper have been proven in terms of these Isabelle/HOL algebraic theories.
\item
The sets of traces semantic model has also been formalised in Isabelle/HOL 
and shown to satisfy the axioms of the algebraic theories, 
thus establishing the consistency of the theories.
We make no claims for completeness.
\end{itemize}
\refsect{language} introduces our language in terms of 
a small set of primitive commands and a small set of operators.
Following sections cover 
the lattice of commands (\refsect{algebra}),
sequential composition (\refsect{sequential}),
fixed points and iteration (\refsect{iteration}),
tests (\refsect{tests}),
assertions (\refsect{assertions}),
atomic steps commands (\refsect{atomic}),
synchronisation operators, parallel and weak conjunction (\refsect{sync}),
guarantees (\refsect{guarantees}),
frames (\refsect{frames}),
relies (\refsect{relies}),
termination (\refsect{termination}),
partial and total correctness (\refsect{partial-total}),
specification commands (\refsect{specifications}),
stability under interference (\refsect{stability}),
parallel (\refsect{parallel}),
optional atomic steps (\refsect{optional}),
finite stuttering (\refsect{stuttering}),
atomic specifications (\refsect{atomic-spec}),
expressions (\refsect{expressions}),
assignments (\refsect{assignments}),
conditionals ($\If$) (\refsect{conditional}),
recursion (\refsect{recursion}),
and
while loops (\refsect{loop}).
\refsect{ex-remove} provides an example refinement of specification (\ref{ex-rem-from-set-spec}) to code
using the laws derived in this paper.
Parts of this refinement are also used as running examples throughout the paper.
\refsect{Isabelle} discusses the formalisation of the theory in Isabelle/HOL.

\begin{RelatedWork}
Rather than presenting the related work in one section,
paragraphs labeled \emph{Related work} have been included throughout the paper.
This is so that the comparison of the approach used in this paper with related work
can refer to the relevant details of how the individual constructs are handled 
in the different approaches.
\end{RelatedWork}

\iffullpaper
There are two levels at which this paper can be read:
by skipping the proofs, the reader gets an overview of the refinement calculus and its laws,
while delving into the proofs gives a greater insight into 
how the underlying theory supports reasoning about concurrent programs
and the reasons for the provisos and form of the refinement laws,
some of which are quite subtle due the effects of interference.
\fi

\section{Core language}\labelsect{language}

We have previously developed a concurrency theory 
\cite{AFfGRGRACP,Concurrent_Ref_Alg-AFP,FM2016atomicSteps,FMJournalAtomicSteps}
that is used to define commands in a wide-spectrum language,
develop refinement laws, and
and prove them correct.

\subsection{Semantic model}\labelsect{model}

In this section we briefly describe the semantic model for our theory, which is based on that in \cite{DaSMfaWSLwC}.
A command $c$ in our theory is modelled as a prefix-closed set of Aczel traces~\cite{Aczel83,DeRoever01}---denoted $\TS{c}$---where each trace is of the form given in \reffig{rely-guar},
in which a trace is a sequence of program states (each giving values of the program variables)
with transitions between states differentiated as either program steps ($\trans{\sigma}{\pstepd}{\sigma'}$) or 
environment steps ($\trans{\sigma}{\estepd}{\sigma'}$).
To allow for non-terminating computations, traces may be infinite.
Three types of traces are distinguished: terminated ($\terminating$), aborting ($\aborting$), and either incomplete or infinite ($\incomplete$).
For $\tr$ a trace, the following notation is used,
\begin{description}
\item[$\tracestates{\tr}$] gives the non-empty sequence of states of $\tr$,
\item[$\tracekind{\tr}$] gives the sequence of kinds of transitions ($\pstepd$ or $\estepd$) of $\tr$ -- its length is one less than $\tracestates{\tr}$,
\item[$\tracetype{\tr}$] gives the type of the trace ($\terminating$, $\aborting$, or $\incomplete$),
and
\item[$\tracetrans{\tr}$] gives the sequence of transitions of $\tr$, 
i.e.\ $\tracetrans{\tr} = \{ i \mapsto (\xtrans{\tracestates{\tr}_i}{\tracekind{\tr}_i}{\tracestates{\tr}_{i+1}}) \spot i \in \dom \tracekind{\tr} \}$.
\end{description}
A trace, $\tr$, is uniquely characterised by $\tracestates{\tr}$, $\tracekind{\tr}$ and $\tracetype{\tr}$.
For the trace in \reffig{rely-guar}, 
\begin{align*}
  \tracestates{\tr} & = [\sigma_0,\sigma_1,\sigma_2,\sigma_3,\sigma_4,\sigma_5,\sigma_6,\sigma_7] \\
  \tracekind{\tr} & = [\epsilon,\epsilon,\pi,\epsilon,\pi,\pi,\epsilon]\\
  \tracetype{\tr} & = \checkmark, \\
  \tracetrans{\tr} & = [ \etrans{\sigma_0}{\sigma_1}, \etrans{\sigma_1}{\sigma_2}, \ptrans{\sigma_2}{\sigma_3}, \etrans{\sigma_3}{\sigma_4}, 
              \ptrans{\sigma_4}{\sigma_5}, \ptrans{\sigma_5}{\sigma_6}, \etrans{\sigma_6}{\sigma_7} ]
\end{align*}
A \emph{prefix} of a trace $\tr$ is a trace $tp$ such that $\tracetype{tp} = \incomplete$,
sequence $\tracestates{tp}$ is a prefix of $\tracestates{\tr}$,
and sequence $\tracekind{tp}$ is a prefix of $\tracekind{\tr}$, 
where prefixes are not required to be strict.
An \emph{extension} of a trace $\tr$ is a trace $tx$ of any type such that 
$\tracestates{\tr}$ is a prefix of $\tracestates{tx}$,
and $\tracekind{\tr}$ is a prefix of $\tracekind{tx}$.
The set of traces $\TS{c}$ of a command $c$ satisfies three healthiness conditions:
\begin{description}
\item[prefix closure]
if $tr$ is a trace in $\TS{c}$, all prefixes of $tr$
(representing its incomplete behaviours because they have type $\incomplete$)
are also in $\TS{c}$;
\item[abort closure]
if $tr$ is an aborting trace of $\TS{c}$, i.e.\ $\tracetype{tr} = \aborting$,
all possible extensions of $tr$ are also in $\TS{c}$; and
\item[magic closed]
$\TS{c}$ contains all trivial incomplete traces consisting of an initial state $\sigma_0$ and no transitions;
these are the traces of the command $\Magic$ introduced below.
\end{description}
A set of traces is \emph{closed} if it is prefix, abort and magic closed.
The semantics of commands satisfies the following properties, in which $c$ and $d$ are commands.
\begin{itemize}
\item
The lattice partial order $c \refsto d$ represents that $c$ is refined (or implemented) by $d$.
In the semantic model refinement corresponds to superset-inclusion, that is, $\TS{c} \supseteq \TS{d}$.
\item
The command $\Magic$ is the least command in the lattice (i.e. every command is refined by $\Magic$).
It is infeasible in every initial state and
in the semantic model it is represented by the set of all incomplete traces that consist of just an initial state and no transitions.
\item
The command $\Abort$ (Dijkstra's \textsf{abort}) is the greatest command in the lattice (i.e.\ every command is a refinement of $\Abort$).
It allows any behaviour whatsoever;
in the semantic model it is represented by the set of all possible valid traces.
\item
The lattice meet $c \sconj d$, with identity $\Abort$,
represents a strong conjunction of $c$ and $d$, and $\Sconj C$ represents the strong conjunction of a set of commands $C$.
In the semantic model $\TS{c \sconj d} = \TS{c} \inter \TS{d}$ and $\TS{\Sconj C} = \Inter_{c \in C} \TS{c}$ and hence $\Sconj \emptyset = \Abort$.
\item
The lattice join $c \nondet d$, with identity $\Magic$,
represents a non-deterministic choice between $c$ and $d$,
and $\Nondet C$ represents a non-deterministic choice over a set of commands $C$.
In the semantic model $\TS{c \nondet d} = \TS{c} \union \TS{d}$ and $\TS{\Nondet C} = \Union_{c \in C} \TS{c}$,
for a non-empty set of commands $C$, and $\Nondet \emptyset = \Magic$. 
\item
$c \Seq d$ represents sequential composition of commands.
In the semantic model $\TS{c \Seq d}$ consists of the following traces:
\begin{enumerate}
\item\label{seq-term}
if there is a terminating trace $tc \in \TS{c}$,
then the trace $tc$ concatenated with any trace of $\TS{d}$ whose initial state matches the final state of $tc$
--- in the concatenation, the final state of $tc$ and the initial state of $td$ are merged into a single state;
\item
all incomplete or non-terminating (infinite) traces in $\TS{c}$;
and
\item
if there is an aborting trace $tc \in \TS{c}$, 
then that aborting trace $tc$ along with all possible extensions of $tc$, so as to preserve abort closure.
\end{enumerate}
\item
$c \together d$ represents the weak conjunction of commands.
In the semantic model $\TS{c \together d}$ consists of traces that are either:
\begin{itemize}
\item
traces of both $c$ and $d$ (i.e.\ in $\TS{c} \inter \TS{d}$);
\item
if $tc$ is an aborting trace in $\TS{c}$ and the incomplete trace corresponding to $tc$ is also a trace in $\TS{d}$,
then $tc$ and all possible extensions of $tc$
(so as to preserve abort closure);
or
\item
if $td$ is an aborting trace in $\TS{d}$ and the incomplete trace corresponding to $td$ is also a trace in $\TS{c}$,
then $td$ and all possible extensions of $td$.
\end{itemize}
\item
$c \parallel d$ represents the parallel composition of $c$ and $d$.
In the semantic model $\TS{c \parallel d}$ is defined in terms of the $match$ relation on traces
that matches a program step of one thread with an environment step of the other to give a program step of their composition,
and matches environment steps of both to give an environment step of their composition.
For any traces $tr$, $tc$ and $td$ the relation $match(tr,tc,td)$ holds if and only if,
\begin{itemize}
\item
$tr$, $tc$, and $td$ are all the same length (but not necessarily the same type),
\item
if $\tracetrans{\tr}_i$ is the program transition $\trans{\sigma}{\pstepd}{\sigma'}$ from $\sigma$ to $\sigma'$ of $tr$ then either,
\begin{itemize}
\item
$\tracetrans{tc}_i = (\trans{\sigma}{\pstepd}{\sigma'})$ and 
$\tracetrans{td}_i = (\trans{\sigma}{\estepd}{\sigma'})$, or
\item
$\tracetrans{td}_i = (\trans{\sigma}{\pstepd}{\sigma'})$ and
$\tracetrans{tc}_i = (\trans{\sigma}{\estepd}{\sigma'})$, and
\end{itemize}
\item
if $\tracetrans{\tr}_i$ is the environment transition $(\trans{\sigma}{\estepd}{\sigma'})$ 
then $\tracetrans{\tr}_i = \tracetrans{tc}_i = \tracetrans{td}_i$.
\end{itemize}
Using relation $match$, we have that $\TS{c \parallel d}$ consists of traces, $tr$, such that either,
\begin{itemize}
\item 
there exist traces $tc \in \TS{c}$ and $td \in \TS{d}$, for which $match(tr, tc, td)$ holds 
and $\tracetype{\tr} = \tracetype{tc} = \tracetype{td}$ and the traces are either terminating or incomplete, 
i.e.\ $\tracetype{\tr} \in \{ \terminating,\incomplete \}$,
or
\item 
there exists an aborting trace $tc \in \TS{c}$ and a trace $td \in \TS{d}$ and a trace $tr'$
such that $match(\tr', tc, td)$ and $\tr$ is either $\tr'$ or some extension of $tr'$,
or
\item
there exists an aborting trace $td \in \TS{d}$ and a trace $tc \in \TS{c}$ and a trace $tr'$
such that $match(\tr', tc, td)$ and $\tr$ is either $\tr'$ or some extension of $tr'$.
\end{itemize}
\end{itemize}
The above operators preserve closure on the sets of traces.

\paragraph{Syntactic precedence of operators}
Unary operators and function application have higher precedence than binary operators.
Amongst the binary operators, 
framing ($:$) has the highest precedence,
followed by sequential composition ($\,\Seq\,$).
Non-deterministic choice ($\nondet$) has the lowest precedence.
Otherwise no assumptions about precedence are made
and parentheses are used to resolve syntactic ambiguity.

\subsection{Function abstraction, application and fixed points} \labelsect{functions}

We use the usual notation for lambda abstraction and function application. 
Least and greatest fixed points of a function $f$ are denoted by $\mu f$ and $\nu f$, respectively. 
Following convention, we abbreviate $\mu(\lambda x \spot c)$ by $(\mu x \spot c)$
and $\nu(\lambda x \spot c)$ by $(\nu x \spot c)$.

\subsection{Relational notation}\labelsect{notation}

We briefly describe the notation used for relations in this paper
which is based on that of VDM~\cite{Jones90a} and Z~\cite{Hayes93,Woodcock96}.
Given a set $s$ and binary relations $r$, $r_1$ and $r_2$, 
$\dom{r}$ is the domain of the relation $r$ \refdef{dom},
$s \dres r$ is $r$ restricted so that its domain is contained in the set $s$ \refdef{dom-restrict},
$r \rres s$ is $r$ restricted so that its range is contained in $s$ \refdef{range-restrict},
$r \limg s \rimg$ is the image of $s$ through $r$ \refdef{image},
$r_1 \semi r_2$ is the relational composition of $r_1$ and $r_2$ \refdef{composition},
and
$\Finrel{r}$ is the reflexive transitive closure of $r$ \refdef{transitive-closure},
which is defined as a least fixed point ($\mu$) on the lattice of relations.
The universal relation over a state space $\Sigma$ is represented by $\universalrel$ \refdef{univ}, and
$\overline{r}$ is the set complement of $r$ with respect to $\universalrel$ \refdef{rel-negation}. The identity relation is represented by $\id{}$ \refdef{id}.
\begin{eqnarray}
  \dom{r} & \defs & \Comprehension{}{(\exists \sigma' \spot (\sigma,\sigma') \in r)}{\sigma} \labeldef{dom} \\
  s \dres r & \defs & \Comprehension{}{\sigma \in s \land (\sigma, \sigma') \in r}{(\sigma,\sigma')} \labeldef{dom-restrict} \\
  r \rres s & \defs & \Comprehension{}{(\sigma, \sigma') \in r \land \sigma' \in s}{(\sigma,\sigma')} \labeldef{range-restrict} \\
  r \limg s \rimg & \defs & \Comprehension{}{(\exists \sigma \in s \spot (\sigma,\sigma') \in r)}{\sigma'} \labeldef{image} \\
  r_1 \semi r_2 & \defs & \Comprehension{}{(\exists \sigma'' \spot (\sigma,\sigma'') \in r_1 \land (\sigma'',\sigma') \in r_2)}{(\sigma,\sigma')}  \labeldef{composition} \\
  \Finrel{r} & \defs & \mu x \spot \id{} \union r \semi x \labeldef{transitive-closure} \\
  \universalrel & \defs & \Sigma \times \Sigma \labeldef{univ} \\
\overline{r} & \defs & \Comprehension{}{(\sigma, \sigma')\in \universalrel \land (\sigma, \sigma') \notin r)}{(\sigma, \sigma')} \labeldef{rel-negation}\\
  \id{} & \defs & \Comprehension{}{\sigma = \sigma'}{(\sigma,\sigma')}  \labeldef{id}
\end{eqnarray}
When dealing with states over a set of variables,
if $X$ is a set of variables,
$\id{X}$ is the identity relation on just the variables in $X$ \refdef{id-X}.
For $\id{X}$, $\sigma$ and $\sigma'$ are mappings from variables to their values,
noting that mappings are a special case of binary relations,
and hence one can apply the domain restriction operator.
\begin{eqnarray}
  \id{X} & \defs & \Comprehension{}{\dom \sigma' = \dom \sigma \land X \dres \sigma = X \dres \sigma'}{(\sigma,\sigma')}  \labeldef{id-X} 
\end{eqnarray}

\subsection{Primitive commands}\labelsect{primtives}

Let $\Sigma$ be the (non-empty) program state space,
where a state $\sigma \in \Sigma$ gives the values of the program's variables.
Given that $r$ is a binary relation on states (i.e.\ $r \subseteq \Sigma \times \Sigma$) 
and $p$ is a set of states (i.e.\ $p \subseteq \Sigma$),
the primitive commands are defined as follows.
\begin{description}
\item[$\cpstep{r}$]
represents an \emph{atomic program step} command
that can perform the transition $\trans{\sigma}{\pstepd}{\sigma'}$ and terminate, 
if the two states are related by $r$ (i.e.~$(\sigma,\sigma') \in r$).
\item[$\cestep{r}$]
represents an \emph{atomic environment step} command
that can perform the transition $\trans{\sigma}{\estepd}{\sigma'}$ and terminate, 
if the two states are related by $r$.
\item[$\cgd{p}$]
represents an \emph{instantaneous test} command that succeeds and terminates immediately if its initial state is in $p$,
otherwise it is infeasible.
\end{description}
For example, $\cpstep{\id{}}$,
where $\id{}$ is the identity relation on states, 
represents a command that performs a (stuttering) program transition ($\trans{\sigma}{\pstepd}{\sigma}$) that does not change the state and terminates;
it differs from the command $\cgd{\Sigma}$, which terminates immediately without performing any program or environment transitions.

If $\cpstep{r}$ (or $\cestep{r}$) can make no transition for some state $\sigma$
(i.e. $\sigma$ is not in the domain of the relation $r$)
it is infeasible from that state;
in the semantic model the only trace of $\cpstep{r}$ with such an initial state $\sigma$ is the incomplete trace with no transitions.
Similarly, for an initial state $\sigma$ not in $p$, the only trace of $\tau{p}$ with initial state $\sigma$ has no transitions,
representing failure of the test.
That gives the following special cases: $\cpstep{\emptyset} = \cestep{\emptyset} = \cgd{\emptyset} = \Magic$,
where we use $\emptyset$ for both the empty set of states and the empty relation on states.

We define the atomic step command, $\cstep{r}$, that can perform either a program step or an environment step,
provided the step satisfies $r$ \refdef{cstep}.
Given that $\universalrel = \Sigma \times \Sigma$ is the universal relation on states,
the command $\cpstepd$ (note the bold font) can perform any program step \refdef{pgm-univ},
$\cestepd$ can perform any environment step \refdef{env-univ},
$\cstepd$ can perform any program or environment step \refdef{step-univ},
and
$\cgdd$ always succeeds and terminates immediately from any state \refdef{test-univ}.\\[-3ex]
\begin{minipage}[t]{0.5\textwidth}
\begin{eqnarray}
  \cstep{r} & \defs & \cpstep{r} \nondet \cestep{r}\labeldef{cstep} \\
  \cpstepd & \defs & \cpstep{\universalrel} \labeldef{pgm-univ} \\
  \cestepd & \defs & \cestep{\universalrel} \labeldef{env-univ}
\end{eqnarray}
\end{minipage}%
\begin{minipage}[t]{0.5\textwidth}
\begin{eqnarray}
  \cstepd & \defs & \cstep{\universalrel} \labeldef{step-univ} \\
  \cgdd & \defs & \cgd{\Sigma} \labeldef{test-univ}
\end{eqnarray}
\end{minipage}

\subsection{Axiomatisation and composite commands}\labelsect{composite}

Axioms of the core language are summarised in \reffig{axioms}, and
explained and explored in the coming sections. Throughout the paper we
introduce composite commands in terms of the primitives. For
convenience these are summarised in \reffig{commands}.

\begin{figure}\small
\begin{flushleft}
The set of all commands, $\mathcal{C}$, forms a completely distributive lattice \cite{DaveyPriestley02},
with least element $\Magic$, 
greatest element $\Abort$,
join ($\Nondet$) representing non-deterministic choice, 
and 
meet ($\Sconj$) representing strong conjunction of commands.
\\[1ex]
Sequential composition ($\Seq$)\\[-2ex]
\begin{minipage}[t]{0.5\textwidth}
\begin{align}
  c_1 \Seq (c_2 \Seq c_3) & = (c_1 \Seq c_2 ) \Seq c_3 \labelax{seq-assoc} \\
  c \Seq \Nil & = c = \Nil \Seq c  \labelax{seq-identity}
\end{align}
\end{minipage}%
\begin{minipage}[t]{0.5\textwidth}
\begin{align}
  (\Nondet C) \Seq d & = \Nondet_{c \in C} (c \Seq d)  \labelax{Nondet-seq-distrib-right} \\
  c \Seq (\Nondet D) & = \Nondet_{d \in D} (c \Seq d) \hspace{1em}\mbox{if $D \neq \emptyset$} \labelax{Nondet-seq-distrib-left}
\end{align}
\end{minipage}
\\[1ex]
Tests: the function $\cgd{}$ forms an isomorphism from the boolean algebra of sets of states to test commands ($\mathcal{T} \subseteq \mathcal{C}$).\\[-2ex]
\begin{minipage}[t]{0.5\textwidth}
\begin{eqnarray}
  \Nondet_{p \in P} (\cgd{p}) & = & \cgd{(\Union P)} \labelax{Nondet-test} \\
  \Sconj_{p \in P} (\cgd{p}) & = & \cgd{(\Inter P)} \hspace{1em}\mbox{if $P \neq \emptyset$} \labelax{Sconj-test} 
\end{eqnarray}
\end{minipage}%
\begin{minipage}[t]{0.5\textwidth}
\begin{eqnarray}
  \overline{\cgd{p}} & = & \cgd{\overline{p}} \labelax{negate-test} \\
  \cgd{p_1} \Seq \cgd{p_2} & = & \cgd{(p_1 \inter p_2)} \labelax{seq-test-test}
\end{eqnarray}
\end{minipage}
\\[1ex]
Atomic step commands: 
the functions $\cpstep{}$ and $\cestep{}$ form isomorphisms from the boolean algebra of binary relations to program, respectively, environment step commands.\\[-2ex]
\begin{minipage}[t]{0.5\textwidth}
\begin{eqnarray}
  \Nondet_{r \in R} (\cpstep{r}) & = & \cpstep{(\Union R)} \labelax{Nondet-pgm} \\
  \Sconj_{r \in R} (\cpstep{r}) & = & \cpstep{(\Inter R)} \hspace{1em}\mbox{if $R \neq \emptyset$} \labelax{Sconj-pgm} \\
  \cgd{p} \Seq \cpstep{r} & = & \cpstep{(p \dres r)} \labelax{seq-test-pgm} \\
  \cpstep{(r \rres p)} \Seq \cgd{p} & = & \cpstep{(r \rres p)} \labelax{seq-pgm-test} \\
  \overline{\cpstep{r_1} \nondet \cestep{r_2}} & = & \cpstep{\overline{r_1}} \nondet \cestep{\overline{r_2}} \labelax{negate-atomic}
\end{eqnarray}
\end{minipage}%
\begin{minipage}[t]{0.5\textwidth}
\begin{eqnarray}
  \Nondet_{r \in R} (\cestep{r}) & = & \cestep{(\Union R)} \labelax{Nondet-env} \\
  \Sconj_{r \in R} (\cestep{r}) & = & \cestep{(\Inter R)} \hspace{1em}\mbox{if $R \neq \emptyset$} \labelax{Sconj-env} \\
  \cgd{p} \Seq \cestep{r} & = & \cestep{(p \dres r)} \labelax{seq-test-env} \\
  \cestep{(r \rres p)} \Seq \cgd{p} & = & \cestep{(r \rres p)} \labelax{seq-env-test}
\end{eqnarray}
\end{minipage}\\[1ex]
Weak conjunction ($\together$)\\[-1ex]
\begin{minipage}{0.5\textwidth}
\begin{align}
  c \together \Chaos & = c = \Chaos \together c \labelax{conj-identity} \\
  c \together \Abort & = \Abort  \labelax{conj-abort} \\
  c \together c & = c \labelax{conj-idempotent}
\end{align}
\end{minipage}%
\begin{minipage}{0.5\textwidth}
\begin{align}
  \cpstep{r_1} \together \cpstep{r_2} & = \cpstep{(r_1 \inter r_2)} \labelax{conj-pgm-pgm} \\
  \cestep{r_1} \together \cestep{r_2} & = \cestep{(r_1 \inter r_2)} \labelax{conj-env-env} \\
  \cpstep{r_1} \together \cestep{r_2} & = \Magic \labelax{conj-pgm-env}
\end{align}
\end{minipage}
\\[1ex]\begin{minipage}{1\textwidth}
Parallel composition ($\parallel$)\\[-1ex]
\begin{minipage}{0.55\textwidth}
\begin{align}
  c \parallel \Skip & = c = \Skip \parallel c \labelax{par-identity} \\
  c \parallel \Abort & = \Abort  \labelax{par-abort} \\
  (c_0 \parallel d_0) \together (c_1 \parallel d_1) & \refsto (c_0 \together c_1) \parallel (d_0 \together d_1) \labelax{conj-par-interchange}
\end{align}
\end{minipage}%
\begin{minipage}{0.45\textwidth}
\begin{align}
  \cpstep{r_1} \parallel \cestep{r_2} & = \cpstep{(r_1 \inter r_2)} \labelax{par-pgm-env} \\
  \cestep{r_1} \parallel \cestep{r_2} & = \cestep{(r_1 \inter r_2)} \labelax{par-env-env} \\
  \cpstep{r_1} \parallel \cpstep{r_2} & = \Magic \labelax{par-pgm-pgm}
\end{align}
\end{minipage}\\[1ex]
Synchronisation: the following axioms hold for $\sync$ either $\parallel$, $\together$, or $\sconj$.
\begin{eqnarray}
   c_1 \sync (c_2 \sync c_3) & = & (c_1 \sync c_2) \sync c_3 \labelax{sync-assoc} \\
   c_1 \sync c_2 & = & c_2 \sync c_1 \labelax{sync-commutes} \\
  (\Nondet C) \sync d & = & \Nondet_{c \in C} (c \sync d)  \hspace*{4em}\mbox{if~} C \neq \emptyset \labelax{Nondet-sync-distrib} \\
  \ata_1 \Seq c_1 \sync \ata_2 \Seq c_2 & = & (\ata_1 \sync \ata_2) \Seq (c_1 \sync c_2) \labelax{sync-atomic-atomic} \\
\Inf{\ata_1} \sync \Inf{\ata_2} & = & \Inf{(\ata_1 \sync \ata_2)} \labelax{sync-inf-inf} \\
  \ata \Seq c \sync \Nil &=& \Magic           \labelax{sync-atomic-nil} \\
  t_1 \sync t_2 & = & t_1 \sconj t_2 \labelax{sync-test-test} \\
  t \Seq c_1 \sync t \Seq c_2 & = & t \Seq (c_1 \sync c_2)  \labelax{sync-test-distrib} \\
  (c_0 \Seq d_0) \sync (c_1 \Seq d_1) & \refsto & (c_0 \sync c_1) \Seq (d_0 \sync d_1) \labelax{sync-seq-interchange}
\end{eqnarray}
\end{minipage}
\end{flushleft}

\caption{Axioms of the core language. 
The naming conventions followed in this paper are:
 $c$ and $d$ are commands;
 $C$ and $D$ are sets of commands;
 $p$ is a set of states;
 $P$ is a set of sets of states;
 $r$, $g$ and $q$ are binary relations on states;
 $R$ is a set of relations;
 $\ata$ is an atomic step command (i.e.\ a command of the form $\cpstep{r_1} \nondet \cestep{r_2})$; 
 $t$ is a test (i.e.\ a command of the form $\cgd{p}$);
 and 
 subscripted forms of the above names follow the same conventions.
 }\labelfig{axioms}
\end{figure}

\begin{figure}
\begin{align}
  \Fin{c} & \defs  (\mu x \spot \cgdd \nondet c \Seq x)  \labeldef{fin-iter} \\
  \Om{c} & \defs  (\nu x \spot \cgdd \nondet c \Seq x) \labeldef{omega-iter} \\
  \Inf{c} & \defs  (\nu x \spot c \Seq x) \labeldef{inf-iter} \\
  \Pre{p} & \defs \Nil \nondet \cgd{\pnegate{p}} \Seq \Abort \labeldef{assert} \\
  \Skip & \defs \Om{\cestepd} \labeldef{skip} \\
  \Chaos & \defs \Om{\cstepd} \labeldef{chaos} \\
  \guar{g} & \defs \Om{(\cpstep{g} \nondet \cestepd)} \labeldef{guar} \\
  \Frame{X}{c} & \defs \guar{\id{\overline{X}}} \together c  \labeldef{frame} \\
  \rely{r} & \defs \Om{(\cstepd \nondet \cestep{\overline{r}} \Seq \Abort)} \labeldef{rely} \\
  \Term & \defs \Fin{\cstepd} \Seq \Om{\cestepd} \labeldef{term} \\
  \PSpec{}{}{q} & \defs \Nondet_{\sigma_0 \in \Sigma} (\cgd{\{\sigma_0\}} \Seq \Chaos \Seq \cgd{(q\limg\{\sigma_0\}\rimg)})  \labeldef{partial-spec} \\
  \Spec{}{}{q} & \defs \PSpec{}{}{q} \together \Term \labeldef{spec} \\
  \Opt{q} & \defs \cpstep{q} \nondet \cgd{(\dom{(q \inter \id{})})} \labeldef{opt} \\
  \Idle & \defs \guar{\id{}} \together \Term \labeldef{idle} \\
  \atomicrel{p,q} & \defs \Idle \Seq \Pre{p} \Seq \Opt{q} \Seq \Idle \labeldef{atomic-spec} \\
  \atomicrel{q} & \defs \atomicrel{\Sigma,q} \labeldef{atomic-spec-no-pre} \\
  \Update{x}{k} & \defs \id{\overline{x}} \rres (\EqEval{x}{k})  \labeldef{update} \\
  x := e & \defs \Nondet_{k \in Val} (\Test{e}_k \Seq \Opt{(\Update{x}{k})} \Seq \Idle) \labeldef{assign} \\
  \If b \Then c \Else d \Fi & \defs (\Test{b}_{\true} \Seq c \nondet \Test{b}_{\false} \Seq d) \Seq \Idle \nondet
    \Nondet_{k \in \overline{\Boolean}} (\Test{b}_{k} \Seq \Abort)  \labeldef{conditional} \\
  \While b \Do c \Od & \defs  \InfFP x \spot \If b \Then c \Seq x \Else \Nil \Fi \labeldef{while}
\end{align}

\caption{Commands defined in terms of primitives,
where the command $\Test{e}_k$ represents evaluating expression $e$ to value $k$ (see \refsect{expressions} for details).}\labelfig{commands}
\end{figure}

\section{Lattice of commands}\labelsect{algebra}

As a foundation of the axiomatisation in \reffig{axioms}, commands form a complete distributive lattice%
\ifmagictop \else%
\footnote{\label{footnote}
In the refinement calculus literature and our earlier papers~\cite{FM2016atomicSteps,FMJournalAtomicSteps} refinement is written $c \sqsubseteq d$ 
but in the program algebra literature (e.g.\ \cite{DBLP:journals/jlp/HoareMSW11}) the reverse ordering $c \refsto d$ is used.
In this paper we use the latter order ($\refsto$) and hence
$\Magic$ is the least element (rather than the greatest),
$\Abort$ as the greatest element (rather than the least),
$\nondet$ (rather than $\sqcap$) is nondeterministic choice,
least (rather than greatest) fixed points give finite iteration, 
and
greatest (rather than least) fixed points give possibly infinite iteration.
While the choice of ordering is arbitrary,
we feel our choice makes working with the algebra simpler because, for example,
finite iteration is now treated in the same way (as a least fixed point)
for both binary relations and commands,
the form of the operators on commands corresponds to those for sets and relations
(e.g.\ $\union$ maps to $\nondet$, and $\inter$ maps to $\sconj$, rather than to the inverted forms).
It also better matches the trace semantics~\cite{DaSMfaWSLwC} as $\refsto$ maps to $\supseteq$,
whereas in the previous work $\sqsubseteq$ mapped to $\supseteq$.
}
\fi
that is ordered by refinement,
$c \refsto d$, representing that $c$ is refined (or implemented) by $d$.
Nondeterministic choice ($\nondet$), or ``choice'' for short, is the lattice join (least upper bound)
and strong conjunction ($\sconj$) is the lattice meet (greatest lower bound).
The everywhere infeasible command $\Magic$ is the least element of the lattice,
and
the immediately aborting command $\Abort$ is the greatest element.
The following lemma allows refinement of a nondeterministic choice over a set $C$
by a choice over $D$,
provided every element of $D$ refines some element of $C$.
Important special cases are if either $C$ or $D$ is a singleton set. 
\begin{lemx*}[refine-choice]{\cite{BackWright98}}
For sets of commands $C$ and $D$,
\begin{align}
  \Nondet C & \refsto \Nondet D & \mbox{~~if~~} \forall d \in D \spot \exists c \in C \spot c \refsto d \\
  \Nondet C & \refsto d & \mbox{~~if~~} \exists c \in C \spot c \refsto d \\
  c &\refsto \Nondet D & \mbox{~~if~~} \forall d \in D \spot c \refsto d
\end{align}
\end{lemx*}

\section{Sequential composition}\labelsect{sequential}

Sequential composition is associative \refax{seq-assoc} and has identity the null command $\Nil$ \refax{seq-identity}
that terminates immediately.
Sequential composition distributes over nondeterministic choice from the right \refax{Nondet-seq-distrib-right} and
over a non-empty nondeterministic choice from the left \refax{Nondet-seq-distrib-left};
$D$ is required to be non-empty because $\Nondet \emptyset = \Magic$ but $\Abort \Seq \Magic = \Abort \neq \Magic$.
The binary versions \refprop{nondet-seq-distrib-right} and \refprop{nondet-seq-distrib-left} 
are derived from the fact that $c \nondet d \defs \Nondet \{c,d\}$.
\begin{align}
  (c_0 \nondet c_1) \Seq d & = c_0 \Seq d \nondet c_1 \Seq d  \labelprop{nondet-seq-distrib-right} \\
  c \Seq (d_0 \nondet d_1) & = c \Seq d_0 \nondet c \Seq d_1 \labelprop{nondet-seq-distrib-left} 
\end{align}

\section{Iteration}\labelsect{iteration}

In the context of a complete lattice, 
we have that least ($\mu$) and greatest ($\nu$) fixed points of monotone functions are well-defined.
Fixed points are used to define finite iteration zero or more times,
$\Fin{c} \defs (\mu x \spot \cgdd \nondet c \Seq x)$ \refdef{fin-iter},
possibly infinite iteration zero or more times,
$\Om{c} \defs (\nu x \spot \cgdd \nondet c \Seq x)$ \refdef{omega-iter},
and infinite iteration, $\Inf{c} \defs (\nu x \spot c \Seq x)$ \refdef{inf-iter}. 
Iteration operators have their usual unfolding (\refprop*{fiter-unfold-left}--\refprop*{iter-unfold}) 
and induction properties (\refprop*{fiter-induction-left}--\refprop*{iter-induction})~\cite{fixedpointcalculus1995}
derived from their definitions as fixed points.
Iteration satisfies the standard decomposition \refprop{decomposition} and isolation \refprop{isolation} properties.
\\[-3ex]
\begin{minipage}[t]{0.5\textwidth}
\begin{align}
   \Fin{c} & = \Nil \nondet c \Seq \Fin{c} \labelprop{fiter-unfold-left}\\
   \Fin{c} & = \Nil \nondet \Fin{c} \Seq c \labelprop{fiter-unfold-right} \\
   \Om{c} & = \Nil \nondet c \Seq \Om{c} \labelprop{iter-unfold} \\
   \Om{(c \nondet d)} & =  \Om{c} \Seq \Om{(d \Seq \Om{c})}  \labelprop{decomposition}
\end{align}
\end{minipage}%
\begin{minipage}[t]{0.5\textwidth}
\begin{align}
   x \refsto \Fin{c} \Seq d & \mbox{\hspace{1.5em}if~} x \refsto d \nondet c \Seq x \labelprop{fiter-induction-left} \\
   x \refsto d \Seq \Fin{c} & \mbox{\hspace{1.5em}if~} x \refsto d \nondet x \Seq c \labelprop{fiter-induction-right} \\   
   \Om{c} \Seq d \refsto x & \mbox{\hspace{1.5em}if~} d \nondet c \Seq x \refsto x \labelprop{iter-induction} \\
   \Om{c} \Seq d & = \Fin{c} \Seq d \nondet \Inf{c} \labelprop{isolation}
\end{align}
\end{minipage}
\\[1ex]
Note that all the above properties of finite iteration are also properties of finite iteration of relations $\Finrel{r}$, 
if $\cgdd$ is replaced by the identity relation $\id{}$ \refdef{id},
nondeterministic choice ($\nondet$) is replaced by union of relations ($\union$), 
sequential composition ($\,\Seq\,$) by relational composition ($\,\semi\,$) 
and 
$\refsto$ by $\supseteq$.
To avoid repeating the properties, 
we use the above properties of  finite iteration for both commands and relations
(with the above replacements made).
Both form Kleene algebras~\cite{Kleen56,conway71}.

\begin{lemx}[absorb-finite-iter]
If $c \refsto d$, then  $\Fin{c} \Seq \Fin{d} = \Fin{c}$.
\end{lemx}

\begin{proof}
The refinement from left to right holds because $\Fin{d} \refsto \Nil$.
For the refinement from right to left we have 
$\Fin{c} = \Fin{c} \Seq \Fin{c} \refsto \Fin{c} \Seq \Fin{d}$,
using the assumption $c \refsto d$ in the last step.
\end{proof}

\section{Tests}\labelsect{tests}

We identify a subset of commands $\mathcal{T}$ that represent instantaneous tests.
$\mathcal{T}$ forms a complete boolean algebra of commands 
(similar to Kozen's Kleene algebra with tests~\cite{kozen97kleene}).
For a state space $\Sigma$ representing the values of the program variables
and $p$ a subset of states ($p \subseteq \Sigma$),
the isomorphism $\tau \in \power \Sigma \fun \mathcal{T}$
maps $p$ to a distinct test $\cgd{p}$,
such that from initial state $\sigma$,
if $\sigma \in p$, $\cgd{p}$ terminates immediately (the null command)
but if $\sigma \not\in p$, $\cgd{p}$ is infeasible.
Every test can be written in the form $\cgd{p}$,
for some $p \subseteq \Sigma$.

The function $\tau$ forms an isomorphism between $\power \Sigma$ and $\mathcal{T}$
that maps set union to nondeterministic choice \refax{Nondet-test};
set intersection to the lattice \sconjname\ \refax{Sconj-test};
set complement to test negation \refax{negate-test};
and sequential composition of tests reduces to a test on the intersection of their sets of states \refax{seq-test-test}.
From these axioms one can deduce the following properties.
\\[-3ex]
\begin{minipage}[t]{0.5\textwidth}
\begin{eqnarray}
  \cgd{p_1} \nondet \cgd{p_2} & = & \cgd{(p_1 \union p_2)} \labelprop{nondet-test-test}  \\
  \cgd{p_1} \sconj \cgd{p_2} & = & \cgd{(p_1 \inter p_2)} \labelprop{conjoin-test-test}
\end{eqnarray}
\end{minipage}%
\begin{minipage}[t]{0.5\textwidth}
\begin{eqnarray}
  \cgd{p_1} \refsto \cgd{p_2} & \mbox{if} & \mbox{$p_1 \supseteq p_2$} \labelprop{test-strengthen} \\
  \Nil & \refsto & \cgd{p} \labelprop{test-intro}
\end{eqnarray}
\end{minipage}%
\\[1ex]\mbox{}%
Note that by \refprop{test-strengthen}, 
$\Nil = \cgd{\Sigma} \refsto \cgd{p} \refsto \cgd{\emptyset} = \Magic$.
Because $\Nil \refsto \cgd{p}$ for any $p$, 
it is a refinement to introduce a test \refprop{test-intro}.

\begin{exax}[test-seq]
Using the notation from \refsect{introduction} to represent sets of states by characteristic predicates,
\(
  \cgd{\Set{x \leq 0}} \Seq \cgd{\Set{x \geq 0}} 
  = \cgd{(\Set{x \leq 0} \inter \Set{x \geq 0})}
  = \cgd{\Set{x = 0}}.
\)
\end{exax}

A choice over a set of states $p$, of a test for a singleton set of states $\{\sigma\}$, succeeds for any state $\sigma$ in $p$ and hence is equivalent to $\cgd{p}$.
\begin{lemx}[Nondet-test-set]
$\Nondet_{\sigma \in p} (\cgd{\{\sigma\}}) = \cgd{p}$
\end{lemx}

\begin{proof}
Using \refax{Nondet-test},~
\(
  \Nondet_{\sigma \in p} (\cgd{\{\sigma\}})
 = 
  \cgd{(\Union_{\sigma \in p} \{ \sigma\})}
 = 
  \cgd{p} .
\)
\end{proof}
A test at the start of a non-deterministic choice restricts the range of the choice.
\begin{lemx}[test-restricts-Nondet]
\(
  \cgd{p} \Seq \Nondet_{\sigma \in \Sigma} (\cgd{\{\sigma\}} \Seq c) = \Nondet_{\sigma \in p} (\cgd{\{\sigma\}} \Seq c) .
\)
\end{lemx}

\begin{proof}
\begin{align*}
&  \cgd{p} \Seq \Nondet_{\sigma \in \Sigma} (\cgd{\{\sigma\}} \Seq c) 
 \Equals*[distribute test \refax{Nondet-seq-distrib-left} as $\Sigma$ is non-empty  and merge tests \refax{seq-test-test}]
  \Nondet_{\sigma \in \Sigma} (\cgd{(p \inter \{\sigma\})} \Seq c) 
 \Equals*[split choice using \reflem{refine-choice}]
  \Nondet_{\sigma \in p} (\cgd{(p \inter \{\sigma\})} \Seq c) \nondet \Nondet_{\sigma_1 \notin p} (\cgd{(p \inter \{\sigma_1\})} \Seq c)
 \Equals*[as $\sigma$ is in $p$ and $\sigma_1$ is not in $p$ and $\cgd{\emptyset} = \Magic$]
  \Nondet_{\sigma \in p} (\cgd{\{\sigma\}} \Seq c) \nondet \Nondet_{\sigma_1 \notin p} (\Magic \Seq c)
 \Equals*[as $\Nondet_{\sigma_1 \notin p} (\Magic \Seq c) = \Nondet_{\sigma_1 \notin p} \Magic = \Magic$ and $\Magic$ is the identity of $\nondet$]
  \Nondet_{\sigma \in p} (\cgd{\{\sigma\}} \Seq c) 
  \qedhere
\end{align*}
\end{proof}

\section{Assertions}\labelsect{assertions}

An \emph{assert} command, $\Pre{p} \defs \Nil \nondet \cgd{\overline{p}} \Seq \Abort$, aborts if $p$ does not hold (i.e.\ for states $\sigma \in \overline{p}$)
but otherwise terminates immediately \refdef{assert}.
It allows any behaviour whatsoever if the state does not satisfy $p$ 
\cite{Wright04}.%
\footnote{An alternative way to encode an assertion $\Pre{p}$ in our theory is as
$\Chaos \nondet \cgd{\overline{p}} \Seq \Abort$.
This version is combined with other commands
using weak conjunction rather than sequential composition.}
It satisfies the following.
\begin{align}
  \Pre{p}     & = \cgd{p} \nondet \cgd{\pnegate{p}} \Seq \Abort \labelprop{assert}
\end{align}

Weakening an assertion is a refinement \refprop{assert-weaken}.
Note that by \refprop{assert-weaken}, 
$\Abort =  \Pre{\emptyset} \refsto \Pre{p} \refsto \Pre{\Sigma} = \Nil$.
Because $\Pre{p} \refsto \Nil$ for any $p$ by \refprop{assert-remove}, 
it is a refinement to remove an assertion.
Tests and assertions satisfy a Galois connection~\cite{Wright04} \refprop{Galois-assert-test}.
Sequential composition of assertions is intersection on their sets of states~\refprop{seq-assert-assert}.
A test dominates an assertion on the same set of states \refprop{seq-test-assert}, 
and
an assertion dominates a test on the same set of states \refprop{seq-assert-test}.
\\[-3ex]
\begin{minipage}[t]{0.55\textwidth}
\begin{eqnarray}
  \Pre{p_1} \refsto \Pre{p_2} & \mbox{if} & p_1 \subseteq p_2  \labelprop{assert-weaken} \\
  \Pre{p} & \refsto & \Nil \labelprop{assert-remove} \\
  \Pre{p} \Seq c \refsto d & \iff & c \refsto \cgd{p} \Seq d \labelprop{Galois-assert-test}
\end{eqnarray}
\end{minipage}%
\begin{minipage}[t]{0.45\textwidth}
\begin{eqnarray}
  \Pre{p_1} \Seq \Pre{p_2} & = & \Pre{p_1 \inter p_2} \labelprop{seq-assert-assert} \\
  \cgd{p} \Seq \Pre{p} & = & \cgd{p}  \labelprop{seq-test-assert}  \\
  \Pre{p} \Seq \cgd{p} & = & \Pre{p}  \labelprop{seq-assert-test} 
\end{eqnarray}
\end{minipage}%
\\[1ex]\mbox{}%

\begin{lemx}[assert-merge]
If $\Pre{p_1} \Seq c \refsto d$ and $\Pre{p_2} \Seq c \refsto d$ then, $\Pre{p_1 \union p_2} \Seq c \refsto d$. 
\end{lemx}

\begin{proof}
Using the Galois connection between assertions and tests \refprop{Galois-assert-test},
the hypotheses are equivalent to $c \refsto \cgd{p_1} \Seq d$ and $c \refsto \cgd{p_2} \Seq d$
and hence by \reflem{refine-choice} 
$c \refsto  \cgd{p_1} \Seq d \nondet \cgd{p_2} \Seq d = (\cgd{p_1} \nondet \cgd{p_2}) \Seq d = \cgd{(p_1 \union p_2)} \Seq d$,
and hence by \refprop{Galois-assert-test}, $\Pre{p_1 \union p_2} \Seq c \refsto d$.
\end{proof}

\section{Atomic step commands}\labelsect{atomic}

We identify a subset of commands, $\mathcal{A}$, that represent atomic steps.%
\footnote{Our atomic step commands are at a similar level of granularity to 
the transitions in an operational semantics, such as that given by Coleman and Jones~\cite{CoJo07}.}
$\mathcal{A}$ forms a complete boolean algebra of commands.
Both $\pi$ and $\epsilon$ are \emph{injective} functions of type $\power(\Sigma \times \Sigma) \fun \mathcal{A}$,
so that distinct relations map to distinct atomic step commands,
and the commands generated by $\pstepd$ and $\estepd$ are distinct except that $\cpstep{\emptyset} = \cestep{\emptyset} = \Magic$.
Every atomic step command can be represented in the form $\cpstep{r_1} \nondet \cestep{r_2}$
for some relations $r_1$ and $r_2$. 
A choice over a set of program step commands
is equivalent to a program step command over the union of the relations \refax{Nondet-pgm}, and 
a strong conjunction over a set of relations corresponds to a program step command over the intersection of the relations \refax{Sconj-pgm}.
A test $\cgd{p}$ preceding a program step command $\cpstep{r}$ is equivalent to a program step command
with its relation restricted so its domain is included in $p$ \refax{seq-test-pgm}.
If a program step command with its relation restricted so that its range is in $p$ is followed by a test of $p$,
that test always succeeds and hence is redundant \refax{seq-pgm-test}.
Note that in general $\cpstep{r} \Seq \cgd{p}$ does not equal $\cpstep{(r \rres p)}$.
Environment step commands satisfy similar axioms (\refax*{Nondet-env}--\refax*{seq-env-test}).
Negating an atomic step command corresponds to 
negating the relations in its program and environment step components \refax{negate-atomic},
for example,
$\overline{\cpstep{r}} = \overline{\cpstep{r} \nondet \cestep{\emptyset}} =\cpstep{\overline{r}} \nondet \cestepd$,
and $\overline{\cpstepd} = \cestepd$.
If $r_1 \supseteq r_2$, both the following hold,
\\[-3ex]
\begin{minipage}[t]{0.5\textwidth}
\begin{eqnarray}
  \cpstep{r_1} & \refsto &\cpstep{r_2} \labelprop{pgm-refine}
\end{eqnarray}
\end{minipage}%
\begin{minipage}[t]{0.5\textwidth}
\begin{eqnarray}
  \cestep{r_1} & \refsto &\cestep{r_2} \labelprop{env-refine}
\end{eqnarray}
\end{minipage}%
\\[1ex]\mbox{}%
and hence by \refprop{pgm-refine} for any relation $r$, 
$\cpstepd = \cpstep{\universalrel} \refsto \cpstep{r} \refsto \cpstep{\emptyrel} = \Magic$,
where $\emptyrel$ is the empty relation,
and similarly by \refprop{env-refine},
$\cestepd = \cestep{\universalrel} \refsto \cestep{r} \refsto \cestep{\emptyrel} = \Magic$.
\begin{exax}[test-pgm-test]
By \refax{seq-test-pgm},
\[
  \cgd{\Set{0 < x}} \Seq \cpstep{\Rel{x \leq x'}}
 =
 \cpstep{(\Set{0 < x} \dres \Rel{x \leq x'})}
 =
 \cpstep{\Rel{0 < x \land x \leq x'}}.
\]
\end{exax}
The following properties follow from \refax{Nondet-pgm} and
\refax{Nondet-env}, 
respectively.\\[-1ex]
\begin{minipage}{0.5\textwidth}
\begin{eqnarray}
  \cpstep{r_1} \nondet \cpstep{r_2} & = & \cpstep{(r_1 \union r_2)} \labelax{nondet-pgm-pgm} 
\end{eqnarray}
\end{minipage}%
\begin{minipage}{0.5\textwidth}
\begin{eqnarray}
  \cestep{r_1} \nondet \cestep{r_2} & = & \cestep{(r_1 \union r_2)} \labelax{nondet-env-env} 
\end{eqnarray}
\end{minipage}\\[1ex]
Weak conjunction ($\together$) is a specification operator, 
such that $c \together d$ behaves as both $c$ and $d$ unless either $c$ or $d$ aborts
in which case $c \together d$ aborts.
The weak conjunction of two program step commands gives a program step over the intersection of their relations \refax{conj-pgm-pgm},
and similarly for environment step commands \refax{conj-env-env}.
A weak conjunction of a program step command with an environment step command is infeasible \refax{conj-pgm-env}.

Parallel composition combines a program step $\cpstep{r_1}$ of one thread 
with an environment step $\cestep{r_2}$ of the other
to form a program step of their composition 
that satisfies the intersection of the two relations \refax{par-pgm-env}.
Parallel combines environment steps of both threads to give an environment step of the composition 
corresponding to the intersection of their relations \refax{par-env-env}.
Because program steps of parallel threads are interleaved, 
parallel combination of two program steps is infeasible \refax{par-pgm-pgm}.
Weak conjunction and parallel satisfy an interchange axiom \refax{conj-par-interchange}.

\begin{exax}[program-parallel-environment]
By \refax{par-pgm-env},
a program step that does not increase $i$ in parallel with an environment step that does not decrease $i$
gives a program step that does not change $i$: 
\(
  \cpstep{\Rel{i \geq i'}} \parallel \cestep{\Rel{i \leq i'}} 
  = \cpstep{(\Rel{i \geq i'} \inter \Rel{i \leq i'})}
  = \cpstep{\Rel{i' = i}}.
\)
\end{exax}

The atomic step command $\cstepd$ is the atomic step identity of weak conjunction \refprop{conj-identity-atomic} and
the atomic step command $\cestepd$ is the atomic step identity of parallel composition \refprop{par-identity-atomic},
in which $\ata$ is any atomic step command (i.e.\ $\ata = \cpstep{g} \nondet \cestep{r}$ for some relations $g$ and $r$).\\[-1ex]
\begin{minipage}{0.5\textwidth}
\begin{eqnarray}
  \ata \together \cstepd & = & \ata \labelprop{conj-identity-atomic}
\end{eqnarray}
\end{minipage}%
\begin{minipage}{0.5\textwidth}
\begin{eqnarray}
  \ata \parallel \cestepd & = & \ata \labelprop{par-identity-atomic}
\end{eqnarray}
\end{minipage}

\section{Synchronisation operators: parallel and weak/strong conjunction}\labelsect{sync}

Parallel composition ($\parallel$) and both weak ($\together$) and strong ($\sconj$) conjunction satisfy similar laws;
the main differences being how they combine pairs of atomic steps,
compare \refax{par-pgm-env}--\refax{par-pgm-pgm} and \refax{conj-pgm-pgm}--\refax{conj-pgm-env};
whether or not they are abort-strict, like parallel \refax{par-abort} and weak conjunction \refax{conj-abort}; and whether or not they are idempotent, like weak conjunction \refax{conj-idempotent} and strong conjunction. 
To bring out the commonality between them we make use of an abstract synchronisation operator, $\sync$,
which is then instantiated to parallel ($\parallel$) and weak ($\together$) and strong ($\sconj$) conjunction~\cite{FMJournalAtomicSteps}.

How the synchronisation operators combine atomic steps, and how they interact with aborting behaviours, influences the identity of each.
Because the command $\cestepd$ \refdef{env-univ} is the identity of parallel for a single atomic step \refprop{par-identity-atomic} 
the command, $\Skip \defs \Om{\cestepd}$ \refdef{skip}, 
allows its environment to do any sequence of steps, including infinitely many, without itself introducing aborting behaviour,
and hence it is defined to be the identity of parallel composition \refax{par-identity}.
Because the command $\cstepd$ 
is the identity of weak conjunction for a single atomic step \refprop{conj-identity-atomic},
the command, $\Chaos \defs \Om{\cstepd}$ \refdef{chaos}, allows any number of 
any program or environment steps but cannot abort, 
and hence it is defined to be the identity of weak conjunction \refax{conj-identity}.
For commands $c$ and $d$ that refine the identity of weak conjunction, weak conjunction simplifies to conjunction~\cite{FMJournalAtomicSteps}:
\begin{eqnarray}
\Chaos \refsto c \land \Chaos \refsto d & \Rightarrow & c \sconj d = c \together d
\labelprop{together-simplified}
\end{eqnarray}
Strong conjunction has, from the lattice axiomatisation, identity $\Abort$. 

A synchronisation operator, $\sync$, is associative \refax{sync-assoc} and commutative \refax{sync-commutes}.
Non-empty non-deterministic choice distributes over a synchronisation operator \refax{Nondet-sync-distrib}.
We exclude the empty choice because $\Nondet \emptyset = \Magic$ 
but for parallel (and weak conjunction but not strong conjunction), 
$(\Nondet \emptyset) \parallel \Abort = \Abort \neq \Magic$.
Initial atomic steps of two commands synchronise before the remainders of the commands synchronise their behaviours \refax{sync-atomic-atomic}.
Infinite iterations of atomic steps synchronise each atomic step \refax{sync-inf-inf}.  
A command that must perform an atomic step cannot synchronise with a command that terminates immediately \refax{sync-atomic-nil}.
Two tests synchronise to give a test that succeeds if both tests succeed \refax{sync-test-test}.
A test distributes over synchronisation \refax{sync-test-distrib}.
Although synchronisation does not satisfy a distributive law with sequential,
it does satisfy the weak interchange axiom \refax{sync-seq-interchange}.
For \refax{sync-seq-interchange}, on the right the steps of $c_0$ synchronise with the steps of $c_1$
and they terminate together, and then the steps of $d_0$ synchronise with those of $d_1$.
That behaviour is also allowed on the left but $(c_0 \Seq d_0)$ synchronising all its steps with $(c_1 \Seq d_1)$
also allows behaviours such as $c_0$ synchronising with the whole of $c_1$ and part of $d_1$,
and $d_0$ synchronising with the rest of $d_1$, and vice versa
(see~\cite{DaSMfaWSLwC,AFfGRGRACP} for more details).
The following laws follow by the interchange axioms
\refax{sync-seq-interchange} and \refax{conj-par-interchange}, respectively.

\begin{lemx*}[sync-seq-distrib]{\cite[Law 6]{FMJournalAtomicSteps}}
If $c \refsto c \Seq c$, 
\[ 
  c \sync (d_0 \Seq d_1) \refsto (c \sync d_0) \Seq (c \sync d_1).
\]
\end{lemx*}

\begin{lemx*}[conj-par-distrib]{\cite[Law 12]{AFfGRGRACP}}
If $c \refsto c \parallel c$, 
\[
  c \together (d_0 \parallel d_1) \refsto (c \together d_0) \parallel (c \together d_1).
\]
\end{lemx*}%

The following laws are also derived from the axioms 
and properties of iterations.
They hold with $\sync$ replaced by any of $\parallel$, $\together$ and $\sconj$.
See~\cite{FMJournalAtomicSteps} for proofs of these properties
(and a range similar properties)
in terms of a synchronous program algebra.
\begin{eqnarray}
  \Om{\ata} \Seq c \sync t & = & c \sync t   \labelprop{test-omega} \\
  \Om{\ata_1} \sync \Om{\ata_2} & = & \Om{(\ata_1 \sync \ata_2)} \labelprop{sync-omega-omega} \\
  \Om{\ata_1} \Seq c_1 \sync \Om{\ata_2} \Seq c_2 & = & 
    \Om{(\ata_1 \sync \ata_2)} \Seq ((\Om{\ata_1} \Seq c_1 \sync c_2) \nondet (c_1 \sync \Om{\ata_2} \Seq c_2))
      \labelprop{sync-iter-iter} \\
  \Om{\ata_1} \Seq c_1 \sync \Fin{\ata_2} \Seq c_2 & = & 
    \Fin{(\ata_1 \sync \ata_2)} \Seq ((\Om{\ata_1} \Seq c_1 \sync c_2) \nondet (c_1 \sync \Fin{\ata_2} \Seq c_2))
      \labelprop{sync-iter-fiter}
\end{eqnarray}

\section{Abort-strict synchronisation operators}\labelsect{abort-strict}

Whether a synchronisation operator is abort strict or not influences its algebraic properties.
\begin{defix}[abort-strict]
A binary operator $\sync$ is \emph{abort strict} if and only if for all commands $c$,
$c \sync \Abort = \Abort$.
\end{defix}
Parallel composition \refax{par-abort} and weak conjunction \refax{conj-abort} are abort strict but strong conjunction is not.
The fact that parallel and weak conjunction are abort-strict influences how they
distribute tests and assertions. For example, for strong conjunction,
we trivially have that an initial test on one side of a conjunction
can be treated as an initial test of the whole synchronisation,
e.g. $c \sconj t \Seq d = t \Seq (c \sconj d)$. For either parallel or
weak conjunction we have, taking $c$ to be $\Abort$ and $t$ to be
$\Magic$ as an example, that this property does not hold:
$\Abort \parallel (t \Seq \Magic) = \Abort \neq t \Seq \Abort = t \Seq (\Abort \parallel \Magic)$.
For arbitrary synchronisation operators (including parallel and weak
conjunction), in \reflem{test-command-sync-command} we require that
the side without the test does not abort immediately, i.e.\ it must
either terminate immediately ($\cgdd$), or do a (non-aborting) step
($\cstepd$) and then any behaviour is allowed, including abort.  A
command $c$ is not immediately aborting if $c \together \Magic =
\Magic$.

\begin{lemx*}[test-command-sync-command]{\cite[Lemma 4]{FMJournalAtomicSteps}}
Given a test $t$, and commands $c$ and $d$,
if $\lnot t \Seq c \together \Magic = \Magic$,%
\footnote{This condition is a slight generalisation of that used in~\cite[Lemma 4]{FMJournalAtomicSteps}
but the proof there generalises straightforwardly with this more general proviso.}
then
\(
  c \sync t \Seq d = t \Seq (c \sync d).     
\)
\end{lemx*}

An initial assertion on one side of an abort-strict synchronisation operator 
can be treated as an initial assertion of the whole synchronisation.

\begin{lemx}[assert-distrib]
If $\sync$ is abort strict, 
$c \sync \Pre{p} \Seq d = \Pre{p} \Seq (c \sync d)$.
\end{lemx}
\begin{proof}
\begin{align*}
& c \sync \Pre{p} \Seq d
\Equals*[case analysis on test $t = \cgd{p}$ using $c = (t \nondet \pnegate{t}) \Seq c = t \Seq c \nondet \pnegate{t}\Seq c$]
\cgd{p} \Seq (c \sync \Pre{p} \Seq d)
\nondet
\cgd{\overline{p}} \Seq (c \sync \Pre{p} \Seq d)
\Equals*[distributivity of test over synchronisation \refax{sync-test-distrib}]
(\cgd{p} \Seq c \sync \cgd{p} \Seq \Pre{p} \Seq d)
\nondet
(\cgd{\overline{p}} \Seq c \sync \cgd{\overline{p}} \Seq \Pre{p} \Seq d)
\Equals*[simplify $\cgd{p} \Seq \Pre{p} = \cgd{p}$ and $\cgd{\overline{p}} \Seq \Pre{p} = \cgd{\overline{p}} \Seq \Abort$ by \refprop{assert}; redistribute test \refax{sync-test-distrib}]
\cgd{p} \Seq (c \sync d)
\nondet
\cgd{\overline{p}} \Seq (c \sync \Abort)
\Equals*[by \refdefi{abort-strict} as $\sync$ is abort strict]
\cgd{p} \Seq (c \sync d)
\nondet
\cgd{\overline{p}} \Seq \Abort
\Equals*[as $\Abort$ is a left annihilator of sequential composition]
\cgd{p} \Seq (c \sync d)
\nondet
\cgd{\overline{p}} \Seq \Abort \Seq (c \sync d)
\Equals*[distributivity of sequential composition \refprop{nondet-seq-distrib-right} and assertion property \refprop{assert}]
\Pre{p} \Seq (c \sync d)
\qedhere
\end{align*}
\end{proof}

Supporting Lemmas \ref{lem-sync-test-assert} to
\ref{lem-test-suffix-test} are used to prove
\reflem{test-suffix-interchange}, which states that tests at the end
of abort-strict synchronisations
can be factored out.

\begin{lemx}[sync-test-assert]
If $\sync$ is abort strict,
$\Nil \sync \Pre{p} = \Pre{p}$.
\end{lemx}

\begin{proof}
From \refax{seq-identity}, \reflem{assert-distrib} as $\sync$ is abort strict, and \refax{sync-test-test} we have that
$ \Nil \sync \Pre{p}
= \Nil \sync \Pre{p} \Seq \Nil
= \Pre{p} \Seq (\Nil \sync \Nil)
= \Pre{p}$.
\end{proof}

\begin{lemx}[test-suffix-assert]
If $\sync$ is abort strict,
\(
  c \sync d \Seq \cgd{p} = (c \sync d \Seq \cgd{p}) \Seq \Pre{p}.
\)
\end{lemx}

\begin{proof}
Refinement from right to left follows as $\Pre{p} \refsto \Nil$ by \refprop{assert-remove}.
The refinement from left to right follows because tests establish assertions \refprop{seq-test-assert}, 
interchanging $\sync$ with sequential composition \refax{sync-seq-interchange}
and \reflem{sync-test-assert} because $\sync$ is abort strict.
\[
  c \sync d \Seq \cgd{p} 
 = c \Seq \Nil \sync d \Seq \cgd{p} \Seq \Pre{p}
 \refsto (c \sync d \Seq \cgd{p}) \Seq (\Nil \sync \Pre{p})
 = (c \sync d \Seq \cgd{p}) \Seq \Pre{p}
 \qedhere
\]
\end{proof}

\begin{lemx}[test-suffix-test]
If $\sync$ is abort strict,
\(
  c \sync d \Seq \cgd{p} = (c \sync d \Seq \cgd{p}) \Seq \cgd{p}.
\)
\end{lemx}

\begin{proof}
The proof uses \reflem*{test-suffix-assert} given that $\sync$ is abort strict, then \refprop{seq-assert-test} 
and then \reflem*{test-suffix-assert} in the reverse direction:
\[
  c \sync d \Seq \cgd{p}
  = (c \sync d \Seq \cgd{p}) \Seq \Pre{p}
  = (c \sync d \Seq \cgd{p}) \Seq \Pre{p} \Seq \cgd{p}
  = (c \sync d \Seq \cgd{p}) \Seq \cgd{p}
  \qedhere
\]
\end{proof}

\begin{lemx}[test-suffix-interchange]
If $\sync$ is abort strict,
\(
  c \sync d \Seq \cgd{p} = (c \sync d) \Seq \cgd{p}.
\)
\end{lemx}

\begin{proof}
The refinement from left to right interchanges $\sync$ and sequential composition \refax{sync-seq-interchange} 
after adding a $\Nil$, and uses \refax{sync-test-test} to show 
$\Nil \sync \cgd{p} = \cgd{\Sigma} \sconj \cgd{p} = \cgd{(\Sigma \inter p)} = \cgd{p}$.
\[
  c \sync d \Seq \cgd{p} 
  = c \Seq \Nil \sync d \Seq \cgd{p}
  \refsto (c \sync d) \Seq (\Nil \sync \cgd{p})
  = (c \sync d) \Seq \cgd{p}
\]
The refinement from right to left adds a test by \refprop{test-intro} and then uses \reflem{test-suffix-test}, given that $\sync$ is abort strict.
\[
  (c \sync d) \Seq \cgd{p} 
  \refsto (c \sync d \Seq \cgd{p}) \Seq \cgd{p}
  = c \sync d \Seq \cgd{p}
  \qedhere
\]
\end{proof}

\section{Guarantees}\labelsect{guarantees}

A command $c$ satisfies a guarantee condition $g$, where $g$ is a binary relation on states,
if every program step of $c$ satisfies $g$ \cite{Jones81d,Jones83a,Jones83b}.
The command, $\guar{g} \defs \Om{(\cpstep{g} \nondet \cestepd)}$, is the most general command 
that satisfies the guarantee relation $g$ for every program step and puts no constraints on its environment \refdef{guar}.
The command, $\guar{g} \together c$, behaves as both $\guar{g}$ and as $c$,
unless at some point $c$ aborts,
in which case $\guar{g} \together c$ aborts;
note that $\guar{g}$ cannot abort.

The term \emph{law} is used for properties that are likely to be used in developing programs,
while \emph{lemma} is used for supporting properties used within proofs.
To make it easier to locate laws/lemmas, they share a single numbering sequence.
If a lemma/law has been proven elsewhere,
a citation to the relevant publication follows the name of the lemma/law.

A guarantee command, $\guar{g_0}$ ensures all program steps satisfy the relation $g_0$.
For relation $g_1$ such that $g_0 \supseteq g_1$, the command $\guar{g_1}$ ensures all program steps satisfy $g_1$
and hence every program step also satisfies $g_0$; hence $\guar{g_0}$ is refined by $\guar{g_1}$.
\begin{lawx*}[guar-strengthen]{\cite[Lemma 23]{FMJournalAtomicSteps}}
If $g_1 \supseteq g_2$, $\guar{g_1} \refsto \guar{g_2}$.
\end{lawx*}
Weak conjoining a guarantee to a command $c$ constrains its behaviour so that all program steps satisfy the guarantee,
and hence is a refinement.
\begin{lawx}[guar-introduce]
$c \refsto \guar{g} \together c$.
\end{lawx}

\begin{proof}
The command $\Chaos$ is the identity of weak conjunction \refax{conj-identity},
and $\Chaos$ corresponds to a guarantee of the universal relation ($\universalrel$),
and hence using \reflaw{guar-strengthen}:
\[
  c = \Chaos \together c = \guar{\universalrel} \together c \refsto \guar{g} \together c. \qedhere
\]
\end{proof}
Two guarantee commands weakly conjoined together ensure both relations $g_1$ and $g_2$ are satisfied by every program step,
i.e.\ their intersection is satisfied by all program steps.
\begin{lawx*}[guar-merge]{\cite[Lemma 24]{FMJournalAtomicSteps}}
$\guar{g_1} \together \guar{g_2} = \guar{(g_1 \inter g_2)}$
\end{lawx*}

Two guarantees in parallel produce program steps that satisfy either guarantee.
\begin{lemx}[par-guar-guar]
$\guar{g_1} \parallel \guar{g_2} = \guar{(g_1 \union g_2)}$
\end{lemx}

\begin{proof}
\begin{align*}
&  \guar{g_1} \parallel \guar{g_2}
 \Equals*[using the definition of a guarantee \refdef{guar}]
  \Om{(\cpstep{g_1} \nondet \cestepd)} \parallel \Om{(\cpstep{g_2} \nondet \cestepd)}
 \Equals*[by \refprop{sync-omega-omega}]
  \Om{((\cpstep{g_1} \nondet \cestepd) \parallel (\cpstep{g_2} \nondet \cestepd))}
 \Equals*[distributing \refax{Nondet-sync-distrib} twice and using \refax{par-pgm-pgm}, \refax{par-env-env} and \refax{par-pgm-env}]
  \Om{(\cpstep{(g_1 \union g_2)} \nondet \cestepd)}
 \Equals*[by definition of a guarantee \refdef{guar}]
  \guar{(g_1 \union g_2)}
  \qedhere
\end{align*}
\end{proof}
A guarantee command weakly conjoined with a sequential composition $(c \Seq d)$ 
ensures all program steps of both $c$ and $d$ satisfy the guarantee,
and similarly for a guarantee weakly conjoined with a parallel composition $(c \parallel d)$.
\ihin{Stronger equality lemma would be better}
\begin{lawx}[guar-seq-distrib]
\(
  \guar{g} \together (c \Seq d) \refsto (\guar{g} \together c) \Seq (\guar{g} \together d)
\)
\end{lawx}

\begin{proof}
The proof follows by \reflem{sync-seq-distrib} for $\sync$ weak conjunction
because from definition \refdef{guar}, a guarantee is of the form $\Om{c}$, and 
$\Om{c} = \Om{c} \Seq \Om{c}$ for any $c$.
\end{proof}

\begin{lawx}[guar-par-distrib]
\(
  \guar{g} \together (c \parallel d) \refsto (\guar{g} \together c) \parallel (\guar{g} \together d) 
\)
\end{lawx}

\begin{proof}
The proof follows from \reflem{conj-par-distrib} 
using \reflem{par-guar-guar} to show its proviso: $\guar{g} = \guar{g} \parallel \guar{g}$.
\end{proof}
A guarantee combined with a test reduces to the test.
\begin{lawx}[guar-test]
$\guar{g} \together \cgd{p} = \cgd{p}$
\end{lawx}

\begin{proof}
The proof follows from the definition of a guarantee as an iteration \refdef{guar} 
using \refprop{test-omega} and \refax{sync-test-test}:
\(
  \guar{g} \together \cgd{p}
 =
  \Om{(\cpstep{g} \nondet \cestepd)} \Seq \cgdd \together \cgd{p}
 =
  \cgdd \together \cgd{p}
 =
  \cgd{p} .
\)
\end{proof}

Guarantees combine with program steps to enforce the guarantee for the step.
\begin{lawx}[guar-pgm]
$\guar{g} \together \cpstep{r} = \cpstep{(g \inter r)}$
\end{lawx}

\begin{proof}
The proof follows from the definition of a guarantee as an iteration \refdef{guar}, by unfolding the iteration \refprop{iter-unfold}, distributing and eliminating infeasible choices, and conjoining program steps \refax{conj-pgm-pgm}:
\(
  \guar{g} \together \cpstep{r}
=
  ((\cpstep{g} \nondet \cestepd) \Seq \Om{(\cpstep{g} \nondet \cestepd)} \nondet \Nil)
  \together \cpstep{r}
=
  \cpstep{g} \together \cpstep{r}
=
  \cpstep{(g \inter r)} .
\)
\end{proof}

An assertion $\Pre{p}$ satisfies any guarantee 
because it makes no program steps at all
unless it aborts, in which case the conjunction aborts.
\begin{lawx}[guar-assert]
\(
  \guar{g} \together \Pre{p} = \Pre{p}
\)
\end{lawx}

\begin{proof}
The proof uses \reflem{assert-distrib} and \reflaw{guar-test}.
\begin{align*}&
  \guar{g} \together \Pre{p}
 =
  \guar{g} \together \Pre{p} \Seq \Nil
 = 
  \Pre{p} \Seq (\guar{g} \together \Nil) 
 = 
  \Pre{p}
  \qedhere
\end{align*}
\end{proof}

\section{Frames}\labelsect{frames}

A frame $X$ is a set of variables that a command $c$ may modify.
To restrict a command $c$ to only modify variables in the set $X$,
the command, $\Frame{X}{c} \defs \guar{\id{\overline{X}}} \together c$, is introduced \refdef{frame}.
It is defined using a guarantee that all variables other than $X$, i.e.\ $\overline{X}$,
are not changed by any program steps.
Recall that for a set of variables $X$, 
$\id{\overline{X}}$ is the identity relation on all variables other than $X$; see \refdef{id-X}.
The binary operator ``$:$'' for frames has the highest precedence,
in particular, it has higher precedence than sequential composition.
Because frames are defined in terms of guarantees, 
they may be distributed over operators using \refax{Nondet-sync-distrib} for nondeterministic choice,
\reflaw{guar-seq-distrib}, \reflaw{guar-par-distrib} and 
the fact that weak conjunction is associative, commutative and idempotent.
\begin{lawx}[distribute-frame]
$\Frame{X}{(c \Seq d)} \refsto \Frame{X}{c} \Seq \Frame{X}{d}$
\end{lawx}

\begin{proof}
The proof follows from the definition of a frame \refdef{frame} and \reflaw*{guar-seq-distrib}. 
\end{proof}
Reducing the frame of a command corresponds to strengthening its guarantee.
\begin{lawx}[frame-reduce]
For sets of identifiers $X$ and $Y$,
$\Frame{(X \union Y)}{c}  \refsto \Frame{Y}{c}$.
\end{lawx}

\begin{proof}
Expanding both sides using \refdef{frame} the proof follows by \reflaw{guar-strengthen}
because $\id{\overline{X \union Y}} \supseteq \id{\overline{Y}}$
as $\overline{X \union Y} \subseteq \overline{Y}$, i.e.\ $X \union Y \supseteq Y$.
\end{proof}

\section{Relies}\labelsect{relies}

In the rely/guarantee approach, 
the allowable interference on a thread $c$ is represented by a rely condition, a 
relation $r$ that is assumed to hold for any 
atomic step taken by the environment of $c$ \cite{Jones81d,Jones83a,Jones83b}.
A rely condition is an \emph{assumption} that every step of the environment satisfies $r$.
The command, $\rely{r} \together c$, is required to behave as $c$,
unless the environment makes a step not satisfying $r$,
in which case it allows any behaviour from that point (i.e.\ it aborts).
The command, $\rely{r} \defs \Om{(\cstepd \nondet \cestep{\overline{r}} \Seq \Abort)}$, 
allows any program or environment steps 
(i.e.\ $\cstepd$ steps)
but if the environment makes a step not satisfying $r$ 
(i.e.\ a step of $\cestep{\overline{r}})$
it aborts \refdef{rely}.
A rely command, $\rely{r_1}$, is satisfied by its environment (technically, it does not abort) if all environment steps satisfy relation $r_1$.
If all environment steps satisfy $r_1$, then for a relation $r_2$ that contains $r_1$,
all environment steps will also satisfy $r_2$, and hence $\rely{r_2}$ is a refinement of $\rely{r_1}$.
\begin{lawx*}[rely-weaken]{\cite[Lemma 25]{FMJournalAtomicSteps}}
If $r_1 \subseteq r_2$, then $\rely{r_1} \refsto \rely{r_2}$.
\end{lawx*}
\noindent
The ultimate weakening is to the universal relation $\universalrel$, which removes the rely altogether.
\begin{lawx}[rely-remove]
$\rely{r} \together c \refsto c$.
\end{lawx}

\begin{proof}
Using \reflaw{rely-weaken} and noting that $\Chaos$ is the identity of weak conjunction \refax{conj-identity} 
and
$\cestep{\overline{\universalrel}} = \cestep{\emptyset} = \Magic$.
\[
  \rely{r} \together c 
 \refsto 
  \rely{\universalrel} \together c 
 =
  \Om{(\cstepd \nondet \cestep{\emptyset} \Seq \Abort)} \together c
 =
  \Om{\cstepd} \together c
 =
  \Chaos \together c 
 =
  c
  \qedhere
\]
\end{proof}
A rely of $r_1$ assumes all environment steps satisfy $r_1$, and
a rely of $r_2$ assumes all environment steps satisfy $r_2$,
and hence their weak conjunction corresponds to assuming
all environment steps satisfy both $r_1$ and $r_2$, i.e.\ $r_1 \inter r_2$.
\begin{lawx*}[rely-merge]{\cite[Lemma 26]{FMJournalAtomicSteps}}
\(
  \rely{r_1} \together \rely{r_2} = \rely{(r_1 \inter r_2)}
\)
\end{lawx*}

Rely conditions may be distributed into a sequential composition.
\ihin{Stronger equality lemma would be better}
\begin{lawx}[rely-seq-distrib]
\(
  \rely{r} \together (c \Seq d) \refsto (\rely{r} \together c) \Seq (\rely{r} \together d).
\)
\end{lawx}

\begin{proof}
The proof follows from \reflem{sync-seq-distrib} with $\sync$ weak conjunction,
where the lemma's proviso that $\rely{r} \refsto \rely{r} \Seq \rely{r}$ follows from the definition \refdef{rely}
and the property of iterations that $\Om{c} = \Om{c} \Seq \Om{c}$ for any $c$.
\end{proof}

A sequential composition within the context of a rely can be refined
by refining one of its components in the context of the rely.
\begin{lawx}[rely-refine-within]
If $\rely{r} \together c_1 \refsto \rely{r} \together d$,~
\[
  \rely{r} \together c_0 \Seq c_1 \Seq c_2 \refsto \rely{r} \together c_0 \Seq d \Seq c_2.
\]
\end{lawx}%

\begin{proof}
\begin{align*}
&  \rely{r} \together c_0 \Seq c_1 \Seq c_2
 \Refsto*[duplicate rely as $\together$ is idempotent and apply \reflaw{rely-seq-distrib} twice]
  \rely{r} \together (\rely{r} \together c_0) \Seq (\rely{r} \together c_1) \Seq (\rely{r} \together c_2) 
 \Refsto*[from the assumption and using \reflaw{rely-remove} to remove three relies]
  \rely{r} \together c_0 \Seq d \Seq c_2 
  \qedhere
\end{align*}
\end{proof}

In the parallel composition $\rely{r} \parallel \guar{r}$,
the guarantee on the right does not break the rely assumption on the left,
but as the rely command allows any behaviour, including program steps that satisfy $r$,
its behaviour subsumes that which can be generated by the guarantee,
and hence their parallel combination reduces to the rely.
\begin{lawx*}[rely-par-guar]{\cite[Lemma 27]{FMJournalAtomicSteps}}
\(
  \rely{r} \parallel \guar{r} = \rely{r}
\)
\end{lawx*}

Relies and guarantees often appear conjoined together;
\reflem*{conj-rely-guar} provides an expansion of their conjunction useful in a later proof.
\begin{lemx}[conj-rely-guar]
$\rely{r} \together \guar{g} = \Om{(\cpstep{g} \nondet \cestep{r})} \Seq (\Nil \nondet \cestep{\overline{r}} \Seq \Abort)$.
\end{lemx}

\begin{proof}
\begin{align*}
&  \rely{r} \together \guar{g}
 \Equals*[from definitions of rely \refdef{rely} and guarantee \refdef{guar}]
  \Om{(\cpstepd \nondet \cestep{r} \nondet \cestep{\overline{r}} \Seq \Abort)} 
    \together \Om{(\cpstep{g} \nondet \cestepd)}
 \Equals*[decomposition property of iterations \refprop{decomposition}: $\Om{(c \nondet d)} = \Om{c} \Seq \Om{(d \Seq \Om{c})}$]
   \Om{(\cpstepd \nondet \cestep{r})} \Seq
     \Om{(\cestep{\overline{r}} \Seq \Abort \Seq \Om{(\cpstepd \nondet \cestep{r})})}
    \together \Om{(\cpstep{g} \nondet \cestepd)}
 \Equals*[as $\Abort$ is a left annihilator]
   \Om{(\cpstepd \nondet \cestep{r})} \Seq
     \Om{(\cestep{\overline{r}} \Seq \Abort)}
    \together \Om{(\cpstep{g} \nondet \cestepd)}
\end{align*}
\begin{align*}&
 \Equals*[unfold iteration \refprop{iter-unfold} and $\Abort$ is an annihilator]
  \Om{(\cpstepd \nondet \cestep{r})} \Seq (\Nil \nondet \cestep{\overline{r}} \Seq \Abort)
    \together \Om{(\cpstep{g} \nondet \cestepd)} \Seq \Nil
 \Equals*[by \refprop{sync-iter-iter} as $(\cpstepd \nondet \cestep{r}) \together (\cpstep{g} \nondet \cestepd) = \cpstep{g} \nondet \cestep{r}$]
  \Om{(\cpstep{g} \nondet \cestep{r})} \Seq 
   \left(\begin{array}{l}
    (\Om{(\cpstepd \nondet \cestep{r})} \Seq (\Nil \nondet \cestep{\overline{r}} \Seq \Abort) \together \Nil) \nondet {} \\
    ((\Nil \nondet \cestep{\overline{r}} \Seq \Abort) \together \Om{(\cpstep{g} \nondet \cestepd)})
   \end{array}\right)
 \Equals*[unfolding iterations \refprop{iter-unfold} and simplifying using \refprop{test-omega} and \refax{sync-atomic-atomic}]
  \Om{(\cpstep{g} \nondet \cestep{r})} \Seq 
  (\Nil \nondet \cestep{\overline{r}} \Seq \Abort)
   \qedhere
\end{align*}
\end{proof}
A rely conjoined with a parallel composition, $\rely{r} \together (c \parallel d)$, 
represents an assumption that every environment step of the whole parallel composition satisfies $r$
but environment steps of $c$ are either environment steps of the whole composition (assumed to satisfy $r$)
or program steps of $d$, which do not necessarily satisfy $r$,
but will if one imposes a guarantee on $d$.
\begin{lawx}[rely-par-distrib]
\[
  \rely{r} \together (c \parallel d) \refsto (\rely{r} \together \guar{r} \together c) \parallel (\rely{r} \together \guar{r} \together d)
\]
\end{lawx}%

\begin{proof}
\begin{align*}
&  \rely{r} \together (c \parallel d)
\Equals*[duplicate the rely condition as $\together$ is idempotent; \reflaw{rely-par-guar}]
   (\rely{r} \parallel \guar{r}) \together (\guar{r} \parallel \rely{r}) \together (c \parallel d)
 \Refsto*[interchanging $\together$ and $\parallel$ by \refax{conj-par-interchange}]
 ((\rely{r} \together \guar{r}) \parallel (\rely{r} \together \guar{r})) \together (c \parallel d)
  \Refsto*[interchanging $\together$ and $\parallel$ by \refax{conj-par-interchange}]
  (\rely{r} \together \guar{r} \together c) \parallel (\rely{r} \together \guar{r} \together d)
  \qedhere
\end{align*}
\end{proof}

\section{Termination}\labelsect{termination}

A command is considered to terminate if it performs only a finite number of program steps.
However, that does not preclude a terminating command being pre-empted by its environment forever.
The command, $\Term \defs \Fin{\cstepd} \Seq \Om{\cestepd}$, can perform only a finite number of program steps
but it does not constrain its environment \refdef{term}.
At first sight it may appear odd that a terminating command allows infinite behaviours
but it should be emphasised that the infinite behaviour ends in an infinite sequence
of environment steps, i.e.\ the thread is never scheduled from some point onwards.
To avoid such pre-emption, fair execution can be incorporated;
the reader is referred to~\cite{FM2018fairness} for a treatment of fairness in our approach.
The command $\Term$ satisfies the following properties.

\begin{lawx}[seq-term-term]
~~~$\Term \Seq \Term = \Term$
\end{lawx}

\begin{proof}
We start by expanding the definition of $\Term$ \refdef{term} on the left.
\begin{align*}
&  \Fin{\cstepd} \Seq \Om{\cestepd} \Seq \Fin{\cstepd} \Seq \Om{\cestepd}
 \Equals*[expand $\Om{\cestepd} \Seq \Fin{\cstepd} \Seq \Om{\cestepd}$ using $\Om{c} \Seq d = \Fin{c} \Seq d \nondet \Inf{c}$ \refprop{isolation}]
  \Fin{\cstepd} \Seq (\Fin{\cestepd} \Seq \Fin{\cstepd} \Seq \Om{\cestepd} \nondet \Inf{\cestepd})
 \Equals*[distributing \refprop{nondet-seq-distrib-left}]
  \Fin{\cstepd} \Seq \Fin{\cestepd} \Seq \Fin{\cstepd} \Seq \Om{\cestepd} \nondet \Fin{\cstepd} \Seq \Inf{\cestepd}
 \Equals*[as $\cstepd \refsto \cestepd$ by \reflem{absorb-finite-iter} $\Fin{\cstepd} \Seq \Fin{\cestepd} = \Fin{\cstepd}$ and $\Fin{c} \Seq \Fin{c} = \Fin{c}$]
  \Fin{\cstepd} \Seq \Om{\cestepd} \nondet \Fin{\cstepd} \Seq \Inf{\cestepd}
 \Equals*[by \refprop{nondet-seq-distrib-left} and $\Om{\cestepd} \nondet \Inf{\cestepd} = \Om{\cestepd}$ by \refprop{isolation}]
  \Fin{\cstepd} \Seq \Om{\cestepd}
 \Equals*[by the definition of $\Term$ \refdef{term}]
  \Term
  \qedhere
\end{align*}
\end{proof}

Parallel composition of two terminating commands terminates.
\begin{lawx*}[par-term-term]{\cite[Lemma 20]{FMJournalAtomicSteps}}
~~~$\Term \parallel \Term = \Term$
\end{lawx*}

\section{Partial and total correctness}\labelsect{partial-total}

In Hoare logic~\cite{Hoare69a} the triple, $\{ p_1\}\, c \;\{ p_2\}$, represents the partial correctness assertion
that if command $c$ is started in a state satisfying predicate $p_1$ and $c$ terminates, then the state on termination satisfies $p_2$.
A total correctness interpretation of the triple requires, in addition, that $c$ terminates from initial states satisfying $p_1$.
Our algebraic characterisations of partial and total correctness are influenced by our ability to express and differentiate terminating, non-terminating and aborting program behaviours.
We use weak correctness assertions, as introduced by von~Wright~\cite{Wright04}, as a basis for both.

The \emph{weak correctness} of the Hoare triple $\{ p_1\}\, c \;\{ p_2\}$
corresponds to either of the following two equivalent algebraic conditions.
\begin{eqnarray}
  \cgd{p_1} \Seq c \Seq \cgd{p_2} & = & \cgd{p_1} \Seq c   \labeldef{weak-correctness1} \\
  c \Seq \cgd{p_2} & \refsto & \cgd{p_1} \Seq c   \labeldef{weak-correctness2}
\end{eqnarray}
A weak correctness assertion $\{ p_1\}\, c \;\{ p_2\}$ is not necessarily preserved by refinement, e.g. if  $c \refsto d$, then it does not follow that $\{ p_1\}\, d \;\{ p_2\}$ is also weakly correct, because the assertion permits $c$ to abort from initial states in which $p_1$ holds, and $\Abort$ may be refined by any possible behaviour, including behaviours that terminate in states violating $p_2$. 
Given that a program that aborts from initial state $p_1$ provides no guarantees about its behaviour after failure (e.g. it may terminate in a state that does not satisfy $p_2$), our definition of \emph{partial correctness} adds to the definition of weak correctness the requirement that $c$ does not abort from states satisfying $p_1$, which can be formulated in either of the two equivalent (by \refprop{Galois-assert-test}) forms,  \\[-1ex]
\begin{minipage}{0.5\textwidth}
\begin{eqnarray}
  \Pre{p_1} \Seq \Chaos & \refsto & c  \labeldef{partial-correctness1} 
\end{eqnarray}
\end{minipage}%
\begin{minipage}{0.5\textwidth}
\begin{eqnarray}
  \Chaos & \refsto & \cgd{p_1} \Seq c  \labeldef{partial-correctness2} 
\end{eqnarray}
\end{minipage} \\[1ex]
\emph{Total correctness} adds the even stronger requirement that $c$ terminates from states satisfying $p_1$, i.e. either of the following equivalent properties holds.\footnote{Our termination requirement for total correctness differs from von Wright~\cite{Wright04}, 
who defines the total correctness assertion $\{ p_1\}\, c \;\{ p_2\}$ to hold when $\cgd{p_1} \Seq c \Seq \cgd{\overline{p_2}} = \Magic$. 
This is because in von Wright's sequential theory an (everywhere) terminating command $c$ satisfies, $c \Seq \Magic = \Magic$,
but this does not hold in our theory, e.g. we do not have that $\Term \Seq \Magic = \Magic$.} \\[-1ex]
\begin{minipage}{0.5\textwidth}
\begin{eqnarray}
  \Pre{p_1} \Seq \Term & \refsto & c  \labeldef{total-correctness1}
\end{eqnarray}
\end{minipage}%
\begin{minipage}{0.5\textwidth}
\begin{eqnarray}
  \Term & \refsto & \cgd{p_1} \Seq c  \labeldef{total-correctness2} 
\end{eqnarray}
\end{minipage} \\[1ex]
Unlike weak correctness assertions, both partial and total correctness assertions are preserved by refinement.
Rather than a single-state postcondition, Coleman and Jones~\cite{CoJo07} make use of a relational postcondition, $q$,
for their quintuple verification rules, giving rise to the following notion of a command $c$ 
being weakly correct with respect to a relation $q$,
in which for a state $\sigma$, $q \limg \{ \sigma \} \rimg$ is the relational image of the singleton set $\{\sigma\}$ through $q$ \refdef{image}.
\begin{defix}[weakly-correct]
A command $c$ is \emph{weakly correct} with respect to a relation $q$ if and only if
\[
  c = \Nondet_{\sigma\in \Sigma} (\cgd{\{\sigma\}} \Seq c \Seq \cgd{(q \limg \{ \sigma \} \rimg)}) .
\]
\end{defix}%

\begin{lemx}[weakly-correct]
To show that command $c$ is weakly correct with respect to relation $q$ it is enough to show that 
$c \Seq \cgd{(q \limg \{ \sigma \} \rimg)} \refsto \cgd{\{\sigma\}} \Seq c$ for all $\sigma \in \Sigma$.
\end{lemx}

\begin{proof}
By \reflem{Nondet-test-set}, $c = \Nondet_{\sigma \in \Sigma} \cgd{\{\sigma\}} \Seq c$, 
and hence it is sufficient to show that for all $\sigma$,
$\cgd{\{\sigma\}} \Seq c = \cgd{\{\sigma\}} \Seq c \Seq \cgd{(q \limg \{ \sigma \} \rimg)}$. 
This refinement holds from left to right because it is just adding a test \refprop{test-intro},
and refinement from right to left holds by the assumption that 
$c \Seq \cgd{(q \limg \{ \sigma \} \rimg)} \refsto \cgd{\{\sigma\}} \Seq c$ and \refax{seq-test-test}.
\end{proof}

We can also define partial and total correctness of a command $c$ with respect to a relation $q$.
\begin{defix}[partially-correct]
  A command $c$ is \emph{partially correct} with respect to a relation $q$ if and only if it is weakly correct with respect to $q$ and $\Chaos \refsto c$.
\end{defix}%
\begin{defix}[totally-correct]
  A command $d$ is \emph{totally correct} with respect to a relation $q$ if and only if it is weakly correct with respect to $q$ and $\Term \refsto c$.
\end{defix}%

Having a theory that supports all three notions of correctness is advantageous. First, in many cases proving either the absence of catastrophic failure or termination is straightforward and hence proofs of either partial or total correctness, respectively, can be simplified by focusing on weak correctness first. In addition, some concurrent algorithms (e.g.\ spin lock) do not guarantee termination, and so require a partial correctness specification, instead of a total-correctness one. 
In the remainder of this section we present lemmas useful for establishing that commands satisfy weak-correctness assertions (remembering that proofs for partial and total correctness can be decomposed).

A refinement of the form
$c \Seq \cgd{p_1} \refsto \cgd{p_0} \Seq c$
asserts the weak correctness condition that when $c$ is started in a state in $p_0$, if $c$ terminates normally (i.e. not as a consequence of failure), 
the termination state is in $p_1$.
So-called \emph{commutativity conditions}
of this form allow a post-state test of a sequential composition to be 
replaced by progressively earlier tests,
e.g.\ if $c_2 \Seq \cgd{p_2} \refsto \cgd{p_1} \Seq c_2$ and $c_1 \Seq \cgd{p_1} \refsto \cgd{p_0} \Seq c_1$ then, 
$c_1 \Seq c_2 \Seq \cgd{p_2} \refsto c_1 \Seq \cgd{p_1} \Seq c_2 \refsto \cgd{p_0} \Seq c_1 \Seq c_2$.
An important special case is when the test corresponds to an invariant, i.e. $p_0 = p_1$,
because rules of this form can be applied to iterations.

A single program or environment step, $\cpstep{r}$ or $\cestep{r}$, 
establishes postcondition $p_1$ if started in a state satisfying $p_0$,
if the image \refdef{image} of $p_0$ under $r$ is in $p_1$.
\begin{lemx}[atomic-test-commute]
If $r \limg p_0 \rimg \subseteq p_1$, then both the following hold.
\begin{eqnarray}
  \cpstep{r} \Seq \cgd{p_1} & \refsto & \cgd{p_0} \Seq \cpstep{r}  \labelprop{pgm-test-commute} \\
  \cestep{r} \Seq \cgd{p_1} & \refsto & \cgd{p_0} \Seq \cestep{r}  \labelprop{env-test-commute}
\end{eqnarray}
\end{lemx}

\begin{proof}
The assumption is equivalent to $p_0 \dres r \rres p_1 = p_0 \dres r$.
The proof of \refprop{pgm-test-commute} uses 
\refprop{pgm-refine}, \refax{seq-pgm-test} and \refax{seq-test-pgm}. 
\[
    \cpstep{r} \Seq \cgd{p_1}
  ~\refsto~
    \cpstep{(p_0 \dres r \rres p_1)} \Seq \cgd{p_1}
  ~=~
    \cpstep{(p_0 \dres r \rres p_1)}
  ~=~ 
    \cpstep{(p_0 \dres r)}
  ~=~
    \cgd{p_0} \Seq \cpstep{r}
\]
The proof for \refprop{env-test-commute} is similar but uses 
\refprop{env-refine}, \refax{seq-env-test} and \refax{seq-test-env}.
\end{proof}

\begin{lemx}[nondet-test-commute]
For tests $t_0$ and $t_1$, and commands $c$ and $d$, 
if $c \Seq t_1 \refsto t_0 \Seq c$ and $d \Seq t_1 \refsto t_0 \Seq d$,
\begin{eqnarray*}
  (c \nondet d) \Seq t_1 & \refsto & t_0 \Seq (c \nondet d) .
\end{eqnarray*}
\end{lemx}

\begin{proof}
From both assumptions 
$(c \nondet d) \Seq t_1 = c \Seq t_1 \nondet d \Seq t_1 \refsto t_0 \Seq c \nondet t_0 \Seq d = t_0 \Seq (c \nondet d)$.
\end{proof}

\begin{lemx}[iteration-test-commute]
For any test $t$ and command $c$,
if $c \Seq t \refsto t \Seq c$, then both the following hold.
\begin{eqnarray}
  \Om{c} \Seq t \refsto t \Seq \Om{c} \labelprop{iter-test-commute} \\
  \Fin{c} \Seq t \refsto t \Seq \Fin{c} \labelprop{fiter-test-commute}
\end{eqnarray}
\end{lemx}

\begin{proof}
\refprop{iter-test-commute} holds by $\omega$-induction \refprop{iter-induction} if
$t \nondet c \Seq t \Seq \Om{c} ~\refsto~t \Seq \Om{c}$,
which is proven using the assumption and $\omega$-folding \refprop{iter-unfold}:
\(
    t \nondet c \Seq t \Seq \Om{c}
  ~\refsto~
    t \nondet t \Seq c \Seq \Om{c}  
  ~=~
    t \Seq (\Nil \nondet c \Seq \Om{c})
  ~=~
    t \Seq \Om{c}.
\)
\refprop{fiter-test-commute} holds by $\star$-induction \refprop{fiter-induction-right} if
$\Fin{c} \Seq t ~~\refsto~~ t \nondet \Fin{c} \Seq t \Seq c$,
which is proven using the assumption and $\star$-folding \refprop{fiter-unfold-right}:
\(
    t \nondet \Fin{c} \Seq t \Seq c
  ~\refines~
    t \nondet \Fin{c} \Seq c \Seq t
  ~=~
    (\Nil \nondet \Fin{c} \Seq c) \Seq t
  ~=~
    \Fin{c} \Seq t.
\)
\end{proof}

\section{Specification commands}\labelsect{specifications}

One can define a partial (correctness) specification command, 
\[
  \PSpec{}{}{q} \defs \Nondet_{\sigma_0 \in \Sigma} (\cgd{\{\sigma_0\}} \Seq \Chaos \Seq \cgd{(q \limg \{ \sigma_0 \} \rimg)}) .
\]
It requires that if started in a state $\sigma_0$, 
then if it terminates, its final state is related to $\sigma_0$ by $q$, 
i.e.\ it is in the relational image of $\{\sigma_0\}$ through $q$ \refdef{partial-spec}.
A total (correctness) specification command, $\Spec{}{}{q} \defs \PSpec{}{}{q} \together \Term$, requires termination \refdef{spec}.%
\footnote{A partial specification command $\PSpec{}{}{q}$ is the greatest command that is partially correct with respect to $q$ 
(from \reflem{partially-correct}), 
and similarly a total correctness specification $\Spec{}{}{q}$ is the greatest command that is totally correct with respect to $q$ 
(from \reflem{totally-correct}). 
Note that a weak specification command defined in this way would be uninteresting, 
because the greatest command that is weakly correct with respect to any $q$ is $\Abort$.}

\begin{RelatedWork}
The semantics of Brookes~\cite{Brookes-full-abstraction} makes use of a ``stuttering'' equivalence relation on commands
that considers two commands equivalent if their sets of traces are equal 
when all finite sequences of stuttering steps are removed from every trace of both.
Because the focus of this paper is on refining from a specification,
an alternative approach is used whereby specification commands implicitly allow finite stuttering,
i.e.\ they are closed under finite stuttering.

Brookes also makes use of ``mumbling'' equivalence
that allows two consecutive program steps $(\cpstep{r_1} \Seq \cpstep{r_2})$ 
to be replaced by a single program step $\cpstep{(r_1 \semi r_2)}$ with the same overall effect,
where ``$\semi$'' is relational composition \refdef{composition}.
Again, specification commands implicitly allow all mumbling equivalent implementations.
Our postcondition specification command \refdef{spec} is defined in such a way 
that if a specification $\Post{q}$ refines to a command $c$, 
and $d$ is semantically equivalent to $c$ modulo finite stuttering and mumbling,
then $\Post{q}$ is also refined by $d$.
In general, it does not require that $c$ and $d$ are refinement equivalent.
\end{RelatedWork}

The following lemma allows a command $c$ synchronised using an abort strict operator $\sync$ to be distributed into a choice 
that resembles the structure used in a specification command \refdef{partial-spec} when $d$ is $\Chaos$.
\begin{lemx}[sync-distribute-relation]
If $\sync$ is abort strict,
\[
  c \sync \Nondet_{\sigma \in \Sigma} (\cgd{\{\sigma\}} \Seq d \Seq \cgd{(q \limg \{ \sigma \} \rimg)}) =  \Nondet_{\sigma \in \Sigma} (\cgd{\{\sigma\}} \Seq (c \sync d) \Seq \cgd{(q \limg \{ \sigma \} \rimg)})
\]
\end{lemx}%

\begin{proof}
The application of \reflem*{test-suffix-interchange} in the last step requires that $\sync$ is abort strict.
\begin{align*}
&  c \sync \Nondet_{\sigma \in \Sigma} (\cgd{\{\sigma\}} \Seq d \Seq \cgd{(q \limg \{ \sigma \} \rimg)})
 \Equals*[by \reflem{Nondet-test-set}]
  \Nondet_{\sigma_1 \in \Sigma} (\cgd{\{\sigma_1\}} \Seq (c \sync \Nondet_{\sigma \in \Sigma} (\cgd{\{\sigma\}} \Seq d \Seq \cgd{(q \limg \{ \sigma \} \rimg)})))
 \Equals*[distribute test \refax{sync-test-distrib} and \reflem{test-restricts-Nondet}]
  \Nondet_{\sigma_1 \in \Sigma} (\cgd{\{\sigma_1\}} \Seq c \sync \Nondet_{\sigma \in \{\sigma_1\}} (\cgd{\{\sigma\}} \Seq d \Seq \cgd{(q \limg \{ \sigma \} \rimg)}))
 \Equals*[as $\sigma_1$ is the only choice for $\sigma$]
  \Nondet_{\sigma_1 \in \Sigma} (\cgd{\{\sigma_1\}} \Seq c \sync \cgd{\{\sigma_1\}} \Seq d \Seq \cgd{(q \limg \{ \sigma_1 \} \rimg)})
 \Equals*[distribute test \refax{sync-test-distrib} in reverse and \reflem{test-suffix-interchange}] 
  \Nondet_{\sigma_1 \in \Sigma} (\cgd{\{\sigma_1\}} \Seq (c \sync d) \Seq \cgd{(q \limg \{ \sigma_1 \} \rimg)})
  \qedhere
\end{align*}
\end{proof}

\begin{lemx}[spec-distribute-sync]
If $\sync$ is abort strict, 
$c \sync (d \together \PSpec{}{}{q}) = (c \sync d) \together \PSpec{}{}{q}$.
\end{lemx}

\begin{proof}
Note that $\Chaos$ is the identity of $\together$, and 
$\together$ is abort strict as required for the first and last applications of \reflem{sync-distribute-relation}.
\begin{align*}&
  c \sync (d \together \PSpec{}{}{q})
 \Equals*[by definition \refdef{partial-spec} and \reflem{sync-distribute-relation} for $\together$]
  c \sync \Nondet_{\sigma \in \Sigma} (\cgd{\{\sigma\}} \Seq d \Seq \cgd{(q \limg \{ \sigma \} \rimg)})
 \Equals*[by \reflem{sync-distribute-relation} as $\sync$ is abort strict]
  \Nondet_{\sigma \in \Sigma} (\cgd{\{\sigma\}} \Seq (c \sync d ) \Seq \cgd{(q \limg \{ \sigma \} \rimg)})
 \Equals*[by \reflem{sync-distribute-relation} for $\together$ (in reverse)]
  (c \sync d ) \together \Nondet_{\sigma \in \Sigma} (\cgd{\{\sigma\}} \Seq \Chaos \Seq \cgd{(q \limg \{ \sigma \} \rimg)})
\end{align*}
\begin{align*}&
 \Equals*[by the definition of a specification \refdef{partial-spec}]
  (c \sync d) \together \PSpec{}{}{q}
  \qedhere
\end{align*}
\end{proof}

A partial specification with postcondition of the universal relation only guarantees not to abort (i.e.\ $\Chaos$)
and a total specification only guarantees to terminate (i.e.\ $\Term$).
\begin{lemx}[spec-univ]
Both $\PSpec{}{}{\universalrel} = \Chaos$ and 
$\Spec{}{}{\universalrel} = \Term$.
\end{lemx}

\begin{proof}
The proof expands the definition of a partial specification command \refdef{partial-spec},
uses the fact that $\universalrel\limg\{\sigma_0\} \rimg = \Sigma$ and 
applies \refax{Nondet-seq-distrib-right} and \reflem{Nondet-test-set}.
\[
  \PSpec{}{}{\universalrel}
 = 
  \Nondet_{\sigma_0 \in \Sigma} (\cgd{\{\sigma_0\}} \Seq \Chaos \Seq \cgd{(\universalrel\limg\{\sigma_0\} \rimg)})
 = 
  (\Nondet_{\sigma_0 \in \Sigma} \cgd{\{\sigma_0\}}) \Seq \Chaos \Seq \Nil
 = 
  \Chaos
\]
For total correctness, $\Spec{}{}{\universalrel} = \PSpec{}{}{\universalrel} \together \Term = \Chaos \together \Term = \Term$. 
\end{proof}

\begin{lawx}[spec-strengthen]
If $q_1 \supseteq q_2$, then both $\PSpec{}{}{q_1} \refsto \PSpec{}{}{q_2}$ and $\Post{q_1} \refsto \Post{q_2}$.
\end{lawx}

\begin{proof}
The proof follows directly from the definition of either specification command because if $q_1 \supseteq q_2$, then
$\cgd{(q_1\limg\{\sigma_0\}\rimg)} \refsto \cgd{(q_2\limg\{\sigma_0\}\rimg)}$ by \refprop{test-strengthen}. 
\end{proof}

\begin{lemx}[spec-introduce]
Both
\( 
  \Chaos \refsto \PSpec{}{}{q}
\)
and
\( 
  \Term \refsto \Post{q}.
\)
\end{lemx}

\begin{proof}
Using \reflem*{spec-univ} and \reflaw{spec-strengthen},
\(
  \Chaos = \PSpec{}{}{\universalrel} \refsto \PSpec{}{}{q}
\)
and
\(
\Term = \Spec{}{}{\universalrel} \refsto \Spec{}{}{q}.
\)
\end{proof}

The definitions of weak correctness, partial correctness and total correctness can be reformulated using the specification commands. 

\begin{lemx}[weakly-correct-spec]
A command $c$ is weakly correct with respect to a relation $q$ if and only if $c \together \PSpec{}{}{q} = c$.
\end{lemx}

\begin{proof}
The proof reduces the equality $c \together \PSpec{}{}{q} = c$ to \refdefi{weakly-correct}.
\begin{align*}
&  c \together \PSpec{}{}{q} = c
 \IFF*[by \refdef{partial-spec}]
  c \together \Nondet_{\sigma \in \Sigma} (\cgd{\{\sigma\}} \Seq \Chaos \Seq \cgd{(q \limg \{ \sigma \} \rimg)}) = c
 \IFF*[by \reflem{sync-distribute-relation} and $\Chaos$ is the identity of $\together$]
  \Nondet_{\sigma \in \Sigma} (\cgd{\{\sigma\}} \Seq c \Seq \cgd{(q \limg \{ \sigma \} \rimg)}) = c
  \qedhere
\end{align*}
\end{proof}

\begin{lemx}[partially-correct]
A command $c$ is partially correct with respect to a relation $q$ if and only if $\PSpec{}{}{q} \refsto c$.
\end{lemx}

\begin{proof}
By \refdefi{partially-correct}, 
$c$ is \emph{partially correct} with respect to a relation $q$ if and only if it is weakly correct with respect to $q$ and $\Chaos \refsto c$,
or by \reflem{weakly-correct-spec} if $\PSpec{}{}{q} \together c = c \mbox{ and } \Chaos \refsto c$.
\begin{align*}&
  \PSpec{}{}{q} \together c = c \mbox{ and } \Chaos \refsto c
 \IFF*[by \reflem{spec-introduce} and \refprop{together-simplified}]
  \PSpec{}{}{q} \land c = c \mbox{ and } \Chaos \refsto c
 \IFF*[lattice property: $c_1 \land c_2 = c_2$ if and only if $c_1 \refsto c_2$ for any $c_1$ and $c_2$]
\end{align*}
\begin{align*}&
  \PSpec{}{}{q} \refsto c \mbox{ and } \Chaos \refsto c
 \IFF*[by \reflem{spec-introduce}]
  \PSpec{}{}{q} \refsto c
 \qedhere
\end{align*}
\end{proof}
  
\begin{lemx}[totally-correct] 
A command $c$ is totally correct with respect to a relation $q$ if and only if $\Spec{}{}{q} \refsto c$.
\end{lemx}

\begin{proof}
Because $\Chaos \refsto \Term$ and $\Term \refsto \Spec{}{}{q}$ (from \reflem{spec-introduce}), using \refprop{together-simplified} we have that
\begin{align*}
& (\Spec{}{}{q} \refsto c)
\Leftrightarrow 
  (\PSpec{}{}{q} \together \Term \refsto c)
\Leftrightarrow
  (\PSpec{}{}{q} \sconj \Term \refsto c)
\Leftrightarrow 
(\PSpec{}{}{q} \refsto c \mbox{ and } \Term \refsto c)
\end{align*}
which is true from \reflem{partially-correct} if and only if $c$ is weakly correct with respect to $q$, $\Chaos \refsto c$ and $\Term \refsto c$. 
Because $\Chaos \refsto \Term$ this is equivalent to \refdefi{totally-correct}.
\end{proof}

Because specifications are defined in terms of tests, 
laws that combine specifications with tests are useful for manipulating specifications.
These laws have corollaries that show how specifications combine with assertions.
Recall that $p \dres q$ is the relation $q$ with its domain restricted to $p$ \refdef{dom-restrict}.
Total specification commands ensure termination.
Below we give proofs of only the partial specification command properties of the lemmas;
the proofs of total specification commands just add $\Term$ on each side.

\begin{lemx}[test-restricts-spec]
Both
\(
  \cgd{p} \Seq \PSpec{}{}{p \dres q} = \cgd{p} \Seq \PSpec{}{}{q}
\)
and
\(
  \cgd{p} \Seq \Spec{}{}{p \dres q} = \cgd{p} \Seq \Spec{}{}{q}.
\)
\end{lemx}

\begin{proof}
\begin{align*}
&  \cgd{p} \Seq \PSpec{}{}{q}
 \Equals*[definition of a  specification \refdef{partial-spec} and \reflem{test-restricts-Nondet}]
  \Nondet_{\sigma_0 \in p} (\cgd{\{\sigma_0\}} \Seq \Chaos \Seq \cgd{(q\limg\{\sigma_0\}\rimg)})
 \Equals*[as $\forall \sigma_0 \in p \spot q \limg \{\sigma_0\}\rimg = (p \dres q)\limg \{\sigma_0\}\rimg$]
  \Nondet_{\sigma_0 \in p} (\cgd{\{\sigma_0\}} \Seq \Chaos \Seq \cgd{((p \dres q)\limg\{\sigma_0\}\rimg)})
 \Equals*[by \reflem{test-restricts-Nondet} and definition of a specification \refdef{partial-spec}]
  \cgd{p} \Seq \PSpec{}{}{p \dres q}
  \qedhere
\end{align*}
\end{proof}

\begin{lemx}[assert-restricts-spec]
$\Pre{p} \Seq \PSpec{}{}{p \dres q} = \Pre{p} \Seq \PSpec{}{}{q}$
and
$\Pre{p} \Seq \Post{p \dres q} = \Pre{p} \Seq \Post{q}$.
\end{lemx}

\begin{proof}
The proof applies \reflem*{test-restricts-spec} using the fact that $\Pre{p} \Seq \cgd{p} = \Pre{p}$ by \refprop{seq-assert-test}.
\[
  \Pre{p} \Seq \PSpec{}{}{p \dres q}
 =
  \Pre{p} \Seq \cgd{p} \Seq \PSpec{}{}{p \dres q}
 =
  \Pre{p} \Seq \cgd{p} \Seq \PSpec{}{}{q}
 =
  \Pre{p} \Seq \PSpec{}{}{q}  
  \qedhere
\]
\end{proof}%
Frames are included in the following law to make it more useful in practice.
\begin{lawx}[spec-strengthen-under-pre]
Let $X$ be a set of variables,
if $p \dres q_2 \subseteq q_1$ both, 
\(
  \Pre{p} \Seq \PSpec{X}{}{q_1} \refsto \Pre{p} \Seq \PSpec{X}{}{q_2}, 
\)
and
\(
  \Pre{p} \Seq \Spec{X}{}{q_1} \refsto \Pre{p} \Seq \Spec{X}{}{q_2}.
\)
\end{lawx}

\begin{proof}
\begin{align*}
&  \Pre{p} \Seq \PSpec{X}{}{q_1} 
 \Equals*[by the definition of a frame \refdef{frame} and \reflem{assert-distrib}]
  (\Pre{p} \Seq \PSpec{}{}{q_1}) \together \guar{\id{\overline{X}}}
 \Refsto*[by \reflaw{spec-strengthen} using assumption $p \dres q_2 \subseteq q_1$]
  (\Pre{p} \Seq \PSpec{}{}{p \dres q_2}) \together \guar{\id{\overline{X}}}
 \Equals*[by \reflem{assert-restricts-spec}]
  (\Pre{p} \Seq \PSpec{}{}{q_2}) \together \guar{\id{\overline{X}}}
 \Equals*[by \reflem{assert-distrib} and the definition of a frame \refdef{frame}]
  \Pre{p} \Seq \PSpec{X}{}{q_2}
  \qedhere
\end{align*}
\end{proof}

A test can be used to restrict the final state of a specification.
Recall that $q \rres p$ is the relation $q$ with its range restricted to $p$ \refdef{range-restrict}.
\begin{lemx}[spec-test-restricts]
Both
\(
  \PSpec{}{}{q} \Seq \cgd{p} = \PSpec{}{}{q \rres p}
\)
and
\(
  \Spec{}{}{q} \Seq \cgd{p} = \Spec{}{}{q \rres p}.
\)
\end{lemx}

\begin{proof}
\begin{align*}
&  \PSpec{}{}{q} \Seq \cgd{p}
 \Equals*[definition of a specification \refdef{partial-spec} and distribute test \refax{Nondet-seq-distrib-right}]
  \Nondet_{\sigma_0 \in \Sigma} (\cgd{\{\sigma_0\}} \Seq \Chaos \Seq \cgd{(q\limg\{\sigma_0\}\rimg)} \Seq \cgd{p})
 \Equals*[merging tests \refax{seq-test-test} and $q\limg\{\sigma_0\}\rimg \inter p = (q \rres p)\limg\{\sigma_0\}\rimg$]
  \Nondet_{\sigma_0 \in \Sigma} (\cgd{\{\sigma_0\}} \Seq \Chaos \Seq \cgd{((q \rres p)\limg\{\sigma_0\}\rimg)} )
 \Equals*[definition of a specification \refdef{partial-spec}]
  \PSpec{}{}{q \rres p}
  \qedhere
\end{align*}
\end{proof}

\begin{lemx}[spec-assert-restricts]
$\PSpec{}{}{q \rres p} \Seq \Pre{p} = \PSpec{}{}{q \rres p}$
and
$\Post{q \rres p} \Seq \Pre{p} = \Post{q \rres p}$.
\end{lemx}

\begin{proof}
The proof applies \reflem*{spec-test-restricts} using the fact that $\cgd{p} \Seq \Pre{p} = \cgd{p}$ by \refprop{seq-test-assert}.
\[
  \PSpec{}{}{q \rres p} \Seq \Pre{p} 
 = 
  \PSpec{}{}{q} \Seq \cgd{p} \Seq \Pre{p}
 =
  \PSpec{}{}{q} \Seq \cgd{p}
 =
  \PSpec{}{}{q \rres p} 
  \qedhere
\]
\end{proof}%
A specification command $\Spec{}{}{q}$ achieves a postcondition of $q\limg p \rimg$ 
from any initial state in $p$.
\begin{lemx}[spec-test-commute]
$\PSpec{}{}{q} \Seq \cgd{(q \limg p \rimg)} \refsto \cgd{p} \Seq \PSpec{}{}{q}$
and
$\Spec{}{}{q} \Seq \cgd{(q \limg p \rimg)} \refsto \cgd{p} \Seq \Spec{}{}{q}$.
\end{lemx}

\begin{proof}
\begin{align*}
&   \PSpec{}{}{q} \Seq \cgd{(q \limg p \rimg)}
  \Equals*[by \reflem{spec-test-restricts}]
   \PSpec{}{}{q \rres (q \limg p \rimg)}
  \Refsto*[by \reflaw{spec-strengthen} as $p \dres q \subseteq q \rres (q \limg p \rimg)$]
   \PSpec{}{}{p \dres q}
  \Refsto*[introducing $\cgd{p}$ \refprop{test-intro} and \reflem{test-restricts-spec}]
   \cgd{p} \Seq \PSpec{}{}{q}
   \qedhere
\end{align*}
\end{proof}

A specification with a post condition that is the composition \refdef{composition} of two relations $q_1$ and $q_2$
may be refined by a sequential composition of one specification command satisfying $q_1$ 
and a second satisfying $q_2$.
\begin{lawx}[spec-to-sequential]
Both
\(
  \PSpec{}{}{q_1 \semi q_2} \refsto \PSpec{}{}{q_1} \Seq \PSpec{}{}{q_2}
\)
and
\(
  \Spec{}{}{q_1 \semi q_2} \refsto \Spec{}{}{q_1} \Seq \Spec{}{}{q_2}.
\)
\end{lawx}

\begin{proof}
From relational algebra, $(q_1 \semi q_2)\limg p \rimg = q_2 \limg q_1 \limg p \rimg \rimg$. This allows the proof to use
two applications of \reflem{spec-test-commute} to show that $\PSpec{}{}{q_1} \Seq \PSpec{}{}{q_2}$ establishes post-condition $(q_1 \semi q_2)\limg \{\sigma_0\} \rimg$ from initial state $\sigma_0$, if it terminates. 
\begin{align*}
&  \PSpec{}{}{q_1 \semi q_2}
 \Equals*[definition of a specification command \refdef{partial-spec} and $(q_1 \semi q_2)\limg \{ \sigma_0 \} \rimg = q_2 \limg q_1 \limg  \{ \sigma_0 \} \rimg \rimg$]
  \Nondet_{\sigma_0 \in \Sigma} (\cgd{\{\sigma_0\}} \Seq \Chaos \Seq \cgd{(q_2 \limg q_1 \limg \{ \sigma_0 \} \rimg \rimg)})
 \Refsto*[as $\Chaos = \Chaos \Seq \Chaos$ and \reflem{spec-introduce} twice]
  \Nondet_{\sigma_0 \in \Sigma} (\cgd{\{\sigma_0\}} \Seq \PSpec{}{}{q_1} \Seq \PSpec{}{}{q_2} \Seq \cgd{(q_2 \limg q_1 \limg \{ \sigma_0 \} \rimg \rimg)})
 \Refsto*[by \reflem{spec-test-commute}]
  \Nondet_{\sigma_0 \in \Sigma} (\cgd{\{\sigma_0\}} \Seq \PSpec{}{}{q_1} \Seq \cgd{(q_1 \limg \{ \sigma_0 \} \rimg)} \Seq \PSpec{}{}{q_2})
\Refsto*[by \reflem{spec-test-commute}]
  \Nondet_{\sigma_0 \in \Sigma} (\cgd{\{\sigma_0\}} \Seq \cgd{\{ \sigma_0 \}}) \Seq \PSpec{}{}{q_1}\Seq \PSpec{}{}{q_2}
 \Equals*[merging tests \refax{seq-test-test} and apply \reflem{Nondet-test-set}]
  \PSpec{}{}{q_1}\Seq \PSpec{}{}{q_2}
\end{align*}%
The total-correctness version uses \reflaw{seq-term-term}, i.e.\ $\Term = \Term \Seq \Term$. 
\end{proof}

The above lemmas can be combined to give a law for splitting a specification
into a sequential composition with an intermediate assertion.
To make the law more useful in practice, we include a frame specifying the variables that are allowed to be modified.
\begin{lawx}[spec-seq-introduce]
For a set of variables $X$, sets of states $p_1$ and $p_2$, and relations $q$, $q_1$ and $q_2$, provided
\(
  p_1 \dres ((q_1 \rres p_2) \semi q_2) \subseteq q,
\)
both
\begin{align*}
&  \Pre{p_1} \Seq \PSpec{X}{}{q} \refsto \Pre{p_1} \Seq \PSpec{X}{}{q_1 \rres p_2} \Seq \Pre{p_2} \Seq \PSpec{X}{}{q_2} \mbox{~~and} \\
&  \Pre{p_1} \Seq \Spec{X}{}{q} \refsto \Pre{p_1} \Seq \Spec{X}{}{q_1 \rres p_2} \Seq \Pre{p_2} \Seq \Spec{X}{}{q_2} .
\end{align*}
\end{lawx}

\begin{proof}
\begin{align*}
&  \Pre{p_1} \Seq \PSpec{X}{}{q}
 \Refsto*[by \reflaw{spec-strengthen-under-pre} and assumption]
  \Pre{p_1} \Seq \PSpec{X}{}{(q_1 \rres p_2) \semi q_2}
 \Refsto*[by \reflaw{spec-to-sequential}]
  \Pre{p_1} \Seq \Frame{X}(\PSpec{}{}{q_1 \rres p_2} \Seq \PSpec{}{}{q_2})
 \Refsto*[by \reflem{spec-assert-restricts} and \reflaw{distribute-frame}]
  \Pre{p_1} \Seq \PSpec{X}{}{q_1 \rres p_2} \Seq \Pre{p_2} \Seq \PSpec{X}{}{q_2} 
  \qedhere
\end{align*}
\end{proof}

\begin{exax}[spec-seq-introduce]
The following uses two applications of \reflaw{spec-seq-introduce} to refine a specification to a sequence of three specifications.
\begin{align*}
  & \RSpec{nw,pw,w}{}{w \supset w' \lor i' \not\in w'} \\
 \refsto~ &\Why{by \reflaw{spec-seq-introduce} -- see justification below} \\
  & \RSpec{nw,pw,w}{}{w \supseteq pw' \land pw' \supseteq w'} \Seq \numberthis\label{yread-w} \\
  & \SPre{pw \supseteq w} \Seq \RSpec{nw,pw,w}{}{pw \supset w' \lor i' \notin w'} \numberthis\label{yrem-i}
\end{align*}
The proof obligation for the application of \reflaw*{spec-seq-introduce} above can be shown as follows.
The intermediate assertion $\Set{pw \supseteq w}$ is ensured by the postcondition of the first command.
\begin{align*}
  & \Rel{w \supseteq pw' \land pw' \supseteq w'} \semi \Rel{pw \supset w' \lor i' \notin w'} \\
  \subseteq~ & (\Rel{w \supseteq pw' \land pw' \supseteq w'} \semi \Rel{pw \supset w'}) \union \Rel{i' \notin w'} \\
  \subseteq~ & \Rel{w \supset w' \lor i' \notin w'}
\end{align*}
For the refinement of (\ref{yrem-i}),
$nw$ is used to hold the value of $pw$ with $i$ removed.
\begin{align*}
\hspace*{-1ex}
 (\ref{yrem-i})
 \refsto & \Why{by \reflaw{spec-seq-introduce} -- see justification below} \\
 & \SPre{pw \supseteq w} \Seq \RSpec{nw,pw,w}{}{nw' = pw -\{i\} \land pw' = pw \land pw' \supseteq w' \land i' = i} \Seq 
 \\
 & \SPre{pw \supseteq w \land nw = pw - \{i\}} \Seq \RSpec{nw,pw,w}{}{pw \supset w' \lor i' \not\in w'} 
\end{align*}
The proof obligation for the application of \reflaw*{spec-seq-introduce} above can be shown as follows.
The intermediate assertion $\Set{pw \supseteq w \land nw = pw - \{i\}}$ is ensured by the postcondition of the first command.
\begin{align*}
 & \Rel{nw' = pw - \{ i \} \land pw' = pw \land pw' \supseteq w' \land i' = i} \semi \Rel{pw \supset w' \lor i' \notin w'} \\
 \subseteq~ & (\Rel{pw' = pw} \semi \Rel{pw \supset w'}) \union \Rel{i' \notin w'} \\
 \subseteq~ & \Rel{pw \supset w' \lor i' \notin w'}
\end{align*}
\end{exax}

The next lemma is important for introducing a parallel composition or weak conjunction of specifications
to refine a single specification in \refsect{parallel}.
\begin{lemx}[sync-spec-spec]
For $\sync$ either $\parallel$ or $\together$, both
\(
  \PSpec{}{}{q_0} \sync \PSpec{}{}{q_1}   = \PSpec{}{}{q_0 \inter q_1} 
\)
and
\(
  \Post{q_0} \sync \Post{q_1}   = \Post{q_0 \inter q_1} . 
\)
\end{lemx}

\begin{proof}
The application of \reflem*{sync-distribute-relation} requires the assumption that $\sync$ is abort strict.
\begin{align*}
& \PSpec{}{}{q_0} \sync \PSpec{}{}{q_1}
\Equals*[definition of $\PSpec{}{}{q_1}$ from \refdef{partial-spec} and \reflem{sync-distribute-relation}]
\Nondet_{\sigma \in \Sigma}  (\cgd{\{\sigma\}} \Seq \PSpec{}{}{q_0} \Seq \cgd{(q_1\limg\{\sigma\}\rimg)})
\Equals*[by \reflem{test-restricts-spec} and \reflem{spec-test-restricts}]
\Nondet_{\sigma \in \Sigma}  (\cgd{\{\sigma\}} \Seq \PSpec{}{}{\{\sigma\} \dres q_0 \rres (q_1 \limg \{\sigma\}\rimg)})
\Equals*[simplify relation]
\Nondet_{\sigma \in \Sigma}  (\cgd{\{\sigma\}} \Seq \PSpec{}{}{\{\sigma\} \dres (q_0 \inter q_1)})
\Equals*[by \reflem{test-restricts-spec}]
(\Nondet_{\sigma \in \Sigma}  \cgd{\{\sigma\}}) \Seq \PSpec{}{}{q_0 \inter q_1}
\end{align*}
\begin{align*}&
\Equals*[by \reflem{Nondet-test-set}]
\PSpec{}{}{q_0 \inter q_1}
\end{align*}%
The property for a total specification follows from that for a partial specification.
\begin{align*}
&  \Spec{}{}{q_0} \sync \Spec{}{}{q_1}
 \Equals*[by definition of a total specification \refdef{spec}]
  (\PSpec{}{}{q_0} \together \Term) \sync (\PSpec{}{}{q_1} \together \Term)
 \Equals*[by \reflem{spec-distribute-sync} twice as $\sync$ is abort strict]
  \PSpec{}{}{q_0} \together \PSpec{}{}{q_1} \together (\Term \sync  \Term)
 \Equals*[by either \reflaw{par-term-term} for parallel or that $\together$ is idempotent]
  (\PSpec{}{}{q_0 \inter q_1} \together \Term
 \Equals*[by definition of a total specification \refdef{spec}]
  \Spec{}{}{q_0 \inter q_1}
  \qedhere
\end{align*}
\end{proof}

\section{Stability under interference}\labelsect{stability}

Stability of a property $p$ over the execution of a command is an important property
and, in the context of concurrency, stability of a property over interference from the
environment is especially important~\cite{ColemanVSTTE08,DBLP:conf/esop/WickersonDP10}.
This section examines stability properties that are useful for later laws.
\begin{defix}[stable]
A set of states $p$ is \emph{stable under} a binary relation $r$ if and only if $r \limg p \rimg \subseteq p$.
An equivalent expression of the property is that $p \dres r \rres p = p \dres r$.
If $p$ is stable under $r$, we also say that the test, $\cgd{p}$, is \emph{stable under} $r$.
\end{defix}
\begin{exax}[stable-pred]
The set of states $\Set{pw \supseteq w}$ is stable under the relation $\Rel{w \supseteq w' \land pw' = pw}$.
\end{exax}%

\begin{lemx}[stable-transitive]
If $p$ is stable under $r$ then $p$ is stable under the reflexive, transitive closure of $r$, $\Finrel{r}$ \refdef{transitive-closure}.
In fact, because $\id{} \subseteq \Finrel{r}$ one has $\Finrel{r}\limg p \rimg = p$.
\end{lemx}

\begin{proof}
The second step of the proof below uses the relational equivalent of \refprop{fiter-induction-right}, i.e.
\begin{equation}
  q \semi \Finrel{r} \subseteq x \mbox{\hspace{2em} if}~ q \union x \semi r \subseteq x . \labelprop{fiter-induction-right-rel}
\end{equation}
A general property of relational image is
\begin{eqnarray}
  r \limg p_0 \rimg \subseteq p_1 & \iff & p_0 \dres r \subseteq r \rres p_1 . \labelprop{rel-image-containment}
\end{eqnarray}
To show $p$ is stable under $\Finrel{r}$, 
\refdefi{stable} requires one to show $\Finrel{r} \limg p \rimg \subseteq p$, or using \refprop{rel-image-containment},
\begin{align*}
&  p \dres \Finrel{r} \subseteq \Finrel{r} \rres p
 \IFF*[property of relational algebra]
  (p \dres \id{}) \semi \Finrel{r} \subseteq \Finrel{r} \rres p
 \ImpliedBy*[by least $\star$-induction \refprop{fiter-induction-right-rel}]
  (p \dres \id{}) \union (\Finrel{r} \rres p) \semi r \subseteq \Finrel{r} \rres p
 \IFF*[properties of relational algebra]
  (\id{} \rres p) \union \Finrel{r} \semi (p \dres r) \subseteq \Finrel{r} \rres p
 \ImpliedBy*[as $p$ is stable under $r$, $p \dres r \subseteq r \rres p$]
  (\id{} \rres p) \union \Finrel{r} \semi r \rres p \subseteq \Finrel{r} \rres p
 \IFF*[distribution of range restriction]
  ({\id{}} \union \Finrel{r} \semi r) \rres p \subseteq \Finrel{r} \rres p
\end{align*}%
The final containment holds by unfolding as $\Finrel{r} = {\id{}} \union \Finrel{r} \semi r$ by \refprop{fiter-unfold-right} for relations. 
\end{proof}

\begin{lemx}[interference-before]
If $p \dres (r \semi q) \subseteq q$ and $p$ is stable under $r$, then $p \dres \Finrel{r} \semi q \subseteq q$.
\end{lemx}

\begin{proof}
The second step of the proof below uses the relational equivalent of \refprop{fiter-induction-left}, i.e.
\begin{equation}
  \Finrel{r} \semi q \subseteq x \mbox{\hspace{2em} if}~ q \union r \semi x \subseteq x . \labelprop{fiter-induction-left-rel}
\end{equation}
The proof follows.
\begin{align*}
&  p \dres \Finrel{r} \semi q \subseteq q
 \ImpliedBy*[as $p$ is stable under $r$]
  \Finrel{(p \dres r)} \semi q \subseteq q
 \ImpliedBy*[by $*$-induction \refprop{fiter-induction-left-rel}]
  q \union (p \dres r \semi q) \subseteq q
\end{align*}%
The latter holds from assumption $p \dres (r \semi q) \subseteq q$. 
\end{proof}

\begin{lemx}[interference-after]
If $p \dres (q \semi r) \subseteq q$, then 
$p \dres q \semi \Finrel{r} \subseteq q$.
\end{lemx}

\begin{proof}
The property is equivalent to $(p \dres q) \semi \Finrel{r} \subseteq p \dres q$,
which holds by $*$-induction \refprop{fiter-induction-right-rel} if
$(p \dres q) \union (p \dres q \semi r) \subseteq p \dres q$,
which follows from the assumption $p \dres (q \semi r) \subseteq q$. 
\end{proof}

\begin{lemx}[guar-test-commute-under-rely]
If $p$ is stable under both $r$ and $g$,
\[
  \rely{r} \together \guar{g} \Seq \cgd{p} ~~\refsto~~ \rely{r} \together \cgd{p} \Seq \guar{g}.
\]
\end{lemx}

\begin{proof}
First note that because $p$ is stable under both $r$ and $g$, 
by \refdefi{stable} $ r\limg p \rimg \subseteq p$ and $g \limg p \rimg \subseteq p$, and hence
by \reflem{atomic-test-commute} and \reflem{nondet-test-commute}.
\begin{equation}
  (\cpstep{g} \nondet \cestep{r})  \Seq \cgd{p} \refsto  \cgd{p} \Seq (\cpstep{g} \nondet \cestep{r})  \labelprop{atomic-test-commute}
\end{equation}
The main proof follows.
\begin{align*}
&  \rely{r} \together \guar{g} \Seq \cgd{p}
 \Equals*[\reflem{test-suffix-interchange}]
  (\rely{r} \together \guar{g}) \Seq \cgd{p}
 \Equals*[by \reflem{conj-rely-guar}]
  \Om{(\cpstep{g} \nondet \cestep{r})} \Seq (\Nil \nondet \cestep{\overline{r}} \Seq \Abort) \Seq \cgd{p}
 \Equals*[distributing the final test \refprop{nondet-seq-distrib-right}]
  \Om{(\cpstep{g} \nondet \cestep{r})} \Seq (\cgd{p} \nondet \cestep{\overline{r}} \Seq \Abort \Seq \cgd{p})
 \Refsto*[as $\Abort \Seq \cgd{p} = \Abort$ and introducing $\cgd{p}$ by \refprop{test-intro}]
  \Om{(\cpstep{g} \nondet \cestep{r})} \Seq (\cgd{p} \nondet \cgd{p} \Seq \cestep{\overline{r}} \Seq \Abort)
 \Equals*[factor out $\cgd{p}$ using \refprop{nondet-seq-distrib-left}]
  \Om{(\cpstep{g} \nondet \cestep{r})} \Seq \cgd{p} \Seq (\Nil \nondet \cestep{\overline{r}} \Seq \Abort)
 \Refsto*[by \reflem{iteration-test-commute} and \refprop{atomic-test-commute}]
  \cgd{p} \Seq \Om{(\cpstep{g} \nondet \cestep{r})} \Seq (\Nil \nondet \cestep{\overline{r}} \Seq \Abort)
 \Equals*[by \reflem{conj-rely-guar}]
  \cgd{p} \Seq (\rely{r} \together \guar{g})
 \Equals*[by \reflem{test-command-sync-command} for $\together$]
  \rely{r} \together \cgd{p} \Seq \guar{g}
  \qedhere
\end{align*}
\end{proof}

Coleman and Jones~\cite{CoJo07} recognised that the combination of a guarantee $g$ and a rely condition $r$
is sufficient to deduce that the overall postcondition $\Finrel{(r \union g)}$ holds on termination
because each step is either assumed to satisfy $r$ (environment step) or guarantees to satisfy $g$ (program step).
That property is made explicit in the following lemmas. 
\begin{lemx}[spec-trade-rely-guar]
$\rely{r} \together \PSpec{}{}{\Finrel{(r \union g)}}  \refsto \rely{r} \together \guar{g}$
\end{lemx}

\begin{proof}
Because $\rely{r} \refsto \rely{r} \together \guar{g}$ by
\reflaw{guar-introduce}, it is enough, by \reflem{weakly-correct-spec}, 
to show that $\rely{r} \together \guar{g}$ is weakly correct
with respect to relation $\Finrel{(r \union g)}$,
which holds by \reflem{weakly-correct} if for any state $\sigma$,
\begin{align*}
& (\rely{r} \together \guar{g}) \Seq \cgd{(\Finrel{(r \union g)}\limg\{\sigma\}\rimg)}
\Equals*[by \reflem{test-suffix-interchange} for weak conjunction]
\rely{r} \together \guar{g} \Seq \cgd{(\Finrel{(r \union g)}\limg\{\sigma\}\rimg)}
\Refsto*[by \reflem*{guar-test-commute-under-rely} as $\Finrel{(r \union g)}\limg\{\sigma\}\rimg$ is stable under both $r$ and $g$]
\rely{r} \together \cgd{(\Finrel{(r \union g)}\limg\{\sigma\}\rimg)} \Seq \guar{g}
\Refsto*[as $\sigma \in \Finrel{(r \union g)}\limg\{\sigma\}\rimg$ follows from reflexivity of $\Finrel{(r \union g)}$]
\rely{r} \together \cgd{\{\sigma\}} \Seq \guar{g}
 \Equals*[by \reflem{test-command-sync-command} for $\together$]
\cgd{\{\sigma\}} \Seq (\rely{r} \together \guar{g})
\qedhere
\end{align*}
\end{proof}

\begin{lawx}[spec-trading]
$\rely{r} \together \guar{g} \together \Post{\Finrel{(r \union g)} \inter q} = \rely{r} \together \guar{g} \together \Post{q}$.
\end{lawx}

\begin{proof}
The refinement from right to left holds by \reflaw{spec-strengthen}
and that from left to right as follows.
\begin{align*}
&  \rely{r} \together \guar{g} \together \Spec{}{}{\Finrel{(r \union g)} \inter q}
 \Equals*[by the definition of a specification \refdef{spec} and \reflem{sync-spec-spec}]
 \rely{r} \together \guar{g} \together \PSpec{}{}{\Finrel{(r \union g)}} \together \PSpec{}{}{q} \together \Term
 \Refsto*[by \reflem{spec-trade-rely-guar}; and definition of a specification \refdef{spec}]
  \rely{r} \together \guar{g} \together \Spec{}{}{q}
  \qedhere
\end{align*}
\end{proof}

\begin{RelatedWork}
In Jones' thesis~\cite[Sect.\ 4.4.1]{Jones81d} 
the parallel introduction law made use of a \emph{dynamic invariant} 
that is a relation between the initial state of a parallel composition
and all successor states (both intermediate states and the final state).
A dynamic invariant, $DINV$, is required to be reflexive and satisfy 
$DINV \semi r \subseteq DINV$, where $r$ is the rely condition, and 
for all threads $i$, satisfy $DINV \semi g_i \subseteq DINV$, where $g_i$ is the guarantee for thread $i$.
$DINV$ is conjoined with the conjunction of the postconditions of all the parallel components  
to show the resulting postcondition holds,
thus allowing a stronger overall postcondition based of the extra information in $DINV$.
If one lets $g$ stand for the union of all the guarantee relations of the individual threads,
i.e.\ $g =\Union_i g_i$,
the conditions on $DINV$ show that it contains $\Finrel{(r \union g)}$.
Hence $\Finrel{(r \union g)}$ can be seen as the smallest relation 
satisfying the properties for $DINV$.
The two-branch parallel introduction rule of Coleman and Jones~\cite{CoJo07} 
uses $\Finrel{(r \union g)}$ in place of $DINV$.
In both~\cite{Jones81d} and~\cite{CoJo07} the dynamic invariant was only used as part of the parallel introduction law,
but in~\cite{HayesJonesColvin14TR} it was recognised that the dynamic invariant could be decoupled
from the parallel introduction law leading to a law similar to \reflaw{spec-trading}.
Here we go one step further to factor out the more basic \reflem{spec-trade-rely-guar}
from which \reflaw{spec-trading} can be derived.
\reflem{spec-trade-rely-guar} is also useful in the proof of \reflaw{rely-idle} below.
\end{RelatedWork}

In the context of a rely condition $r$ and guarantee condition $g$, 
the strengthening of a postcondition can also assume the transitive closure of the union of the rely and guarantee.
In addition, a frame consisting of a set of variables $X$ corresponds to an additional guarantee of $\id{\overline{X}}$.
\begin{lawx}[spec-strengthen-with-trading]
If $p \dres (\Finrel{(r \union (g \inter \id{\overline{X}}))} \inter q_2) \subseteq q_1$, 
\begin{align*}
&  \rely{r} \together \guar{g} \together \Pre{p} \Seq \Spec{X}{}{q_1} \refsto \rely{r} \together \guar{g} \together \Pre{p} \Seq \Spec{X}{}{q_2}.
\end{align*}
\end{lawx}

\begin{proof}
\begin{align*}
&  \rely{r} \together \guar{g} \together \Pre{p} \Seq \Spec{X}{}{q_1}
 \Refsto*[by \reflaw{spec-strengthen-under-pre} using the assumption]
   \rely{r} \together \guar{g}
   \together
   \Pre{p} \Seq \Spec{X}{}{\Finrel{(r \union (g \inter \id{\overline{X}}))} \inter q_2}
\Equals*[definition of a frame \refdef{frame} and \reflaw{guar-merge}]
   \rely{r} \together \guar{(g \inter \id{\overline{X}})}
   \together
   \Pre{p} \Seq \Spec{}{}{\Finrel{(r \union (g \inter \id{\overline{X}}))} \inter q_2}
\end{align*}
\begin{align*}&
 \Equals*[by \reflaw{spec-trading}]
   \rely{r} \together \guar{(g \inter \id{\overline{X}})}
   \together
   \Pre{p} \Seq \Spec{}{}{q_2}
 \Equals*[by \reflaw{guar-merge} in reverse and definition of a frame \refdef{frame}]
    \rely{r} \together \guar{g}
    \together
    \Pre{p} \Seq \Spec{X}{}{q_2}
  \qedhere
\end{align*}
\end{proof}

\begin{exax}[loop-body]
The following application of \reflaw*{spec-strengthen-with-trading} strengthens
a postcondition under the assumption of both the precondition and the
reflexive, transitive closure of the rely and guarantee. 
After the strengthening, the precondition is weakened using \refprop{assert-weaken}.
\begin{align*}
  & \Rrely{w \supseteq w' \land i' = i} \together \Rguar{w \supseteq w' \land w - w' \subseteq \{i\}} \together {} \\
  & \t1 \SPre{w \subseteq \Subrange{0}{N-1} \land i \in \Subrange{0}{N-1} \land k \supseteq w} \Seq \\
  & \t1 \RSpec{w}{}{w' \subseteq \Subrange{0}{N-1} \land i' \in \Subrange{0}{N-1} \land (k \supset w' \lor i' \notin w')} \\
 \refsto~ 
  & \Rrely{w \supseteq w' \land i' = i} \together \Rguar{w \supseteq w' \land w - w' \subseteq \{i\}} \together {} \\
 & \t1 \SPre{w \subseteq \Subrange{0}{N-1} \land i \in \Subrange{0}{N-1}} \Seq \\
 & \t1 \RSpec{w}{}{w \supset w' \lor i' \not\in w'}
\end{align*}%
We have
$\Finrel{(r \union (g \inter \id{\overline{w}}))} \subseteq \Rel{w \supseteq w' \land i' = i}$, 
and so it is sufficient to show the following, which is straightforward.
\begin{align*}
  & \Rel{
    \underbrace{w \subseteq \Subrange{0}{N-1} \land i \in \Subrange{0}{N-1} \land k \supseteq w}_{p} \land
    \underbrace{w \supseteq w' \land i' = i}_{\mbox{}\supseteq \Finrel{(r \union (g \inter \id{\overline{w}}))}} \land
    \underbrace{(w \supset w' \lor i' \not\in w')}_{q_2}
  }  \\
  \subseteq~ 
  & \Rel{\underbrace{w' \subseteq \Subrange{0}{N-1} \land i' \in \Subrange{0}{N-1} \land (k \supset w' \lor i' \notin w')}_{q_1}}
\end{align*}
\end{exax}

If a rely ensures that a set of variables $Y$, that is not in the frame of a specification, is unchanged,
that is sufficient to ensure $Y$ is unchanged in the postcondition of the specification.
\begin{lawx}[frame-restrict]
For sets of variables $X$, $Y$ and $Z$, if $Z \subseteq X$ and $Y \subseteq \overline{Z}$
and $r \subseteq \id{Y}$ then,
\(
  \rely{r} \together \Spec{X}{}{\id{Y} \inter q} \refsto \rely{r} \together \Spec{Z}{}{q}.
\)
\end{lawx}%

\begin{proof}
Because $Y \subseteq \overline{Z}$, $\id{\overline{Z}} \subseteq \id{Y}$
and hence $\Finrel{(r \union \id{\overline{Z}})} \subseteq \Finrel{(\id{Y} \union \id{Y})} = \id{Y}$.
\begin{align*}
&  \rely{r} \together \Spec{X}{}{\id{Y} \inter q}
\Refsto*[by \reflaw{frame-reduce} using assumption $Z \subseteq X$]
   \rely{r} \together \Spec{Z}{}{\id{Y} \inter q}
\Refsto*[by \reflaw{spec-strengthen-with-trading} as $\Finrel{(r \union \id{\overline{Z}})} \inter q \subseteq \id{Y} \inter q$ ]
  \rely{r} \together \Spec{Z}{}{q}
\end{align*}
The application of \reflaw{spec-strengthen-with-trading} uses the implicit guarantee of $\guar{\universalrel}$ (i.e. $\Chaos$), noting that $\universalrel \inter \id{\overline{Z}} = \id{\overline{Z}}$.
\end{proof}

\begin{exax}[frame-restrict]
The following example refinement applies \reflaw{frame-reduce} to the first and third sequentially-composed specifications and \reflaw{frame-restrict} to the second to restrict their frames.
For the application of \reflaw{frame-restrict} to the second specification,
$X$ is $\{nw,pw,w\}$, $Y$ is $\{pw,i\}$ and $Z$ is $\{nw\}$,
and the rely ensures $\Rel{pw' = pw \land i' = i}$.   
\begin{align*}
  & \Rrely{w \supseteq w' \land i' = i \land nw' = nw \land pw' = pw} \together {} \\
  & \t1 \RSpec{nw,pw,w}{}{w \supseteq pw' \land pw' \supseteq w'} \Seq \\
  & \t1 \SPre{pw \supseteq w} \Seq \RSpec{nw,pw,w}{}{nw' = pw -\{i\} \land pw' = pw \land pw' \supseteq w' \land i' = i} \Seq   \\
  & \t1 \SPre{pw \supseteq w \land nw = pw - \{i\}} \Seq \RSpec{nw,pw,w}{}{pw \supset w' \lor i' \not\in w'} \\
 \refsto~ & \Why{by \reflaw{frame-reduce}, \reflaw{frame-restrict} and \reflaw*{frame-reduce}} \\
  & \Rrely{w \supseteq w' \land i' = i \land nw' = nw \land pw' = pw} \together {} \\
  & \t1 \RSpec{pw}{}{w \supseteq pw' \land pw' \supseteq w'} \Seq  \\
  & \t1 \SPre{pw \supseteq w} \Seq \RSpec{nw}{}{nw' = pw -\{i\} \land pw' \supseteq w'} \Seq   \\
  & \t1 \SPre{pw \supseteq w \land nw = pw - \{i\}} \Seq \RSpec{w}{}{pw \supset w' \lor i' \not\in w'} 
  \qedhere
\end{align*}
\end{exax}

\section{Parallel introduction}\labelsect{parallel}
  
A core law for rely/guarantee concurrency is introducing a parallel composition.
The following law is taken from our earlier paper~\cite[Sect.\ 8.3]{FMJournalAtomicSteps}. 
Because it is a core rely/guarantee concurrency law we repeat it here for completeness.
The parallel introduction law is an abstract version of that of Jones~\cite{Jones83b}. 
The main difference from Jones is that it is expressed based on our synchronous algebra primitives
and hence an algebraic proof is possible (see~\cite[Sect.\ 8.3]{FMJournalAtomicSteps}).
\begin{lawx}[spec-introduce-par]
\begin{equation*}
  \rely{r} \together \Spec{}{}{q_0 \inter q_1} \refsto 
    (\rely{(r \union r_0)} \together \guar{r_1} \together \Spec{}{}{q_0}) \parallel
    (\rely{(r \union r_1)} \together \guar{r_0} \together \Spec{}{}{q_1})
\end{equation*}
\end{lawx}

By monotonicity, any preconditions and guarantees can be carried over from 
the left side of an application of \reflaw*{spec-introduce-par} to the right side 
and then distributed into the two branches of the parallel.

\section{Refining to an (optional) atomic step}\labelsect{optional}

The optional atomic step command, $\Opt{q} \defs \cpstep{q} \nondet \cgd{(\dom{(q \inter \id{})})}$, 
performs an atomic program step satisfying $q$, or 
if $q$ can be satisfied by not changing the state, it can also do nothing \refdef{opt}.
The set $\dom{(q \inter \id{})}$ represents the set of all states from which $q$ is satisfied 
by not changing the state, i.e.\ $\Comprehension{}{(\sigma,\sigma) \in q}{\sigma}$.
The optional atomic step command is used in the definition of an atomic specification command (\refsect{atomic-spec})
and in the definition of an assignment command (\refsect{assignments})
to represent the step that atomically updates the variable.
The definition allows an assignment with no effect, such as $x := x$,
to be implemented by either doing an assignment that assigns to $x$ its current value or 
doing nothing.
\begin{lawx}[opt-strengthen-under-pre]
If $p \dres q_2 \subseteq q_1$, then $\Pre{p} \Seq \Opt{q_1} \refsto \Pre{p} \Seq \Opt{q_2}$.
\end{lawx}

\begin{proof}
\begin{align*}
&  \Pre{p} \Seq \Opt{q_1}
 \Equals*[by definition of $\Opt{}$ \refdef{opt}; distribution]
  \Pre{p} \Seq \cpstep{q_1} \nondet \Pre{p} \Seq \cgd{(\dom{(q_1 \inter \id{})})}
 \Refsto*[by \refprop{pgm-refine} and \refprop{test-strengthen} as $p \dres q_2 \subseteq q_1$ and $\dom(p \dres q_1 \inter \id{}) = p \inter \dom(q_1 \inter \id{})$]
  \Pre{p} \Seq \cpstep{(p \dres q_2)} \nondet \Pre{p} \Seq \cgd{(p \inter \dom{(q_2 \inter \id{})})}
 \Equals*[by \refax{seq-test-pgm} and \refax{seq-test-test} and \refprop{seq-assert-test} and \refdef{opt}]
  \Pre{p} \Seq \Opt{q_2}
  \qedhere
\end{align*}
\end{proof}

\begin{lemx}[spec-to-pgm]
\(
  \Post{q} \refsto \cpstep{q}
\)
\end{lemx}

\begin{proof}
Because $\Term \refsto \cpstep{q}$, using \reflem{totally-correct} it is sufficient to show that
$\cpstep{q} \Seq \cgd{(q\limg\{\sigma\}\rimg)} \refsto \cgd{\{\sigma\}} \Seq \cpstep{q}$ for all $\sigma$,
which follows directly using \reflem{atomic-test-commute}.
\end{proof}

\begin{lemx}[spec-to-test]
$\Post{q} \refsto \cgd{(\dom{(q \inter \id{})})}$
\end{lemx}

\begin{proof}
Because $\Term \refsto \Nil \refsto \cgd{(\dom{(q \inter \id{})})}$, by \reflem{totally-correct} it is sufficient to show
that $\cgd{(\dom{(q \inter \id{})})}$ is weakly correct with respect to relation $q$. That is, for all $\sigma_0$ it is enough to show:
\begin{align*}
  & \cgd{(\dom{(q \inter \id{})})} \Seq \cgd{(q\limg\{\sigma_0\}\rimg)}
    \refsto
    \cgd{\{\sigma_0\}} \Seq \cgd{(\dom{(q \inter \id{})})}
  \IFF*[merging tests \refax{seq-test-test} and \refprop{test-strengthen}]
    \dom{(q \inter \id{})} \inter q \limg \{ \sigma_0 \} \rimg
    \supseteq
    \dom{(q \inter \id{})} \inter \{ \sigma_0 \}
  \IFF*[expanding the definitions of domain and relational image]
    \Comprehension{}{(\sigma,\sigma) \in q \land (\sigma_0,\sigma) \in q}{\sigma}
    \supseteq
    \Comprehension{}{(\sigma,\sigma) \in q \land \sigma_0 =\sigma}{\sigma}
  \IFF*[set-theoretical reasoning]
    \true
   \qedhere
\end{align*}
\end{proof}

\begin{lawx}[spec-to-opt]
$\Post{q} \refsto \Opt{q}$.
\end{lawx}

\begin{proof}
The proof follows from the definition of $\Opt{}$ \refdef{opt} 
by \reflem{spec-to-pgm} and \reflem{spec-to-test}.
\end{proof}

A guarantee $g$ on an optional step satisfying $q$,
strengthens the optional's relation to satisfy $g$.
\begin{lawx}[guar-opt]
If $g$ is reflexive,
$\guar{g} \together \Opt{q} = \Opt{(g \inter q)}$.
\end{lawx}

\begin{proof}
Because $g$ is reflexive, $g \inter \id{} = \id{}$.
\begin{align*}
&  \guar{g} \together \Opt{q} 
  \Equals*[from the definition of an optional step \refdef{opt}; distribute]
  (\guar{g} \together \cpstep{q})
   \nondet
  (\guar{g} \together \cgd{(\dom{(q \inter \id{})})})
  \Equals*[from  \reflaw{guar-pgm} and \reflaw{guar-test}]
   \guar{(g \inter q)}
   \nondet
   \cgd{(\dom{(q \inter \id{})})}
 \Equals*[as $q \inter \id{} = g \inter q \inter \id{}$ because $g$ is reflexive; definition of $\Opt{}$ \refdef{opt}]
   \Opt{(g \inter q)}
  \qedhere
\end{align*}%
\end{proof}

\begin{lawx}[spec-guar-to-opt]
If $g$ is reflexive, 
$\guar{g} \together \Spec{x}{}{q} \refsto \Opt{(\id{\overline{x}} \inter g \inter q)}$.
\end{lawx}

\begin{proof}
The proof uses the definition of a frame \refdef{frame}, \reflaw{guar-merge}, \reflaw{spec-to-opt} and \reflaw{guar-opt} as $g$ is reflexive.
\begin{align*}&
  \guar{g} \together \Spec{x}{}{q}
 =
  \guar{(\id{\overline{x}} \inter g)} \together \Spec{}{}{q}
 \refsto
  \guar{(\id{\overline{x}} \inter g)} \together \Opt{q}
 =
  \Opt{(\id{\overline{x}} \inter g \inter q)}
 \qedhere
\end{align*}
\end{proof}

\section{Handling stuttering steps}\labelsect{stuttering}

The command, $\Idle \defs \guar{\id{}} \together \Term$, allows only a finite number of stuttering program steps
that do not change the state;
$\Idle$ does not constrain its environment \refdef{idle}.
Two $\Idle$ commands in sequence is equivalent to a single $\Idle$.
\begin{lemx}[seq-idle-idle]
$\Idle = \Idle \Seq \Idle$
\end{lemx}

\begin{proof}
Refinement from right to left holds because $\Idle \refsto \Nil$.
For refinement from left to right, 
the proof makes use of \reflaw{seq-term-term} and \reflaw{guar-seq-distrib}:
$\Idle
= \guar{\id{}} \together \Term
= \guar{\id{}} \together \Term \Seq \Term
\refsto (\guar{\id{}} \together \Term) \Seq (\guar{\id{}} \together \Term)
= \Idle \Seq \Idle .
$
\end{proof}

A reflexive guarantee combined with the $\Idle$ command is $\Idle$.
\begin{lemx}[guar-idle]
If $g$ is reflexive, $\guar{g} \together \Idle = \Idle$.
\end{lemx}

\begin{proof}
Because $g$ is reflexive $g \inter \id{} = \id{}$. 
The proof then follows from \refdef{idle} using \reflaw{guar-merge}.
\[
  \guar{g} \together \Idle
 =
  \guar{(g \inter \id{})} \together \Term
 =
  \guar{\id{}} \together \Term
 =
  \Idle
  \qedhere
\]
\end{proof}

If $p$ is stable under $r$
then $p$ is stable over the command $\rely{r} \together \Idle$ 
because it only performs stuttering program steps that do not change the state
and the environment steps are assumed to maintain $p$.

\begin{lemx}[rely-idle-stable]
If $p$ is stable under $r$,
\[
  \rely{r} \together \Idle \Seq \cgd{p} ~~\refsto~~ \rely{r} \together \cgd{p} \Seq \Idle.
\]
\end{lemx}%

\begin{proof}
Note that any property $p$ is stable under the identity relation $\id{}$.
\begin{align*}&
  \rely{r} \together \Idle \Seq \cgd{p}
 \Equals*[by definition of $\Idle$ \refdef{idle}]
  \rely{r} \together (\guar{\id{}} \together \Term) \Seq \cgd{p}
 \Equals*[by \reflem{test-suffix-interchange}]
 \rely{r} \together \guar{\id{}} \Seq \cgd{p} \together \Term
 \Refsto*[by \reflem{guar-test-commute-under-rely} as $p$ is stable under both $r$ and $\id{}$]
  \rely{r} \together \cgd{p} \Seq \guar{\id{}} \together \Term
 \Equals*[by \reflem{test-command-sync-command}]
  \rely{r} \together \cgd{p} \Seq (\guar{\id{}} \together \Term)
 \Equals*[by definition of $\Idle$ \refdef{idle}]
  \rely{r} \together \cgd{p} \Seq \Idle
 \qedhere
\end{align*}
\end{proof}

\begin{lemx}[rely-idle-stable-assert]
If $p$ is stable under $r$ then, 
$\rely{r} \together \Pre{p} \Seq \Idle \refsto \rely{r} \together \Idle \Seq \Pre{p}$.
\end{lemx}

\begin{proof}
The proof introduces a test, $\cgd{p}$, 
which establishes $p$ as an assertion \refprop{seq-test-assert}, 
then applies \reflem{rely-idle-stable}, 
applies \refprop{seq-assert-test} to elide the test,
and finally removes an assertion \refprop{assert-remove}:
$\rely{r} \together \Pre{p} \Seq \Idle 
\refsto \rely{r} \together \Pre{p} \Seq \Idle \Seq \cgd{p} \Seq \Pre{p} 
\refsto \rely{r} \together \Pre{p} \Seq \cgd{p} \Seq \Idle \Seq \Pre{p}
\refsto \rely{r} \together \Idle \Seq \Pre{p} .
$
\end{proof}

The following lemma is used as part of refining a specification (of a restricted form) to an expression evaluation.
The command $\Idle$ refines a specification with postcondition $\Finrel{r}$ in a rely context of $r$.
In addition, if $p$ is stable under $r$, $\Idle$ maintains $p$.
A special case of the law is if $p$ is $\Sigma$, 
i.e.\ $\rely{r} \together \Spec{}{}{\Finrel{r}} \refsto \rely{r} \together \Idle$.
\begin{lawx}[rely-idle]
If $p$ is stable under $r$, then~~~
\(
  \rely{r} \together \Pre{p} \Seq \Spec{}{}{\Finrel{r} \rres p}
  \refsto
  \rely{r} \together \Idle.
\)
\end{lawx}

\begin{proof}
All environment steps of the right side are assumed to satisfy $r$ and
all program steps satisfy the identity relation, 
and hence by \reflem{stable-transitive} the right side 
maintains $p$ and satisfies $\Finrel{(\id{} \union r)} = \Finrel{r}$.
\begin{align*}
&  \rely{r} \together \Pre{p} \Seq \Spec{}{}{\Finrel{r} \rres p}
 \Refsto*[by \reflaw{spec-strengthen-under-pre}; $\Finrel{r} \limg p \rimg \subseteq p$ by \reflem*{stable-transitive}; \refprop{assert-remove}]
 \rely{r} \together \Spec{}{}{\Finrel{r}}
 \Equals*[by the definition of a total-correctness specification command \refdef{spec}]
 \rely{r} \together \PSpec{}{}{\Finrel{r}} \together \Term
 \Refsto*[by \reflem{spec-trade-rely-guar} as $\Finrel{(r \union \id{})} = \Finrel{r}$]
  \rely{r} \together \guar{\id{}} \together \Term
 \Equals*[definition of $\Idle$ \refdef{idle}]
  \rely{r} \together \Idle
  \qedhere
\end{align*}
\end{proof}

If a specification $\Pre{p} \Seq \Spec{}{}{q}$ is placed in a context that allows interference satisfying $r$ before and after it,
the overall behaviour may not refine the specification.
If the precondition $p$ holds initially, it must hold after any interference steps satisfying $r$,
i.e.\ $p$ must the stable under $r$. 
If the specification is preceded by an interference step satisfying $r$, 
then a step satisfying $r$ followed by a sequence of steps that satisfies $q$ should also satisfy $q$ 
-- this leads to condition \refprop{q-tolerates-r-before}, which also assumes $p$ holds initially. 
Condition \refprop{q-tolerates-r-after} is similarly required to handle an interference step following the specification.
\begin{defix}[tolerates-interference]
Given a set of states $p$ and relations $q$ and $r$,
$q$ \emph{tolerates} $r$ \emph{from} $p$ if,
$p$ is stable under $r$ and
\begin{eqnarray}
  p \dres (r \semi q) & \subseteq & q \labelprop{q-tolerates-r-before} \\
  p \dres (q \semi r) & \subseteq & q .  \labelprop{q-tolerates-r-after}
\end{eqnarray}
\end{defix}

\begin{exax}[tolerates]
The relation $\Rel{pw \supset w' \lor i' \notin w'}$ 
tolerates the rely relation $\Rel{w \supseteq w' \land i'=i \land nw'=nw \land pw'=pw}$
from states in $\Set{pw \supseteq w \land nw = pw - \{i\}}$
because 
$\Rel{pw \supseteq w \land nw = pw - \{i\}}$ is stable under 
the rely 
and 
\begin{align*}
&  \Rel{\underbrace{pw \supseteq w \land nw = pw - \{i\}}_{p} \land \underbrace{w \supseteq w' \land i'=i \land nw'=nw \land pw'=pw}_{r}} \semi \\
 & \t1 \Rel{\underbrace{pw \supset w' \lor i' \notin w'}_{q}}
 \Subseteq
  \Rel{pw \supseteq w \land w \supseteq w' \land i'=i \land pw'=pw} \semi \Rel{pw \supset w' \lor i' \notin w'}
 \Subseteq
  (\Rel{pw' = pw} \semi \Rel{pw \supset w'}) \union \Rel{i' \notin w'}
 \Subseteq
  \Rel{pw \supset w' \lor i' \notin w'}
\end{align*}%
and
\begin{align*}
&  \Rel{\underbrace{pw \supseteq w \land nw = pw - \{i\}}_{p} \land \underbrace{(pw \supset w' \lor i' \notin w')}_{q}} \semi \\
& \t1 \Rel{\underbrace{w \supseteq w' \land i'=i \land nw'=nw \land pw'=pw}_{r}}
 \Subseteq
  \Rel{pw \supset w' \lor i' \notin w'} \semi \Rel{w \supseteq w' \land i' = i}
 \Subseteq
  \Rel{pw \supset w' \lor i' \notin w'} .
  \qedhere
\end{align*}
\end{exax}%

\begin{RelatedWork}
\refdefi{stable} and \refprop{q-tolerates-r-before} correspond respectively to 
conditions \textbf{PR-ident} and \textbf{RQ-ident} used by Coleman and Jones~\cite[Sect. 3.3]{CoJo07}, 
in which $r$ is assumed to be reflexive and transitive, and
condition \refprop{q-tolerates-r-after} is a slight generalisation of their condition \textbf{QR-ident}
because \refprop{q-tolerates-r-after} includes the restriction to the set $p$.
The conditions are also related to the the concept of stability of $p$ and $q$ in the sense of 
Wickerson et al.~\cite{Wickerson10-TR,DBLP:conf/esop/WickersonDP10}, 
although in that work post conditions are treated to single-state predicates rather than relations.
\end{RelatedWork}

\begin{lemx}[tolerates-transitive]
If $q$ tolerates $r$ from $p$ then,~~
\(
  p \dres \Finrel{r} \semi q \semi \Finrel{r} \subseteq q. 
\)
\end{lemx}

\begin{proof}
Two auxiliary properties are derived from \refdefi{tolerates-interference}.
\begin{eqnarray}
  p \dres \Finrel{r} \semi q \subseteq q &  & \mbox{if \refprop{q-tolerates-r-before} and $p$ is stable under $r$} \labelprop{q-tolerates-fin-r-before} \\
  p \dres q \semi \Finrel{r} \subseteq q &  & \mbox{if \refprop{q-tolerates-r-after}} \labelprop{q-tolerates-fin-r-after}
\end{eqnarray}
Properties \refprop{q-tolerates-fin-r-before} and \refprop{q-tolerates-fin-r-after} follow by 
\reflem{interference-before} and \reflem{interference-after}, respectively.
The proof of the main theorem is straightforward using \refprop{q-tolerates-fin-r-before} and then \refprop{q-tolerates-fin-r-after}.
\[
  p \dres \Finrel{r} \semi q \semi \Finrel{r}
 \subseteq
  p \dres q \semi \Finrel{r}
 \subseteq
  q
  \qedhere
\]
\end{proof}

Assuming the environment only performs steps satisfying $r$,
a specification that tolerates $r$ can tolerate $\Idle$ commands before and after it.
\begin{lawx}[tolerate-interference]
If $q$ tolerates $r$ from $p$ then,
\[
  \rely{r} \together \Pre{p} \Seq \Spec{}{}{q} = \rely{r} \together \Idle \Seq \Pre{p} \Seq \Spec{}{}{q} \Seq \Idle.
\]
\end{lawx}%

\begin{proof}
The refinement from right to left follows as $\Idle \refsto \Nil$,
and the refinement from left to right holds as follows.
\begin{align*}
&  \rely{r} \together \Pre{p} \Seq \Spec{}{}{q}
 \Refsto*[by \reflaw{spec-strengthen-under-pre} using \reflem{tolerates-transitive}] 
  \rely{r} \together \Pre{p} \Seq \Spec{}{}{\Finrel{r} \semi q \semi \Finrel{r}}
 \Refsto*[by \reflaw{spec-to-sequential} twice]
  \rely{r} \together \Pre{p} \Seq \Spec{}{}{\Finrel{r}} \Seq \Spec{}{}{q} \Seq \Spec{}{}{\Finrel{r}}
 \Refsto*[by \reflaw{spec-strengthen} and \reflem{spec-assert-restricts}]
  \rely{r} \together \Pre{p} \Seq \Spec{}{}{\Finrel{r} \rres p} \Seq \Pre{p} \Seq \Spec{}{}{q} \Seq \Spec{}{}{\Finrel{r}}
 \Refsto*[by \reflaw{rely-refine-within}; \reflaw{rely-idle} twice with $\Sigma$ for $p$ in second]
  \rely{r} \together \Idle \Seq \Pre{p} \Seq \Spec{}{}{q} \Seq \Idle
  \qedhere
\end{align*}
\end{proof}

The command $\Idle$ plays a significant role in the definition of
expressions because every program step of an expression evaluation
does not change the observable state.  
\reflem{idle-test-idle} below plays a crucial role in
\reflem{eval-single-reference}, which is the main lemma used for
handling expressions (including boolean
conditions). \reflem{par-idle-idle} and \reflem{test-par-idle} are
used in the proof of \reflem{idle-test-idle}.

\begin{lemx}[par-idle-idle]
$\Idle \parallel \Idle = \Idle$
\end{lemx}

\begin{proof}
From $\Idle \refsto \Skip$ and monotonicity of parallel we have, 
$\Idle \parallel \Idle \refsto \Idle \parallel \Skip = \Idle$. 
For refinement in the other direction we show
\begin{align*}
& \Idle 
\Equals*[definition of $\Idle$ \refdef{idle}]
\guar{\id{}} \together \Term
\Equals*[by \reflaw{par-term-term}]
\guar{\id{}} \together (\Term \parallel \Term)
\Refsto*[by \reflaw{guar-par-distrib}]
(\guar{\id{}} \together \Term) \parallel (\guar{\id{}} \together \Term)
\Equals*[definition of $\Idle$ \refdef{idle}]
\Idle \parallel \Idle
\qedhere
\end{align*}
\end{proof}

\begin{lemx}[idle-expanded]
$\Idle = \Fin{(\cpstep{\id{}} \nondet \cestepd)} \Seq \Om{\cestepd}$
\end{lemx}

\begin{proof}
From the definitions of $\Idle$ \refdef{idle}, a guarantee
\refdef{guar}, and $\Term$ \refdef{term}, using 
\refprop{sync-iter-fiter}.
\end{proof}
Finite stuttering either side of a test is equivalent to finite stuttering in parallel;
the skips in the following lemma allow for environment steps corresponding to the parallel $\Idle$ command.
\begin{lemx}[test-par-idle]
$\Idle \Seq t \Seq \Idle = \Skip \Seq t \Seq \Skip \parallel \Idle$ 
\end{lemx}

\begin{proof}
\begin{align*}
& \Skip \Seq t \Seq \Skip \parallel \Idle
\Equals*[by the definition of $\Skip$ \refdef{skip} and \reflem{idle-expanded}]
\Om{\cestepd} \Seq t \Seq \Om{\cestepd} \parallel \Fin{(\cpstep{\id{}} \nondet \cestepd)} \Seq \Om{\cestepd}
\Equals*[by \refprop{sync-iter-fiter} as $\cestepd \parallel (\cpstep{\id{}} \nondet \cestepd) = \cpstep{\id{}} \nondet \cestepd$, and using \reflem{idle-expanded}]
\Fin{(\cpstep{\id{}} \nondet \cestepd)} \Seq
(
((\Om{\cestepd} \Seq t \Seq \Om{\cestepd})\parallel \Om{\cestepd})
\nondet 
(t \Seq \Om{\cestepd} \parallel \Idle)
)
\Equals*[by \reflem{test-command-sync-command}; $\Om{\cestepd}$ is the identity of parallel \refax{par-identity}]
\Fin{(\cpstep{\id{}} \nondet \cestepd)} \Seq
(
\Om{\cestepd} \Seq t \Seq \Om{\cestepd}
\nondet 
t \Seq \Idle
)
\Equals*[by \reflem*{absorb-finite-iter}, $\Fin{(\cpstep{\id{}} \nondet \cestepd)} = \Fin{(\cpstep{\id{}} \nondet \cestepd)}\Seq \Fin{\cestepd}$,
distributivity \refprop{nondet-seq-distrib-left}, and $\Fin{\cestepd} \Seq \Om{\cestepd} = \Om{\cestepd}$]
\Fin{(\cpstep{\id{}} \nondet \cestepd)} \Seq
(
\Om{\cestepd} \Seq t \Seq \Om{\cestepd}
\nondet 
\Fin{\cestepd} \Seq t \Seq \Idle
)
\Equals*[by $\Om{c}\Seq d = \Fin{c}\Seq d \nondet \Inf{c}$ by \refprop{isolation}]
\Fin{(\cpstep{\id{}} \nondet \cestepd)} \Seq
(
\Fin{\cestepd} \Seq t \Seq \Om{\cestepd}
\nondet
\Inf{\cestepd} 
\nondet 
\Fin{\cestepd} \Seq t \Seq \Idle
)
\Equals*[using $\Om{c}\Seq d = \Fin{c}\Seq d \nondet \Inf{c}$ \refprop{isolation}]
\Fin{(\cpstep{\id{}} \nondet \cestepd)} \Seq
(
\Fin{\cestepd} \Seq t \Seq \Om{\cestepd}
\nondet
\Om{\cestepd} \Seq t \Seq \Idle
)
\Equals*[using $\Om{\cestepd} \refsto \Fin{\cestepd}$ and $\Idle \refsto \Om{\cestepd}$ and monotonicity to eliminate the first choice]
\Fin{(\cpstep{\id{}} \nondet \cestepd)} \Seq \Om{\cestepd} \Seq t \Seq \Idle
\Equals*[\reflem{idle-expanded}]
\Idle \Seq t \Seq \Idle
\qedhere
\end{align*}
\end{proof}

\begin{lemx}[idle-test-idle]
$\Idle \Seq t \Seq \Idle \parallel \Idle = \Idle \Seq t \Seq \Idle$
\end{lemx}

\begin{proof}
The proof uses \reflem{test-par-idle}, \reflem{par-idle-idle} and \reflem*{test-par-idle} again.
\[
\Idle \Seq t \Seq \Idle \parallel \Idle
= 
\Skip \Seq t \Seq \Skip \parallel \Idle \parallel \Idle
= 
\Skip \Seq t \Seq \Skip \parallel \Idle
= 
\Idle \Seq t \Seq \Idle 
\qedhere
\]
\end{proof}

\section{Atomic specification commands}\labelsect{atomic-spec}

The atomic specification command, $\atomicrel{p,q} \defs \Idle \Seq \Pre{p} \Seq \Opt{q} \Seq \Idle$,
performs a single atomic program step or test satisfying $q$
under the assumption that $p$ holds in the state in which the step occurs;
it allows finite stuttering before and after the step and
does not constrain its environment \refdef{atomic-spec}.
The default precondition is the set of all states so that $\atomicrel{q} \defs \atomicrel{\Sigma,q}$ \refdef{atomic-spec-no-pre}.

\begin{exax}[CAS]
Below is an atomic specification of a compare-and-swap (CAS) machine instruction.%
\footnote{CAS instructions typically have an additional local boolean variable, $done$, 
that returns whether the update succeeded or not. 
That is not needed for the example used here but is trivial to add to the specification.}
The local variable $pw$ represents the previously sampled value of $w$ and 
local variable $nw$ represents the value $w$ is to be updated to, 
provided $w$ still has the value $pw$,
otherwise $w$ is left unchanged.
Both $pw$ and $nw$ are intended to be local variables. 
 \begin{align*}
 CAS \defs\ 
 &    \Frame{w}{\Ratomicrel{
                                               (w = pw \implies w' = nw) \land 
                                               (w \neq pw \implies w' = w)  \numberthis\labeldef{CAS}
                      }}
\end{align*}
\end{exax}

\begin{RelatedWork}
An atomic specification command can also be used to specify atomic operations on a data structure, 
as used by Dingel~\cite{Dingel02}.
In Dingel's work the semantics of his language considers two commands the same 
if they are equivalent modulo finite stuttering,
whereas our definition \refdef{atomic-spec} does not use such an equivalence
but builds the stuttering into the atomic specification directly using $\Idle$ commands.
Note that in order for $\atomicrel{p,q}$ to be closed under finite stuttering
it is defined in terms of $\Opt{q}$ rather than $\cpstep{q}$
because, for example, 
$\cpstep{\id{}}$ requires a single stuttering step
whereas 
$\Opt{\id{}}$ allows either a single stuttering step or no steps. 
\end{RelatedWork}

The following two laws follow from the definition of an atomic specification command \refdef{atomic-spec},
\refprop{assert-weaken} and \reflaw{opt-strengthen-under-pre}.
\begin{lawx}[atomic-spec-weaken-pre]
If $p_0 \subseteq p_1$ then, 
$\atomicrel{p_0,q} \refsto \atomicrel{p_1,q}$.
\qed
\end{lawx}

\begin{lawx}[atomic-spec-strengthen-post]
If $p \dres q_2 \subseteq q_1$ then, $\atomicrel{p,q_1} \refsto \atomicrel{p,q_2}$.
\qed
\end{lawx}

A reflexive guarantee on an atomic specification requires the specification to satisfy the guarantee.
\begin{lawx}[atomic-guar]
If $g$ is a reflexive relation, $\guar{g} \together \atomicrel{p, q} \refsto \atomicrel{p, g \inter q}$.
\end{lawx}

\begin{proof}
\begin{align*}
&  \guar{g} \together \atomicrel{p, q}
 \Equals*[definition of atomic specification \refdef{atomic-spec}]
  \guar{g} \together \Idle \Seq \Pre{p} \Seq \Opt{q} \Seq \Idle
 \Refsto*[\reflaw{guar-seq-distrib}, \reflem{guar-idle} and \reflaw{guar-assert}]
  \Idle \Seq \Pre{p} \Seq (\guar{g} \together \Opt{q}) \Seq \Idle
 \Equals*[by  \reflaw{guar-opt} as $g$ is reflexive]
  \Idle \Seq \Pre{p} \Seq \Opt{(g \inter q)} \Seq \Idle
 \Equals*[definition of atomic specification \refdef{atomic-spec}]
  \atomicrel{p, g \inter q}
  \qedhere
\end{align*}
\end{proof}

A specification can be refined to an atomic specification that must also satisfy any guarantee.
\begin{lawx}[atomic-spec-introduce]
If $g$ is reflexive, 
and
$q$ tolerates $r$ from $p$ then,
\[
  \rely{r} \together \guar{g} \together \Pre{p} \Seq \Spec{}{}{q} \refsto \rely{r} \together \atomic{p,g \inter q}.
\]
\end{lawx}%

\begin{proof}
\begin{align*}
&  \rely{r} \together \guar{g} \together \Pre{p} \Seq \Spec{}{}{q}
 \Equals*[by \reflaw{tolerate-interference} as $q$ tolerates $r$ from $p$]
  \rely{r} \together \guar{g} \together \Idle \Seq \Pre{p} \Seq \Spec{}{}{q} \Seq \Idle
 \Refsto*[by \reflaw{spec-to-opt} and definition of an atomic specification \refdef{atomic-spec}]
  \rely{r} \together \guar{g} \together \atomicrel{p,q}
 \Refsto*[by \reflaw{atomic-guar} as $g$ is reflexive]
  \rely{r} \together \atomic{p,g \inter q}
  \qedhere
\end{align*}
\end{proof}

\begin{exax}[intro-CAS]
\reflaw{atomic-spec-introduce} allows a specification to be replaced by an atomic specification,
after strengthening the postcondition (with trading).
\begin{align*}
 & \Rguar{w \supseteq w' \land w - w' \subseteq \{i\}} \together 
    \Rrely{w \supseteq w' \land i' = i \land nw' = nw \land pw' = pw} \together {} \\
 & \SPre{pw \supseteq w \land nw = pw - \{i\}} \Seq \RSpec{w}{}{pw \supset w' \lor i' \notin w'} \\
 \refsto & \Why{replace $i' \notin w'$ by $i \notin w'$ using \reflaw{spec-strengthen-with-trading}} \\
 & \Rguar{w \supseteq w' \land w - w' \subseteq \{i\}} \together 
    \Rrely{w \supseteq w' \land i' = i \land nw' = nw \land pw' = pw} \together {} \\
 & \SPre{pw \supseteq w \land nw = pw - \{i\}} \Seq \RSpec{w}{}{pw \supset w' \lor i \notin w'} \\
 \refsto & \Why{by \reflaw{atomic-spec-introduce}} \\
 & \Rrely{w \supseteq w' \land i' = i \land nw' = nw \land pw' = pw} \together {} \\
 & \Frame{w}{\atomicrel{\Set{pw \supseteq w \land nw = pw - \{ i \}}, \Rel{w \supseteq w' \land w - w' \subseteq \{i\} \land (pw \supset w' \lor i \notin w')}}}  \numberthis\label{atomic-rem}
\end{align*}
The law requires that the guarantee is reflexive (which is trivial) and
that $\Rel{pw \supset w' \lor i' \notin w'}$ tolerates 
$\Rel{w \supseteq w' \land i' = i \land nw' = nw \land pw' = pw}$ from $\Set{pw \supseteq w \land nw = pw - \{i\}}$,
as shown in \refexa{tolerates}.
The atomic step may be refined using \reflaw{atomic-spec-strengthen-post}, 
\reflaw{atomic-spec-weaken-pre} and \reflaw{rely-remove}, 
to a form equivalent to the compare-and-swap (CAS) operation \refdef{CAS}.
\begin{align*}
  (\ref{atomic-rem})
 \refsto &~
    \Frame{w}{\Ratomicrel{(w = pw \implies w' = nw) \land (w \neq pw \implies w' = w)}} \numberthis\label{atomic-CAS}
\end{align*}
The proof obligation for the application of \reflaw*{atomic-spec-strengthen-post} can be shown as follows;
the weakenings are straightforward.
\begin{align*}
&  \Set{pw \supseteq w \land nw = pw - \{ i \}} \dres \Rel{(w = pw \implies w' = nw) \land (w \neq pw \implies w' = w)}
 \Equals 
  \Rel{pw \supseteq w \land nw = pw - \{ i \} \land (w = pw \implies w' = nw) \land (w \neq pw \implies w' = w)}
 \Subseteq
  \Rel{w \supseteq w' \land w - w' \subseteq \{i\} \land (pw \supset w' \lor i \notin w')}
\end{align*}
\end{exax}

\section{Expressions under interference}\labelsect{expressions}

In the context of concurrency, 
the evaluation of an expression can be affected by interference 
that modifies shared variables used in the expression.
For fine-grained parallelism, normally simple aspects of programs 
such as expression evaluation in assignments and conditionals
are fraught with unexpected dangers,
for example, an expression like $x-x$ is not guaranteed to be zero
if the value of $x$ can be changed by interference between the two accesses to $x$.%
\footnote{To allow for all possible implementations of expression evaluation,
we allow each reference to a variable in an expression to be fetched from shared memory separately,
so that different references to the same variable may have different values.
If expression evaluation only accessed each variable once,
no matter how many times it appears within an expression,
stronger properties about expression evaluation are possible,
such as $x - x = 0$.
See~\cite{HayesBurnsDongolJones12} for a discussion of different forms of expression evaluators
and their relationships.
}
In our approach, expression evaluation is not considered to be atomic
and programming language expressions are not part of the core language,
rather expression evaluation is defined in terms of constructs in the core language.
Hence laws for reasoning about expressions (including boolean guards for conditionals)
can be proven in terms of the properties of the constructs from which expressions are built.

\begin{RelatedWork}
Issues such as $x-x$ evaluating to a non-zero value can be avoided by assuming expression evaluation is atomic
(as done by Xu et al.\ \cite{XuRoeverHe97}, Prensa Nieto~\cite{PrensaNieto03}, Schellhorn et al.~\cite{SchellhornTEPR14}, 
San{\'a}n et al.\ \cite{Sanan21} and Dingel~\cite{Dingel02})
but that leads to a theory that is less suitable for practical programming languages
because their implementations do not respect such atomicity constraints.
\end{RelatedWork}

\refsect{expr-syntax} defines the semantics of expression evaluation under interference
that may change the value of variables in the expression. 
\refsect{expr-\constant} considers \constant\ expressions that evaluate to the same value 
before and after interference, 
and
\refsect{expr-single-reference} considers the case when 
the evaluation of an expression is equivalent to 
evaluating it in one of the states during the execution of the evaluation.

\subsection{Expressions}\labelsect{expr-syntax}

The syntax of expressions, $e$, includes constants ($\kappa$), program variables ($x$), 
unary operators ($\Ominus$) and binary operators ($\Oplus$).
\begin{equation}
  e ::= \kappa \mid x \mid \Ominus e \mid e_1 \Oplus e_2 \labeldef{expr}
\end{equation}
First, we give the semantics of expression evaluation in a single state;
this corresponds to a side-effect-free expression's semantics in the context of a sequential program.
\begin{defix}[expr-single-state]
The notation $\Eval{e}{\sigma}$ stands for the value of the expression $e$ in the state $\sigma$.
Its definition is the usual inductive definition over the structure of the expression,
where $\UnarySem{\Ominus}{}$ is interpreted as the semantics of the operator $\Ominus$ on values and
$\BinarySem{}{\Oplus}{}$ is interpreted as the semantics of $\Oplus$ on values.\\[-1ex]
\begin{minipage}{0.5\textwidth}
\begin{eqnarray}
  \Eval{\kappa}{\sigma} & = & \kappa  \labelax{eval-const} \\
  \Eval{x}{\sigma} & = & \sigma(x) \labelax{eval-var}
\end{eqnarray}
\end{minipage}%
\begin{minipage}{0.5\textwidth}
\begin{eqnarray}
  \Eval{(\Ominus e)}{\sigma} & = & \UnarySem{\Ominus}{\Eval{e}{\sigma}} \labelax{eval-unary} \\
  \Eval{(e1 \Oplus e2)}{\sigma} & = & \BinarySem{\Eval{e1}{\sigma}}{\Oplus}{\Eval{e2}{\sigma}} \labelax{eval-binary}
\end{eqnarray}
\end{minipage}
\end{defix}
\noindent 

The command $\Test{e}_k$ represents evaluating the expression $e$ to the value $k$.
The evaluation of an expression $e$ to $k$ does not change any variables and
may either succeed or fail.
If the evaluation succeeds in evaluating $e$ to be $k$,
$\Test{e}_k$ terminates but if it fails $\Test{e}_k$ becomes infeasible
(but note that it may contribute some stuttering program steps and environment steps before becoming infeasible).
Because successful expression evaluation terminates and does not change any variables, 
an expression evaluation $\Test{e}_k$ refines $\Idle$, 
the command that does a finite number of stuttering program steps.
In the definition of $\Test{e}_k$ below these stuttering steps are represented by $\Idle$
and allow for updates to variables that are not observable, such as machine registers.
Expression evaluation is often used in a non-deterministic choice over all possible values for $k$,
and hence just one choice of $k$ succeeds for any particular execution.
Here expressions are assumed to be well defined; 
the semantics of Colvin et al.~\cite{DaSMfaWSLwC} provides a more complete definition 
that handles undefined expressions like divide by zero.
The notation $\EqEval{e_1}{e_2}$ stands for the set of states 
in which $e_1$ evaluates to the same value as $e_2$ \refdef{eq-val};
the set may be empty.
\begin{eqnarray}
  \EqEval{e1}{e2} & \defs & \EvalSet{e1}{\Eval{e2}{\sigma}} \labeldef{eq-val}
\end{eqnarray}
\begin{defix}[expr-evaluation]
The semantics of expression evaluation in the context of interference, $\Test{e}_k$,
is defined inductively over the structure of an expression.
A constant $\kappa$ evaluates to a value $k$ if $\kappa = k$ but fails (becomes infeasible) otherwise
\refdef{eval-const}.
A program variable $x$ is similar but the value of $x$ depends on the state 
in which $x$ is accessed \refdef{eval-var}, which may not be the initial state;
it is assumed that the access to $x$ is atomic.
The evaluation $\Eval{x}{\sigma}$ of a variable $x$ in state $\sigma$ is the one place 
in expression evaluation that is dependent on the choice of representation of the state.
The unary expression $\Ominus e$ evaluates to $k$ if $e$ evaluates to a value $k_1$
such that $k = \UnarySem{\Ominus}{k_1}$ \refdef{eval-unary}.
The expression $e_1 \Oplus e_2$ evaluates to $k$ if there exist values $k_1$ and $k_2$
such that $e_1$ evaluates to $k_1$, 
$e_2$ evaluates to $k_2$, and
$k = \BinarySem{k_1}{\Oplus}{k_2}$.
The evaluation of $e_1$ and $e_2$ can be arbitrarily interleaved 
and hence the definition represents their evaluation as a parallel composition \refdef{eval-binary}.
\begin{eqnarray}
  \Test{\kappa}_k & \defs & \Idle \Seq \cgd{(\EqEval{k}{\kappa})} \Seq \Idle \labeldef{eval-const} \\
  \Test{x}_k & \defs & \Idle \Seq \cgd{(\EqEval{k}{x})} \Seq \Idle  \labeldef{eval-var} \\
  \Test{\Ominus e}_k & \defs & \Nondet \Comprehension{k_1}{k = \UnarySem{\Ominus}{k_1}}{\Test{e}_{k_1}} \labeldef{eval-unary} \\
  \Test{e_1 \Oplus e_2}_k & \defs & \Nondet \Comprehension{k_1, k_2}{k = \BinarySem{k_1}{\Oplus}{k_2}}{\Test{e_1}_{k_1} \parallel \Test{e_2}_{k_2}}  \labeldef{eval-binary} 
\end{eqnarray}
\end{defix}
\ihbig{Note}{For \reflem{eval-single-reference} with a rely added on the right to be an equality, rather than a refinement,
we need to add and $\Idle$ at the start of \refdef{eval-unary} and \refdef{eval-binary},
so that if their non-deterministic choices are over the empty set, their evaluations reduce to $\Idle \Seq \Magic$, rather than $\Magic$.}

For a unary operator like absolute value, there may be values of $k$ for which no value of $k_1$ exists,
e.g.\ for $k= -1$, there is no value of $k_1$ such that $-1 = \Abs{k_1}$
because the absolute value cannot be negative;
$\Test{\Abs{e}}_k$ is infeasible for such values of $k$. 
If $k$ is a positive integer, such as 5, 
both $5 =  \Abs{5}$ and $5 = \Abs{-5}$
and hence there may be multiple values of $k_1$ for a single value of $k$ in the choice within \refdef{eval-unary}.
Similarly for binary operators, there may be many pairs of values $k_1$ and $k_2$ such that $k = \BinarySem{k_1}{\Oplus}{k_2}$.
Conditional expressions, including conditional ``and'' and ``or'', are not treated here
but can be easily defined (see~\cite{DaSMfaWSLwC}).%
\footnote{Conditional ``and'' can be defined in terms of a conditional (\refsect{conditional}):
\(
  \Test{e_1 \&\& e_2}_k \defs \If e_1 \Then \Test{e_2}_k \Else \Test{\false}_k \Fi.
\)
}

\begin{lemx}[idle-eval]
For any expression $e$ and value $k$, $\Idle \refsto \Test{e}_k$.
\end{lemx}

\begin{proof}
The proof is by induction over the structure of expressions \refdef{expr}.
For the binary case it relies on \reflem{par-idle-idle}. 
\end{proof}

\begin{lawx}[guar-eval]
If $g$ is reflexive, $\guar{g} \together \Test{e}_k = \Test{e}_k$.
\end{lawx}

\begin{proof}
By \reflem{idle-eval}, $\Idle \refsto \Test{e}_k$ and hence $\Idle \together \Test{e}_k = \Test{e}_k$,
therefore using \reflem{guar-idle} as $g$ is reflexive,
\[
  \guar{g} \together \Test{e}_k
 =
  \guar{g} \together \Idle \together \Test{e}_k
 =
  \Idle \together \Test{e}_k
 =
  \Test{e}_k.
  \qedhere
\]
\end{proof}

\subsection{Expressions that are \constant\ under a rely}\labelsect{expr-\constant}

An expression $e$ is \constant\ under $r$ if the evaluation of $e$ in each of two states
related by $r$ gives the same value.
\begin{defix}[\constant-under-rely]
An expression $e$ is \emph{\constant} under a relation $r$ if and only if for all $\sigma$ and $\sigma'$,
\(
  (\sigma,\sigma') \in r \implies \Eval{e}{\sigma} = \Eval{e}{\sigma'} .
\)
\end{defix}
Obviously, if all variables used in $e$ are unmodified by the interference $r$, 
$e$ is \constant,
but there are other examples for which the expression may be \constant\ even though 
the values of its variables are modified by the interference, for example, 
given integer variables $x$ and $y$,
\begin{itemize}
\item
the absolute value of a variable $x$, $\Abs{x}$, 
is \constant\ under interference that negates $x$ because $\Abs{-x} = \Abs{x}$, 
\item
$\Abs{x} + \Abs{y}$ is \constant\ under interference that may negate either $x$ or $y$,
\item
$even(x)$ is \constant\ under interference that changes $x$ by a value $2*k$ for some integer $k$,
\item
$(x \bmod N)$ is \constant\ under interference that adds $N$ to $x$ because $(x+N) \bmod N = x \bmod N$,
\item
$x-x$ is \constant\ under any interference because each evaluation of $x-x$
is performed in a single state and hence it evaluates to zero in both states,
\item
$x*0$ is \constant\ under any interference 
because its value does not depend on that of $x$,
and
\item
for an array $A$, $A$ indexed by $i$ (i.e.\ $A_i$) is \constant\ under interference that modifies neither $i$ nor $A_i$,
although it may modify other elements within $A$.
\end{itemize}
\begin{RelatedWork}
Coleman~\cite{ColemanVSTTE08} and Wickerson et al.~\cite{Wickerson10-TR} use 
a stronger syntactic property that requires that no variables used within $e$ are modified;
none of the examples above are handled under their definition 
unless all variables are assumed to be unmodified.
Our approach can also handle algorithms in which threads are concurrently accessing separate elements in array,
using a rely that ensures the other thread is not modifying the element being accessed but may be modifying other elements.
The source of the additional generality of our definition is that 
it is defined in terms of the semantics of expressions rather than being based on their syntactic form.
The stronger assumptions of Coleman and Wickerson et al.\ are important special cases
of our more general properties.
\end{RelatedWork}

\begin{lemx}[\constant-expr-stable]
If an expression $e$ is \constant\ under $r$, then for any value $k$, 
$(\EqEval{k}{e})$ is stable under $r$.
\end{lemx}

\begin{proof}
By \refdefi*{\constant-under-rely}, $(\sigma_0,\sigma) \in r \implies \Eval{e}{\sigma_0} = \Eval{e}{\sigma}$ and using \refdefi{stable}.
\begin{align*}
&  r \limg \EqEval{k}{e} \rimg
 \Equals
  \Comprehension{}{\exists \sigma_0 \cdot \sigma_0 \in \EqEval{k}{e} \land (\sigma_0,\sigma) \in r}{\sigma}
 \Subseteq
  \Comprehension{}{\exists \sigma_0 \cdot k = \Eval{e}{\sigma_0} \land \Eval{e}{\sigma_0} = \Eval{e}{\sigma}}{\sigma}
 \Equals
  \Comprehension{}{k = \Eval{e}{\sigma}}{\sigma}
 \Equals
  \EqEval{k}{e}
  \qedhere
\end{align*}
\end{proof}

\subsection{Single-reference expressions}\labelsect{expr-single-reference}

Evaluating an expression in the context of interference may lead to anomalies
because evaluation of an expression such as $x+x$ may 
retrieve different values of $x$ for each of its occurrences 
and hence it is possible for $x+x$ to evaluate to an odd value
even though $x$ is an integer variable.
However, $2*x$ always evaluates to an even value, even if $x$ is subject to modification.
While the expression $x-x$ is \constant\ under any interference $r$
(because evaluating it in any single state always gives 0),
its evaluation under interference that modifies $x$ may use different values of $x$ 
from different states and hence may give a non-zero answer.
This means that normal algebraic identities like $x+x = 2*x$ and $x-x = 0$ are no longer valid.
In fact, these equalities become refinements:%
\footnote{Hence one can define a notion of refinement between expressions $e_1$ and $e_2$ 
as $\forall k \spot \Test{e_1}_k \refsto \Test{e_2}_k$.}
$\Test{x+x}_k \refsto \Test{2*x}_k$ and $\Test{x-x}_k \refsto \Test{0}_k$.
Such anomalies may be reduced if we restrict our attention to expressions
that are \emph{single reference} under a rely condition $r$
because the evaluation of a single-reference expression under interference $r$
is equivalent to calculating its value in one of the states during its evaluation,
as is shown in \reflem{eval-single-reference} below.

\begin{defix}[single-reference-under-rely]
An expression $e$ is \emph{single reference under} a relation $r$ iff $e$ is
\begin{itemize}
\item 
a constant $\kappa$, or
\item
a program variable $x$ and access to $x$ is atomic, or
\item
a unary expression $\Ominus e_1$ and $e_1$ is single reference under $r$, or
\item
a binary expression $e_1 \Oplus e_2$ and 
both $e_1$ and $e_2$ are single reference under $r$, and 
at least one of $e_1$ and $e_2$ is \constant\ under $r$.
\end{itemize}
\end{defix}
Under this definition, the expression $\Abs{x}+y$ is single reference under interference that negates $x$
because both $\Abs{x}$ and $y$ are single-reference expressions and 
$\Abs{x}$ is \constant\ under interference that negates $x$.
Note that an expression being \constant\ under $r$ does not imply it is single reference under $r$,
e.g.\ $x-x$ is \constant\ under any rely but it is not single reference under a rely that allows $x$ to change.
Note that by our definition, 
the expression $0*(x+x)$ is not single reference under a rely that allows $x$ to change 
(because $x+x$ is not single reference) 
but $0*(x+x)$ can be shown to be equivalent to the expression $0$, 
which is single reference under any rely.%
\footnote{To handle this case the definition of a single reference expression could allow 
an alternative for binary operators of the form: $e1$ is single reference and 
$\forall \sigma, v, v' \spot e1_{\sigma} \Oplus v = e1_{\sigma} \Oplus v'$.
For the example $0*(x+x)$, the expression $0$ is trivially single reference and $0*v = 0*v'$
for all values $v$ and $v'$.
We do not feel such an extension is warranted because expressions such as $0*(x+x)$ 
are not useful in practice.
}

\begin{RelatedWork}
Coleman~\cite{ColemanVSTTE08} and Wickerson et al.~\cite{Wickerson10-TR} use 
a stronger \emph{single unstable variable} property that requires at most one variable, $x$, 
within $e$ is modified by the interference and $x$ is only referenced once in $e$.
For example, $\Abs{x}+y$ does not satisfy their single unstable variable property 
under interference that negates $x$.
Overall this gives us more general laws about single-reference expressions, 
which are used to handle expression evaluation within 
assignments (\refsect{assignments}),
conditionals (\refsect{conditional})
and loops (\refsect{loop}).
\end{RelatedWork}

If an expression is single reference under $r$,
then in a context in which all environment steps are assumed to satisfy $r$,
its evaluation is equivalent to its evaluation in the single state 
in which the single-reference variable is accessed.
Evaluating expression $e$ to the value $k$ in a single state can be represented by the test $\cgd{(\EqEval{k}{e})}$,
leading to the following fundamental law that is used in the proofs of laws for
programming language constructs involving single-reference expressions.
\begin{lemx}[eval-single-reference]
If $e$ is a single-reference expression under $r$,
and $k$ is a value,
\begin{eqnarray} \labelprop{eval-single-reference}
  \rely{r} \together \Idle \Seq \cgd{(\EqEval{k}{e})} \Seq \Idle
  & \refsto & 
  \Test{e}_{k}.
\end{eqnarray}
\end{lemx}
\begin{proof}
If $r$ is not reflexive, weaken $r$ to $r \union \id{}$ using \reflaw{rely-weaken}.
The remainder of the proof assumes $r$ is reflexive.
The proof is by induction over the structure of the expression \refdef{expr}.
If the expression $e$ is a constant $\kappa$ or a program variable $x$,
$\Test{e}_k = \Idle \Seq \cgd{(\EqEval{k}{e})} \Seq \Idle$ 
and \refprop{eval-single-reference} holds using \reflaw{rely-remove}.
If the expression $e$ is of the form $\Ominus e_1$ for some expression $e_1$,
then because $e$ is single-reference under $r$, so is $e_1$, and hence
the inductive hypothesis is:
$\rely{r} \together \Idle \Seq \cgd{(\EqEval{k_1}{e_1})} \Seq \Idle ~~\refsto~~ \Test{e_1}_{k_1}$,
for all $k_1$. 
Hence
\begin{align*}
&  \rely{r} \together \Idle \Seq \cgd{(\EqEval{k}{(\Ominus e_1)})} \Seq \Idle \refsto \Test{\Ominus e_1}_{k}
 \IFF*[by the definition of evaluating a unary expression \refdef{eval-unary}]
  \rely{r} \together \Idle \Seq \cgd{(\EqEval{k}{(\Ominus e_1)})} \Seq \Idle \refsto \Nondet \Comprehension{k_1}{k = \UnarySem{\Ominus}{k_1}}{\Test{e_1}_{k_1}}
 \ImpliedBy*[by \reflem{refine-choice}]
  \forall k_1 \spot k = \UnarySem{\Ominus}{k_1} \implies \rely{r} \together \Idle \Seq \cgd{(\EqEval{k}{(\Ominus e_1)})} \Seq \Idle \refsto \Test{e_1}_{k_1}
 \ImpliedBy*[as $k = \UnarySem{\Ominus}{k_1}$ implies $\cgd{(\EqEval{k}{(\Ominus e_1)})} = \cgd{(\EqEval{(\UnarySem{\Ominus}{k_1})}{(\Ominus e_1)})} \refsto \cgd{(\EqEval{k_1}{e_1})}$]
  \forall k_1 \spot \rely{r} \together \Idle \Seq \cgd{(\EqEval{k_1}{e_1})} \Seq \Idle \refsto \Test{e_1}_{k_1}
\end{align*}%
which is the inductive assumption.
Note that in the reasoning in the last step, multiple values of $k_1$ may give the same value of $k$,
so this is not in general an equality, only a refinement.
For example, if $\Ominus$ is absolute value, 
then both the states in which $e_1$ evaluates to $k_1$ and 
the states in which $e_1$ evaluates to $-k_1$ 
satisfy $\EqEval{(\UnarySem{\Ominus}{k_1})}{(\Ominus e_1)}$ 
but only the states in which $e_1$ evaluates to $k_1$ satisfy $\EqEval{k_1}{e_1}$.

If $e$ is of the form $e_1 \Oplus e_2$, then because $e$ is single reference under $r$, so are both $e_1$ and $e_2$,
and hence we may assume the following two inductive hypotheses:
\begin{align}
  \rely{r} \together \Idle \Seq \cgd{(\EqEval{k_1}{e_1})} \Seq \Idle & \refsto \Test{e_1}_{k_1} & \mbox{for all $k_1$} \labelprop{binary-hyp1} \\ 
  \rely{r} \together \Idle \Seq \cgd{(\EqEval{k_2}{e_2})} \Seq \Idle & \refsto \Test{e_2}_{k_2} & \mbox{for all $k_2$} \labelprop{binary-hyp2}
\end{align}
We are required to show
\begin{align*}
&  \rely{r} \together \Idle \Seq \cgd{(\EqEval{k}{(e_1 \Oplus e_2)})} \Seq \Idle \refsto \Test{e_1 \Oplus e_2}_{k}
 \IFF*[by the definition of evaluating a binary expression \refdef{eval-binary}]
  \rely{r} \together \Idle \Seq \cgd{(\EqEval{k}{(e_1 \Oplus e_2)})} \Seq \Idle \refsto
    \Nondet \Comprehension{k_1, k_2}{k = \BinarySem{k_1}{\Oplus}{k_2}}{\Test{e_1}_{k_1} \parallel \Test{e_2}_{k_2}}
 \ImpliedBy*[by \reflem{refine-choice}]
  \forall k_1, k_2 \spot k = \BinarySem{k_1}{\Oplus}{k_2} \implies 
    \rely{r} \together \Idle \Seq \cgd{(\EqEval{k}{(e_1 \Oplus e_2)})} \Seq \Idle \refsto \Test{e_1}_{k_1} \parallel \Test{e_2}_{k_2}
 \ImpliedBy*[as $k = \BinarySem{k_1}{\Oplus}{k_2}$, $\cgd{(\EqEval{k}{(e_1 \Oplus e_2)})} = \cgd{(\EqEval{(\BinarySem{k_1}{\Oplus}{k_2})}{(e_1 \Oplus e_2)})} \refsto \cgd{((\EqEval{k_1}{e_1}) \inter (\EqEval{k_2}{e_2}))}$]
    \forall k_1, k_2 \spot \rely{r} \together \Idle \Seq \cgd{((\EqEval{k_1}{e_1}) \inter (\EqEval{k_2}{e_2}))} \Seq \Idle \refsto \Test{e_1}_{k_1} \parallel \Test{e_2}_{k_2} \numberthis\labelprop{binary-eval-single-reference}
\end{align*}
Let $t_1 = \cgd{(\EqEval{k_1}{e_1})}$ and $t_2 = \cgd{(\EqEval{k_2}{e_2})}$
and recall that $\pnegate{\cgd{p}} = \cgd{\pnegate{p}}$ by \refax{negate-test}.
As $e$ is assumed to be single reference under $r$, 
from \refdefi{single-reference-under-rely} either $e_1$ or $e_2$ is \constant\ under $r$.
By symmetry assume $e_1$ is \constant\ under $r$ 
and hence by \reflem{\constant-expr-stable} both $t_1$ and $\pnegate{t_1}$ are stable under $r$.
Now we show \refprop{binary-eval-single-reference}.
\begin{align*}&
  \Test{e_1}_{k_1} \parallel \Test{e_2}_{k_2}
 \Refines*[by the inductive hypotheses  \refprop{binary-hyp1} and \refprop{binary-hyp2}]
  (\rely{r} \together \Idle \Seq t_1 \Seq \Idle) \parallel (\rely{r} \together \Idle \Seq t_2 \Seq \Idle)
 \Equals*[case analysis on $t_1$, using $c = (t_1 \nondet \pnegate{t_1}) \Seq c = t_1\Seq c \nondet \pnegate{t_1}\Seq c$ and $\parallel$ commutes]
  t_1 \Seq
    ((\rely{r} \together \Idle \Seq t_2 \Seq \Idle) \parallel
     (\rely{r} \together \Idle \Seq t_1 \Seq \Idle)) \nondet {}\\
& \pnegate{t_1} \Seq
    ((\rely{r} \together \Idle \Seq t_1 \Seq \Idle) \parallel
     (\rely{r} \together \Idle \Seq t_2 \Seq \Idle))
 \Refines*[using $\Idle\Seq t \Seq \Idle \refines \Idle$ for any test $t$]
  t_1 \Seq
    ((\rely{r} \together \Idle \Seq t_2 \Seq \Idle) \parallel 
     (\rely{r} \together \Idle)) \nondet {}\\
& \pnegate{t_1} \Seq
    ((\rely{r} \together \Idle \Seq t_1 \Seq \Idle) \parallel
     (\rely{r} \together \Idle))
 \Equals*[by \reflem*{test-command-sync-command} distribute tests into $\parallel$ and $\together$]
  (\rely{r} \together t_1 \Seq \Idle \Seq t_2 \Seq \Idle) \parallel
    (\rely{r} \together \Idle) \nondet {}\\
&  (\rely{r} \together \pnegate{t_1} \Seq \Idle \Seq t_1 \Seq \Idle) \parallel
    (\rely{r} \together \Idle)
 \Refines*[by assumption $t_1$ and $\pnegate{t_1}$ are stable under $r$ and \reflem{rely-idle-stable}]
  (\rely{r} \together \Idle \Seq t_1\Seq t_2 \Seq \Idle) \parallel
  (\rely{r} \together \Idle) \nondet {}\\
&(\rely{r} \together \Idle \Seq \pnegate{t_1} \Seq t_1 \Seq \Idle) \parallel
  (\rely{r} \together \Idle)
\end{align*}
\begin{align*}&
 \Equals*[using $\pnegate{t_1}\Seq t_1 = \Magic \refines t_1\Seq t_2$ and monotonicity]
  (\rely{r} \together \Idle \Seq t_1 \Seq t_2 \Seq \Idle) \parallel
  (\rely{r} \together \Idle)
\Equals*[as $\Idle$ and tests guarantee $\id{}$]
  (\rely{r} \together \guar{\id{}} \together (\Idle \Seq t_1 \Seq t_2 \Seq \Idle)) \parallel
  (\rely{r} \together \guar{\id{}} \together \Idle)
 \Refines*[by \reflaw{rely-par-distrib} and \reflaw*{guar-strengthen} as $\id{} \subseteq r$ as $r$ is reflexive]
  \rely{r} \together ((\Idle \Seq t_1 \Seq t_2 \Seq \Idle) \parallel \Idle)
 \Equals*[expanding abbreviations of tests $t_1$ and $t_2$ and merging the tests \refax{seq-test-test}]
  \rely{r} \together ((\Idle \Seq (\cgd{((\EqEval{k_1}{e1}) \inter (\EqEval{k2}{e2}))}) \Seq \Idle) \parallel \Idle)
 \Equals*[\reflem{idle-test-idle}]
  \rely{r} \together (\Idle \Seq (\cgd{((\EqEval{k_1}{e1}) \inter (\EqEval{k2}{e2}))}) \Seq \Idle)
\qedhere
\end{align*}
\end{proof}

The following lemma allows an expression evaluation
(e.g.\ within an assignment or in guards of conditionals and loops)
to be introduced from a specification.
\begin{lawx}[rely-eval]
For a value $k$, 
expression $e$,
set of states $p$,
and
relations $r$ and $q$,
if 
$e$ is single reference under $r$,
$q$ tolerates $r$ from $p$,
and
$(p \inter \EqEval{k}{e}) \dres \id{} \subseteq q$,
\[
  \rely{r} \together \Pre{p} \Seq \Spec{}{}{q} ~~\refsto~~ \Test{e}_{k}.
\]
\end{lawx}%

\begin{proof}
\begin{align*}
&  \rely{r} \together \Pre{p} \Seq \Spec{}{}{q} 
 \Equals*[by \reflaw{tolerate-interference} as $q$ tolerates $r$ from $p$]
  \rely{r} \together \Idle \Seq \Pre{p} \Seq \Spec{}{}{q} \Seq \Idle
 \Refsto*[by \reflaw*{spec-strengthen-under-pre} using assumption $(p \inter \EqEval{k}{e}) \dres \id{} \subseteq q$; \refprop{assert-remove}]
  \rely{r} \together \Idle \Seq \Spec{}{}{\EqEval{k}{e} \dres \id{}} \Seq \Idle
 \Refsto*[by \reflem{spec-to-test} as $\dom{((\EqEval{k}{e} \dres \id{}) \inter \id{})} = \EqEval{k}{e}$]
  \rely{r} \together \Idle \Seq \cgd{(\EqEval{k}{e})} \Seq \Idle
 \Refsto*[by \reflem{eval-single-reference} as $e$ is single reference under $r$]
  \Test{e}_{k}
  \qedhere
\end{align*}
\end{proof}

The following law is useful for handling boolean expressions used in conditionals and while loops.
\begin{lawx}[rely-eval-expr]
For a value $k$, expression $e$,
sets of states $p$ and $p_0$,
and 
relation $r$,
if 
$e$ is single reference under $r$,
$p$ is stable under $r$,
$p \inter \EqEval{k}{e} \subseteq p_0$,
and
$p_0$ is stable under $(p \dres r)$,
\[
  \rely{r} \together \Pre{p} \Seq \Spec{}{}{\Finrel{r} \rres (p \inter p_0)} ~~\refsto~~ \Test{e}_{k}.
\]
\end{lawx}%

\begin{proof}
Note that because $p$ is stable under $r$, $p_0$ being stable under $(p \dres r)$ is equivalent to $(p \inter p_0)$ being stable under $r$.
The proof uses \reflaw{rely-eval}, taking
$q$ to be $\Finrel{r} \rres (p \inter p_0)$
because this tolerates $r$ from $p$, 
and
$(p \inter \EqEval{k}{e}) \dres \id{} \subseteq \Finrel{r} \rres (p \inter p_0)$
because $\id{} \subseteq \Finrel{r}$ and $p \inter \EqEval{k}{e} \subseteq p \inter p_0$.
\end{proof}

\section{Assignments under interference}\labelsect{assignments}

An assignment (non-atomically) evaluates its expression $e$ to some value $k$ 
and then atomically updates the variable $x$ to be $k$,
as represented by the relation $\Update{x}{k} \defs \id{\overline{x}} \rres (\EqEval{x}{k})$  \refdef{update}.
We repeat its definition \refdef{assign}:
\begin{eqnarray}
  x := e & ~\defs~ & \Nondet_{k \in Val} (\Test{e}_k \Seq \Opt{(\Update{x}{k})} \Seq \Idle). \labeldef{assign2}
\end{eqnarray}
The non-deterministic choice allows $\Test{e}_k$ to evaluate to any value
but only one value succeeds for any particular execution.
Interference from the environment may change the values of variables used within $e$ 
and hence influence the value of $k$.
The command $\Opt{(\Update{x}{k})}$ 
atomically updates $x$ to be $k$ 
but may do nothing if $x$ is already $k$,
so that assignments like $x := x$ can be implemented by doing nothing at all.
Interference may also change the value of $x$ after it has been updated.
The $\Idle$ command at the end allows for both environment steps and 
any hidden (stuttering) steps in the implementation
after the update has been made; 
hidden (stuttering) steps are also allowed by the definition of expression evaluation.

\begin{RelatedWork}
In terms of a trace semantics in \refsect{model}~\cite{DaSMfaWSLwC}, 
any trace that is equivalent to a trace of $x := e$ modulo finite stuttering
is also a trace of $x :=e$,
i.e.\ definition \refdef{assign2} of $x :=e$ is closed under finite stuttering.
This holds because 
(i) expression evaluation is closed under finite stuttering,
(ii) the optional update allows a possible stuttering update step to be eliminated,
and 
(iii) the final $\Idle$ command allows stuttering steps after the update.
We follow this convention for the definition of all constructs that correspond to executable code.
This is in contrast to the usual approach of building finite stuttering 
into the underlying trace semantics~\cite{Brookes-full-abstraction,Dingel02}.

A number of approaches~\cite{XuRoeverHe97,PrensaNieto03,DBLP:conf/esop/WickersonDP10,Sanan21,SchellhornTEPR14}
treat a complete assignment command as a single atomic action,
although they allow for interference before and after the atomic action.
Such approaches do not provide a realistic model for fine-grained concurrency.
Coleman and Jones~\cite{CoJo07} do provide a fine-grained operational semantics
that is closer to the approach used here but the laws they develop are more restrictive.
\end{RelatedWork}

Consider refining a specification of the form 
$\rely{r} \together \guar{g} \together \Pre{p} \Seq \Spec{x}{}{q}$
to an assignment command $x := e$,
where we assume access to $x$ is atomic,
$e$ is a single-reference expression,
and $g$ is reflexive.
In dealing with an assignment to $x$ we make use of a specification augmented with a frame of $x$,
recalling from the definition of a frame \refdef{frame}
that $\Frame{x}{c} = \guar{\id{\overline{x}}} \together c$
and noting that guarantees distribute into other constructs.
\reffig{assignment} gives an overview of the execution of $x := e$,
and the constraints on $q$ and $g$ that are required to show that the assignment satisfies the specification:
\begin{figure}
\begin{center}
\input{assignment}
\end{center}
\caption{ Execution of $x := e$ assuming that access to $x$ is atomic
  and $e$ is a single-reference expression (noting that $\sigma_2$ may
  be $\sigma_3$ if the optional atomic step is instantaneous).  The
  execution is annotated using the assumption that the initial state
  satisfies precondition $p$ and that the environment steps satisfy
  relation $r$, and it includes the constraints on relation $q$ and
  reflexive relation $g$ that are required to show that the assignment
  satisfies guarantee $g$ and postcondition specification $q$ under
  those assumptions. }
\labelfig{assignment}
\end{figure}
\begin{itemize}
\item
end-to-end the execution must satisfy $q$;
\item
because $e$ is single-reference under $r$, 
the evaluation of $e$ to some value $k$ corresponds to 
evaluating it in one of the states ($\sigma_1$) during its evaluation;
\item
the optional program step that atomically updates $x$ between $\sigma_2$ and $\sigma_3$ must satisfy $g$;
\item
the state after the update ($\sigma_3$) satisfies $\EqEval{k}{x}$;
and
\item
all the steps before $\sigma_2$ and after $\sigma_3$ are either
environment steps that satisfy $r$ or program steps that do not modify any variables
and hence any subsequence of these steps satisfies $\Finrel{r}$,
from which one can deduce that $\sigma_2$ is in $\Finrel{r} \limg \EqEval{k}{e} \rimg$.
\end{itemize}
Because the assignment is defined in terms of an optional atomic step command,
the transition from $\sigma_2$ to $\sigma_3$ may be elided, i.e.\ $\sigma_3$ is $\sigma_2$;
in this case 
$q$ must be satisfied by any sequence of steps satisfying $\Finrel{r}$ starting from a state satisfying $p$,
and
$g$ is satisfied because all program steps are stuttering steps and $g$ is assumed to be reflexive.

If $q$ is assumed to tolerate $r$ from $p$ (\refdefi*{tolerates-interference})
then in \reffig{assignment} if $q$ holds between states $\sigma_2$ and $\sigma_3$, $q$ also holds between $\sigma$ and $\sigma'$.
We also have that $p$ is stable under $r$ and hence $p \inter \EqEval{k}{e}$ holds in state $\sigma_1$.
We introduce a set of states $p_1\,k$ parameterised by $k$, 
such that $p_1\,k$ is stable under $r$ and $p \inter \EqEval{k}{e} \subseteq p_1\,k$, 
and hence $p_1\,k$ is established in state $\sigma_1$ and because it is stable under $r$, $p_1\,k$ holds in state $\sigma_2$.
\begin{figure}
\begin{center}
\input{assignment2}
\end{center}
\caption{The simplified constraints on relation $q$ and relation $g$
  that are required to show that $x := e$ satisfies guarantee $g$ and
  postcondition specification $q$ under precondition $p$ and rely $r$,
  assuming that access to $x$ is atomic, $e$ is a single-reference
  expression, $g$ is reflexive, $q$ tolerates $r$ from $p$,
  and $p \inter \EqEval{k}{e} \subseteq p_1\,k$.
}\labelfig{assignment2}
\end{figure}
That allows the constraints on $q$ and $g$ in \reffig{assignment} to be simplified to those in
\reffig{assignment2}, and
that leads to the following general refinement law to introduce an assignment,
from which a number of special case laws are derived.

\begin{lawx}[rely-guar-assign]
Given sets of states $p$,
a set of states $p_1\,k$ parameterised by $k$, 
relations $g$, $r$ and $q$, 
a variable $x$,
and 
an expression $e$,
if
$g$ is reflexive,
$e$ is single reference under $r$,
$q$ tolerates $r$ from $p$,
and for all $k$,
$p_1\,k$ is stable under $r$, and
\begin{eqnarray} 
  p \inter \EqEval{k}{e} & \subseteq & p_1\,k \labelprop{establish-p1} \\
  p_1\,k \dres \Update{x}{k} & \subseteq & g \inter q \labelprop{assignment}
\end{eqnarray}
then
$\rely{r} \together \guar{g} \together \Pre{p} \Seq \Spec{x}{}{q} \refsto x := e$.
\end{lawx}

\begin{proof}
\begin{align*}
&  \rely{r} \together \guar{g} \together \Pre{p} \Seq \Spec{x}{}{q}
 \Equals*[duplicate precondition; \reflaw{tolerate-interference} as $q$ tolerates $r$ from $p$]
  \rely{r} \together \guar{g} \together \Pre{p} \Seq \Idle \Seq \Pre{p} \Seq \Spec{x}{}{q} \Seq \Idle
 \Refsto*[by \reflem{refine-choice} with fresh $k$, \reflem{seq-idle-idle} and \refprop{test-intro}]
  \Nondet_{k \in Val} (\rely{r} \together \guar{g} \together \Pre{p} \Seq \Idle \Seq \cgd{(\EqEval{k}{e})} \Seq \Idle \Seq \Pre{p} \Seq \Spec{x}{}{q} \Seq \Idle)
 \Refsto*[by \reflem{rely-idle-stable-assert}  and \refprop{seq-test-assert}, \refprop{seq-assert-assert} and \refprop{assert-remove}]
  \Nondet_{k \in Val} (\rely{r} \together \guar{g} \together \Idle \Seq \cgd{(\EqEval{k}{e})} \Seq \Pre{p \inter \EqEval{k}{e}} \Seq \Idle \Seq \Spec{x}{}{q} \Seq \Idle)
 \Refsto*[by \reflem{rely-idle-stable-assert} assumption \refprop{establish-p1}; $p_1\,k$ stable under $r$]
  \Nondet_{k \in Val} (\rely{r} \together \guar{g} \together \Idle \Seq \cgd{(\EqEval{k}{e})} \Seq \Idle \Seq \Pre{p_1\,k} \Seq \Spec{x}{}{q} \Seq \Idle)
 \Refsto*[by \reflem{eval-single-reference} as $e$ is single reference under $r$]
  \Nondet_{k \in Val} (\rely{r} \together \guar{g} \together \Test{e}_k \Seq \Pre{p_1\,k} \Seq \Spec{x}{}{q} \Seq \Idle)
 \Refsto*[by \reflaw{guar-seq-distrib}; \reflaw{guar-eval}; \reflem{guar-idle}]
  \Nondet_{k \in Val} (\rely{r} \together \Test{e}_k \Seq (\guar{g} \together \Pre{p_1\,k} \Seq \Spec{x}{}{q}) \Seq \Idle)
 \Refsto*[by \reflem{assert-distrib} and \reflaw{spec-guar-to-opt}]
  \Nondet_{k \in Val} (\rely{r} \together \Test{e}_k \Seq \Pre{p_1\,k} \Seq \Opt{(\id{\overline{x}} \inter g \inter q)} \Seq \Idle)
 \Refsto*[by \reflaw{opt-strengthen-under-pre} and assumption \refprop{assignment}]
  \Nondet_{k \in Val} (\rely{r} \together \Test{e}_k \Seq \Pre{p_1\,k} \Seq \Opt{(\Update{x}{k})} \Seq \Idle)
 \Refsto*[by \reflaw{rely-remove}, \refprop{assert-remove} and definition of an assignment \refdef{assign}]
  x := e
 \qedhere
\end{align*}
\end{proof}

If $e$ is both single reference and \constant\ under $r$ then its evaluation is unaffected by inference satisfying $r$.
\begin{lawx}[local-expr-assign]
Given 
a set of states $p$,
relations $g$, $r$ and $q$, 
variable $x$,
and
an expression $e$ that is both single reference and \constant\ under $r$,
if
$g$ is reflexive,
$q$ tolerates $r$ from $p$,
and for all $k$,
$(p \inter \EqEval{k}{e}) \dres \Update{x}{k} \subseteq g \inter q$,
\[
  \rely{r} \together \guar{g} \together \Pre{p} \Seq \Spec{x}{}{q} \refsto x := e.
\]
\end{lawx}%

\begin{proof}
The proof uses \reflaw{rely-guar-assign} taking $p_1\,k$ to be $p \inter \EqEval{k}{e}$,
which is stable under $r$ by \reflem{\constant-expr-stable} because $e$ is \constant\ under $r$:
$r \limg p \inter \EqEval{k}{e} \rimg \subseteq r \limg p \rimg \inter r \limg \EqEval{k}{e} \rimg \subseteq p \inter \EqEval{k}{e}$.
\end{proof}

\begin{exax}[assign-nw]
\reflaw*{local-expr-assign} is applied to refine a rely/guarantee specification to an assignment
involving variables that are not subject to any interference.
\begin{align*}
&  \Rguar{w \supseteq w' \land w - w' \subseteq \{i\}} \together 
     \Rrely{w \supseteq w' \land i' = i \land nw' = nw \land pw' = pw} \together {} \\
&  \t1 \SPre{pw \supseteq w} \Seq \RSpec{nw}{}{nw' = pw - \{i\} \land pw' \supseteq w'} 
  \Refsto nw := pw - \{i\}
\end{align*}%
The provisos of the law hold as follows:
the expression $pw - \{i\}$ is single reference and \constant\ under the rely; 
the guarantee 
is reflexive;
the postcondition $\Rel{nw' = pw - \{i\} \land pw' \supseteq w'}$ tolerates the rely 
from the precondition $\Set{pw \supseteq w}$;
and for all $k$,
\begin{align*}&
  \Rel{pw \supseteq w \land k = pw - \{ i \} \land w' = w \land pw' = pw \land i' = i \land k = nw'}
 \Subseteq 
  \Rel{w \supseteq w' \land w - w' \subseteq \{i\} \land nw' = pw - \{i\} \land pw' \supseteq w'}.
\end{align*}
\end{exax}

The following law allows the sampling of the value of a single-reference expression $e$.
It assumes that the interference may monotonically decrease $e$ (or monotonically increase $e$) during execution and
hence the sampled value (in $x$) must be between the initial and final values of $e$.
The notation $\GeEval{e1}{e2}$ stands for $\Comprehension{}{e1_\sigma \Transrelationeq e2_\sigma}{\sigma}$.
\begin{lawx}[rely-assign-monotonic]
Given a set of states $p$,
relations $g$ and $r$,
an expression $e$,
and 
a variable $x$,
if 
$g$ is reflexive,
$p$ is stable under $r$,
$x$ is \constant\ under $r$,
$e$ is single-reference under $r$,
and $\Transrelationeq$ is a reflexive, transitive binary relation,
such that for all $k$,
\begin{eqnarray}
  (p \inter \GeEval{k}{e}) \dres \Update{x}{k} & \subseteq & g  \labelprop{monotonic-g} \\
  r & \subseteq & \Comprehension{}{\Eval{e}{\sigma} \Transrelationeq \Eval{e}{\sigma'}}{(\sigma, \sigma')} \labelprop{monotonic-under-r} \\
  p \dres \id{\overline{x}} & \subseteq & \Comprehension{}{\Eval{e}{\sigma} \Transrelationeq \Eval{e}{\sigma'}}{(\sigma, \sigma')}  \labelprop{monotonic-nochange-x}
\end{eqnarray}
then
\(
  \rely{r} \together \guar{g} \together \Pre{p} \Seq \Spec{x}{}{\Comprehension{}{\Eval{e}{\sigma} \Transrelationeq \Eval{x}{\sigma'} \Transrelationeq \Eval{e}{\sigma'}}{(\sigma,\sigma')}} ~\refsto~x := e.
\)
\end{lawx}
For example, the relation $\Transrelationeq$ may be $\geq$ on integers
with postcondition $\Eval{e}{\sigma} \geq \Eval{x}{\sigma'} \geq \Eval{e}{\sigma'}$,
or $\Transrelationeq$ may be $\leq$ on integers
with postcondition $\Eval{e}{\sigma} \leq \Eval{x}{\sigma'} \leq \Eval{e}{\sigma'}$,
or for a set-valued expression, $\Transrelationeq$ may be $\supseteq$
with postcondition $\Eval{e}{\sigma} \supseteq \Eval{x}{\sigma'} \supseteq \Eval{e}{\sigma'}$.

\begin{proof}
In the proof, the idiom, $\Nondet_j \cgd{(\EqEval{j}{e})} \Seq c$, can be thought of as introducing a logical variable $j$ to capture the initial value of $e$,
similar to the construct, $\kw{let}~j = e~\kw{in}~c$.
\begin{align*}&
  \rely{r} \together \guar{g} \together \Pre{p} \Seq \Spec{x}{}{\Comprehension{}{\Eval{e}{\sigma} \Transrelationeq \Eval{x}{\sigma'} \Transrelationeq \Eval{e}{\sigma'}}{(\sigma,\sigma')}}
 \Equals*[fresh $j$, $\Nondet_{j} \cgd{(\EqEval{j}{e})} = \cgd{(\Union_{j} \EqEval{j}{e})} = \cgd{\Sigma} = \Nil$ and \refprop{seq-test-assert}]
  \Nondet_{j} \cgd{(\EqEval{j}{e})} \Seq (\rely{r} \together \guar{g} \together \Pre{p \inter \EqEval{j}{e}} \Seq \Spec{x}{}{\Comprehension{}{\Eval{e}{\sigma} \Transrelationeq \Eval{x}{\sigma'} \Transrelationeq \Eval{e}{\sigma'}}{(\sigma,\sigma')}})
 \Equals*[by \reflaw{spec-strengthen-under-pre} as $j = \Eval{e}{\sigma}$]
  \Nondet_{j} \cgd{(\EqEval{j}{e})} \Seq (\rely{r} \together \guar{g} \together \Pre{p \inter \EqEval{j}{e}} \Seq \Spec{x}{}{\Comprehension{}{j \Transrelationeq \Eval{x}{\sigma'} \Transrelationeq \Eval{e}{\sigma'}}{(\sigma,\sigma')}})
 \Equals*[weaken precondition \refprop{assert-weaken} to $p \inter \GeEval{j}{e}$, which is stable under $r$]
  \Nondet_{j} \cgd{(\EqEval{j}{e})} \Seq (\rely{r} \together \guar{g} \together \Pre{p \inter \GeEval{j}{e}} \Seq \Spec{x}{}{\Comprehension{}{j \Transrelationeq \Eval{x}{\sigma'} \Transrelationeq \Eval{e}{\sigma'}}{(\sigma,\sigma')}})
 \Equals*[by \reflaw{rely-guar-assign} -- see below]
  \Nondet_{j} \cgd{(\EqEval{j}{e})} \Seq x := e
 \Equals*[as $ \Nondet_{j} \cgd{(\EqEval{j}{e})} = \Nil$]
  x := e
\end{align*}
For the application of \reflaw{rely-guar-assign}, $p$ is $p \inter \GeEval{j}{e}$, 
$p_1\,k$ is $p \inter \GeEval{j}{k} \inter \GeEval{k}{e}$,
and $q$ is $\Comprehension{}{j \Transrelationeq \Eval{x}{\sigma'} \Transrelationeq \Eval{e}{\sigma'}}{(\sigma,\sigma')}$.
Property $p_1\,k$ is stable under $r$ because
$r \limg p \inter \GeEval{j}{e} \inter \GeEval{k}{e} \rimg \subseteq r \limg p \rimg \inter r \limg \GeEval{j}{e} \rimg \inter r \limg \GeEval{k}{e} \rimg \subseteq p \inter \GeEval{j}{e} \inter \GeEval{k}{e}$
because $p$, $\GeEval{j}{e}$ and $\GeEval{k}{e}$ are stable under $r$ by \refprop{monotonic-under-r}.
Post condition $q$ tolerates $r$ from $p \inter \GeEval{j}{e}$ because $x$ is invariant under $r$ and \refprop{monotonic-under-r}.
Property $p_1\,k$ is established because 
$p \inter \GeEval{j}{e} \inter \EqEval{k}{e} \subseteq p \inter \GeEval{j}{k} \inter \GeEval{k}{e}$, which follows by set theory and logic.
The left side of \refprop{assignment} is contained in $g$ by \refprop{monotonic-g}
and it is contained in $q$ by the following reasoning,
which relies upon $\Transrelationeq$ being reflexive and transitive.
\begin{align*}&
  (p \inter \GeEval{j}{k} \inter \GeEval{k}{e}) \dres \Update{x}{k}
 \Subseteq*[rewriting as a set comprehension and \refprop{monotonic-nochange-x}]
  \Comprehension{}{j \Transrelationeq k \land k \Transrelationeq \Eval{e}{\sigma} \land \Eval{e}{\sigma} \Transrelationeq \Eval{e}{\sigma'} \land k = \Eval{x}{\sigma'}}{(\sigma,\sigma')}
 \Subseteq*[as $j \Transrelationeq k \land k = \Eval{x}{\sigma'} \implies j \Transrelationeq \Eval{x}{\sigma'}$ and $\Eval{x}{\sigma'} = k \land k \Transrelationeq \Eval{e}{\sigma} \land \Eval{e}{\sigma} \Transrelationeq \Eval{e}{\sigma'} \implies \Eval{x}{\sigma'} \Transrelationeq \Eval{e}{\sigma'}$]
  \Comprehension{}{j \Transrelationeq \Eval{x}{\sigma'} \Transrelationeq \Eval{e}{\sigma'}}{(\sigma,\sigma')}
 \qedhere
\end{align*}
\end{proof}

\begin{exax}[assign-pw]
\reflaw{rely-assign-monotonic} is applied to refine a rely/guarantee specification
to an assignment under interference that may remove elements from $w$.
\begin{align*}
&  \Rguar{w \supseteq w' \land w - w' \subseteq \{i\}} \together 
  \Rrely{w \supseteq w' \land pw' = pw} \together {} \\
&  \t1\RSpec{pw}{}{w \supseteq pw' \land pw' \supseteq w'} 
  \Refsto pw := w
\end{align*}%
The provisos of \reflaw*{rely-assign-monotonic} hold because
the guarantee $\Rel{w \supseteq w' \land w - w' \subseteq \{i\}}$ is reflexive,
$\Set{true}$ is trivially stable under any rely condition,
$pw$ is \constant\ under the rely $\Rel{w \supseteq w' \land pw' = pw}$,
$w$ is single-reference under any rely condition because access to $w$ is atomic, 
$\supseteq$ is a reflexive, transitive relation,
$\id{\overline{pw}}$ ensures the guarantee $\Rel{w \supseteq w' \land w - w' \subseteq \{i\}}$,
and
rely $\Rel{w \supseteq w' \land pw' = pw}$ ensures $\Rel{w \supseteq w'}$,
as does $\id{\overline{pw}}$.
\end{exax}

\begin{RelatedWork}
The assignment law of Coleman and Jones~\cite{CoJo07} requires that none of the variables in $e$ and $x$ 
are modified by the interference, which is overly restrictive.
Wickerson et al.~\cite{DBLP:conf/esop/WickersonDP10} use an atomic assignment statement
and assume the precondition and (single-state) postcondition are stable under the rely condition.
Xu et al.\ \cite{XuRoeverHe97}, Prensa Nieto~\cite{PrensaNieto03}, Schellhorn et al.~\cite{SchellhornTEPR14}, 
San{\'a}n et al.\ \cite{Sanan21} and Dingel~\cite{Dingel02})
also assume assignments are atomic.
\end{RelatedWork}

The laws developed above make the simplifying assumption 
that the expression in an assignment is single reference under the rely condition $r$.
That covers the vast majority of cases one needs in practice.
Although the underlying theory could be used to develop laws to handle expressions that are not single reference,
the cases single reference expressions do not cover can be handled by introducing a sequence of assignments
using local variables for intermediate results, 
such that the expression in each assignment is single reference under $r$.

\section{Conditionals}\labelsect{conditional}

A conditional statement,
\(
  \If b \Then c \Else d \Fi \defs (\Test{b}_{\true} \Seq c \nondet \Test{b}_{\false} \Seq d) \Seq \Idle \nondet
    \Nondet_{k \in \overline{\Boolean}} (\Test{b}_{k} \Seq \Abort),
\)
either evaluates its boolean condition $b$ to true and executes its ``then'' branch $c$,
or evaluates $b$ to false and executes its ``else'' branch $d$ \refdef{conditional}.
Because expressions in the language are untyped the definition includes a third alternative 
that aborts if the guard evaluates to a value other than true or false,
as represented here by the complement of the set of booleans, $\Boolean$.
The third alternative can also be used to cope with the guard expression being undefined,
e.g.~it includes a division by zero.
The $\Idle$ in the definition allows steps that do not modify observable state,
such as branching within an implementation.
\begin{lawx}[guar-conditional-distrib]
For any reflexive relation $g$, 
\[
  \guar{g} \together \If b \Then c \Else d \Fi  \refsto  \If b \Then (\guar{g} \together c) \Else (\guar{g} \together d) \Fi.
\]
\end{lawx}%

\begin{proof}
The proof follows because weak conjunction distributes over non-deterministic choice \refax{Nondet-sync-distrib} and 
guarantees distribute over sequential composition by \reflaw{guar-seq-distrib}.
Finally \reflaw{guar-eval} and \reflem{guar-idle} are applied because expression evaluation and $\Idle$
guarantee any reflexive guarantee.
\end{proof}

To simplify the presentation in this paper, 
when we use a boolean expression $b$ in a position in which a set of states is expected, 
it is taken to mean the corresponding set of states $\EqEval{b}{\true}$.
\\[0.5ex]
\noindent
\begin{minipage}{0.65\textwidth}
An informal motivation for the form of the law for refining to
a conditional is given via the control flow graph at the right
(which ignores the case when $b$ evaluates to a non-boolean). 
Dashed arcs indicate that interference can occur during the transition,
while un-dashed arcs indicate an instantaneous transition.
At entry the precondition $p$ is assumed to hold.
The precondition is assumed to be stable under the interference $r$ while the guard $b$ is evaluated,
and because the guard is assumed to be single reference
its value is that in one of the states during its evaluation,
call this its \emph{evaluation state}, $\sigma_1$.
If $b$ evaluates to true, $p \inter b$ holds in the evaluation state
but although $p$ is stable under $r$, $b$ may not be.
To handle that, it is assumed there is a set of states $b_t$
that is stable under $r$ and such that $p \inter b \subseteq b_t$.
\end{minipage}%
\begin{minipage}{0.35\textwidth}
\begin{center}
\input{if}
\end{center}
\end{minipage}\\[1ex]
If the guard evaluates to true the ``then'' branch of the conditional
is executed and it establishes the postcondition $q$ and terminates.
The postcondition relation $q$ is required to tolerate interference $r$ and hence
$q$ is also established between the initial and final states of the whole conditional.
The ``else'' branch is similar but uses a set of states $b_f$
such that $p \inter \pnegate{b} \subseteq b_f$ and $b_f$ is stable under $r$.

Interference from the environment may affect the evaluation of a boolean test $b$.
While each access to a variable within an expression is assumed to be atomic,
the overall evaluation of an expression is not assumed to be atomic.
Even if $b$ is single reference under $r$, 
it may evaluate to true in the state $\sigma_1$ 
in which the access to the single-reference variable occurs
but the interference may then change the state to a new state $\sigma_2$
in which $b$ no longer holds.
For example, the boolean expression $0 \leq x$ may evaluate to true in $\sigma_1$
but if the interference can decrease $x$ below zero, $0 \leq x$ may be false 
in the later state $\sigma_2$.

As a more complex example, consider $y < x \land y < z$ for $b$,
where interference cannot increase $x$ and leaves $y$ and $z$ unchanged.%
\footnote{This example boolean expression is similar to one required for Owicki's example \cite{Owicki75} to find the least index 
in an array that satisfies some property (see \cite[p.28]{HayesJones18}).}
If $y < x \land y < z$ evaluates to true in $\sigma_1$,
$y < z$ will still evaluate to true in state $\sigma_2$ after interference 
(because its variables are not modified)
but $y < x$ may be invalidated 
(because $x$ may be decreased so that $x \leq y$);
hence $y < z$ can be used for $b_t$.
The negation of the above example is $y \geq x \lor y \geq z$,
which can be used for $b_f$ 
because it is stable under interference that may only decrease $x$ and not change $y$ and $z$.
Note that
\begin{eqnarray*}
  p ~~=~~ (p \inter b) \union (p \inter \pnegate{b}) ~~\subseteq~~ b_t \union b_f
\end{eqnarray*}
but there may be states in which both $b_t$ and $b_f$ hold.
For the above example, taking $b_t$ as $y < z$ and $b_f$ as $y \geq x \lor y \geq z$,
both conditions hold in states satisfying $y < z \land (y \geq x \lor y \geq z)$,
i.e.\ states satisfying $z > y \land y \geq x$.

If the guard of the conditional evaluates to a non-boolean, the conditional aborts.
To avoid this possibility the law for introducing a conditional assumes that
the precondition $p$ ensures that $b$ evaluates to an element of type $\Boolean$.
Using the following definition, the latter is expressed as $p \subseteq \TypeOf{b}{\Boolean}$.
\begin{defix}[type-of]
For expression $e$ and set of values $T$,
\begin{eqnarray*}
  \TypeOf{e}{T} & \defs & \Comprehension{}{\Eval{e}{\sigma} \in T }{\sigma}.
\end{eqnarray*}
\end{defix}

\begin{lawx}[rely-conditional]
For a boolean expression $b$,
sets of states $p$, $b_t$ and $b_f$,
and relation $q$, if 
$b$ is single reference under $r$,
$q$ tolerates $r$ from $p$,
$p \inter b \subseteq b_t$,
$p \inter \pnegate{b} \subseteq b_f$,
$p \subseteq \TypeOf{b}{\Boolean}$,
and 
both $b_t$ and $b_f$ are stable under $p \dres r$,
\begin{align*}
  \rely{r} \together \Pre{p} \Seq \Spec{}{}{q} 
  & \refsto 
  \If b \Then (\rely{r} \together \Pre{b_t \inter p} \Seq \Spec{}{}{q}) \Else (\rely{r} \together \Pre{b_f \inter p} \Seq \Spec{}{}{q}) \Fi.
\end{align*}
\end{lawx}

\begin{proof}
For the application of \reflaw{spec-strengthen-under-pre} below, $p \dres \Finrel{r} \semi q \subseteq q$ 
by \refprop{q-tolerates-fin-r-before} as
$q$ tolerates $r$ from $p$.
The proof begins by duplicating the specification as $\nondet$ is idempotent.
\begin{align*}
&  (\rely{r} \together \Pre{p} \Seq \Spec{}{}{q}) \nondet (\rely{r} \together \Pre{p} \Seq \Spec{}{}{q})
  \Refsto*[by \reflaw{tolerate-interference} as $q$ tolerates $r$ from $p$, and $\Idle \refsto \Nil$]
    (\rely{r} \together \Pre{p} \Seq \Spec{}{}{q} \Seq \Idle) \nondet (\rely{r} \together \Pre{p} \Seq \Spec{}{}{q})
  \Equals*[by \reflaw{rely-seq-distrib} and \reflaw*{rely-remove}; \reflaw{spec-strengthen-under-pre}]
    (\rely{r} \together \Pre{p} \Seq \Spec{}{}{\Finrel{r} \semi q}) \Seq \Idle \nondet (\rely{r} \together \Pre{p} \Seq \Spec{}{}{\Finrel{r} \semi q})
  \Equals*[non-deterministic choice is idempotent]
    ((\rely{r} \together \Pre{p} \Seq \Spec{}{}{\Finrel{r} \semi q}) \nondet 
     (\rely{r} \together \Pre{p} \Seq \Spec{}{}{\Finrel{r} \semi q})) \Seq \Idle \nondet 
    (\rely{r} \together \Pre{p} \Seq \Spec{}{}{\Finrel{r} \semi q})
  \Refsto*[\reflaw{spec-seq-introduce} three times and $\Pre{\emptyset} = \Abort$]
    ((\rely{r} \together \Pre{p} \Seq \Spec{}{}{\Finrel{r} \rres (b_t \inter p)} \Seq \Pre{b_t \inter p} \Seq \Spec{}{}{q}) \nondet {} \\
&  ~(\rely{r} \together \Pre{p} \Seq \Spec{}{}{\Finrel{r} \rres (b_f \inter p)} \Seq \Pre{b_f \inter p} \Seq \Spec{}{}{q})) \Seq \Idle \nondet {} \\
&  (\rely{r} \together \Pre{p} \Seq \Spec{}{}{\Finrel{r} \rres \emptyset} \Seq \Abort)
  \Refsto*[by \reflaw*{rely-refine-within} using \reflaw*{rely-eval-expr} twice and assumptions; see below for third branch]
    (\Test{b}_{\true} \Seq (\rely{r} \together \Pre{b_t \inter p} \Seq \Spec{}{}{q}) \nondet 
    ~\Test{b}_{\false} \Seq (\rely{r} \together \Pre{b_f \inter p} \Seq \Spec{}{}{q})) \Seq \Idle \nondet {} \\
&  ~\Nondet_{k \in \overline{\Boolean}} (\Test{b}_{k} \Seq \Abort)
  \Equals*[definition of conditional \refdef{conditional}]
     \If b \Then (\rely{r} \together \Pre{b_t \inter p} \Seq \Spec{}{}{q}) \Else (\rely{r} \together \Pre{b_f \inter p} \Seq \Spec{}{}{q}) \Fi 
\end{align*}%
The third branch refinement holds as follows.
\begin{align*}
&  \rely{r} \together \Pre{p} \Seq \Spec{}{}{\Finrel{r} \rres \emptyset} \refsto \Nondet_{k \in \overline{\Boolean}} \Test{b}_{k}
 \ImpliedBy*[by \reflem{refine-choice} and $\Finrel{r} \rres \emptyset = \emptyset$]
  \forall k \in  \overline{\Boolean} \spot \rely{r} \together \Pre{p} \Seq \Spec{}{}{\emptyset} \refsto \Test{b}_{k}
 \ImpliedBy*[by \reflaw{rely-eval}]
  \forall k \in  \overline{\Boolean} \spot p \inter \EqEval{k}{b} = \emptyset
\end{align*}%
The latter holds from the assumption $p \subseteq \TypeOf{b}{\Boolean}$ and \refdefi{type-of}. 
\end{proof}

\begin{RelatedWork}
Wickerson et al.~\cite{Wickerson10-TR} develop a similar rule
but instead of $b_t$ and $b_f$ they use $\lceil b \rceil_r$ and $\lceil \lnot {b} \rceil_r$, respectively,
where they define $\lceil b \rceil_r$ as the smallest set $b_t$ 
such that $b \subseteq b_t$ and $b_t$ is stable under $r$.
That corresponds to requiring that $b_t$ in \reflaw{rely-conditional} is the least set 
containing $b$ that is stable under $r$.
\reflaw{rely-conditional} also takes into account that the precondition $p$ may also be assumed to hold
and hence is more flexible than the rule given by Wickerson et al.
As before another difference is that Wickerson et al.\ use postconditions of a single state,
rather than relations.

Coleman~\cite{ColemanVSTTE08} gives a rule for a simple conditional (with no ``else'' part).
The approach he takes is to split the guard expression $b$ into $b_s \land b_u$
in which $b_s$ contains no variables that can be modified by the interference $r$ and 
$b_u$ has a single variable that may be modified by $r$, and that variable only occurs once in $b_u$.
His conditions are strictly stronger than those used in \reflaw{rely-conditional} and hence
his rule is not as generally applicable.

Xu et al.\ \cite{XuRoeverHe97}, Prensa Nieto~\cite{PrensaNieto03}, Schellhorn et al.~\cite{SchellhornTEPR14}, 
San{\'a}n et al.\ \cite{Sanan21} and Dingel~\cite{Dingel02})
assume guard evaluation is atomic.
\end{RelatedWork}

\section{Recursion}\labelsect{recursion}

This section develops a law, \reflaw{well-founded-recursion}, to
handle recursion using well-founded induction to show termination.  It
uses a variant expression $w$ and a well-founded relation
$(\WFrelation{\_}{\_})$, and is applied in \refsect{loop} to verify
refinement laws for while loops, which are defined there using
recursion.

The proof of supporting lemma, \reflem*{well-founded-variant} below,
makes use of well-founded induction,
that for a property $P(k)$ defined on values,
can be stated as follows: if $(\WFrelation{\_}{\_})$ is well founded,
\begin{equation}\labeldef{well-founded-induction}
  \left(\forall k \spot (\forall j \spot \WFrelation{k}{j} \implies P(j)) \implies P(k)\right)
    ~~\implies~~ (\forall k \spot P(k)).
\end{equation}
The notation $\GeEval{e1}{e2}$ abbreviates 
$\Comprehension{}{\WFrelationeq{\Eval{e1}{\sigma}}{\Eval{e2}{\sigma}}}{\sigma}$ and 
$\GtEval{e1}{e2}$ abbreviates 
$\Comprehension{}{\WFrelation{\Eval{e1}{\sigma}}{\Eval{e2}{\sigma}}}{\sigma}$.

\begin{lemx}[well-founded-variant]
For a variant expression $w$,
commands $s$ and $c$,
and 
a well-founded relation $(\WFrelation{\_}{\_})$,
if for fresh $k$
\begin{equation}\labelprop{well-founded-variant-assumption}
  \forall k \spot (\Pre{\GtEval{k}{w}} \Seq s \refsto c) \implies (\Pre{\EqEval{k}{w}} \Seq s \refsto c)
\end{equation}
then $s \refsto c$.
\end{lemx}

\begin{proof}
The notation $\Nondet_{j}^{\WFrelation{k}{j}} c_j$ stands for the nondeterministic choice over all $c_j$
such that $\WFrelation{k}{j}$.
The proof starts from the assumption \refprop{well-founded-variant-assumption}.
\begin{align*}
&  \forall k \spot (\Pre{\GtEval{k}{w}} \Seq s \refsto c) \implies (\Pre{\EqEval{k}{w}} \Seq s \refsto c)
 \IFF*[by Galois connection between tests and assertions \refprop{Galois-assert-test} twice]
  \forall k \spot (s \refsto \cgd{(\GtEval{k}{w})} \Seq c) \implies (s \refsto \cgd{(\EqEval{k}{w})} \Seq c)
 \IFF*[union of tests \refax{Nondet-test} as $\GtEval{k}{w} = \Union_{j}^{\WFrelation{k}{j}} \EqEval{j}{w}$]
  \forall k \spot (s \refsto \Nondet_{j}^{\WFrelation{k}{j}} (\cgd{(\EqEval{j}{w})} \Seq c)) \implies (s \refsto \cgd{(\EqEval{k}{w})} \Seq c)
 \Implies*[by \reflem{refine-choice}]
  \forall k \spot (\forall j \spot \WFrelation{k}{j} \implies (s \refsto \cgd{(\EqEval{j}{w})} \Seq c)) \implies (s \refsto \cgd{(\EqEval{k}{w})} \Seq c)
 \Implies*[by well-founded induction \refdef{well-founded-induction} as $(\WFrelation{\_}{\_})$ is well founded]
  \forall k \spot s \refsto \cgd{(\EqEval{k}{w})} \Seq c
 \Implies*[by \reflem{refine-choice}]
  s \refsto \Nondet_{k} \cgd{(\EqEval{k}{w})} \Seq c
 \IFF*[as $k$ is fresh, $\Nondet_{k} \cgd{(\EqEval{k}{w})} = \cgd{(\Union_k \EqEval{k}{w})} = \cgd{\Sigma} = \Nil$ by \refax{Nondet-test}]
  s \refsto c
  \qedhere
\end{align*}
\end{proof}

\reflaw{well-founded-recursion}
applies \reflem*{well-founded-variant} for $c$ in the form
of the greatest fixed point, $\InfFP f$, 
of a monotone function $f$ on commands.

\begin{lawx}[well-founded-recursion]
For a set of states $p$,
a variant expression $w$, 
a command $s$, 
a well-founded relation $(\WFrelation{\_}{\_})$,
and a monotone function on commands $f$, if 
\begin{eqnarray}
  \Pre{p} \Seq s & \refsto & \InfFP f \labelprop{well-founded-assumption1} \\
  \forall k \spot \Pre{\EqEval{k}{w}} \Seq s & \refsto & f(\Pre{\GtEval{k}{w} \union p} \Seq s) \labelprop{well-founded-assumption2}
\end{eqnarray}
then,
$s \refsto \InfFP f$.
\end{lawx}
The proviso \refprop{well-founded-assumption1} is typically used to handle the case in which
$p$ holding initially ensures $\InfFP f$ does not utilise any recursive calls,
e.g. taking $f$ to be $\lambda x \spot \If b \Then (c \Seq x) \Else \Nil \Fi$, it allows one to handle
the case in which the loop guard $b$ is guaranteed to evaluate to false.
A special case of the law is if $p = \emptyset$,
in which case proviso \refprop{well-founded-assumption1} holds trivially.

\begin{proof}
By \reflem{well-founded-variant} to show $s \refsto \InfFP f$, it suffices to show,
\[
  \forall k \spot (\Pre{\GtEval{k}{w}} \Seq s \refsto \InfFP f) \implies (\Pre{\EqEval{k}{w}} \Seq s \refsto \InfFP f).
\]
To show this for any $k$, assume $\Pre{\GtEval{k}{w}} \Seq s \refsto \InfFP f$ and show
\begin{align*}
& \Pre{\EqEval{k}{w}} \Seq s 
 \Refsto*[by \refprop{well-founded-assumption2}] 
  f(\Pre{\GtEval{k}{w} \union p} \Seq s)
 \Refsto*[by \reflem{assert-merge} as $\Pre{\GtEval{k}{w}} \Seq s \refsto \InfFP f$ and \refprop{well-founded-assumption1}; $f$ is monotone]
  f(\nu f)
 \Equals*[folding fixed point]
  \nu f
  \qedhere
\end{align*}
\end{proof}

\begin{RelatedWork}
Schellhorn et al.~\cite{SchellhornTEPR14} include recursion in their approach.
San{\'a}n et al.\ \cite{Sanan21} allow parameterless procedures and 
make use of a natural number call depth bound to avoid infinite recursion.
The other approaches \cite{CoJo07,Dingel02,PrensaNieto03,DBLP:conf/esop/WickersonDP10,XuRoeverHe97} do not consider recursion,
instead they define the semantics of while loops via an operational semantics,
as do San{\'a}n et al.\ \cite{Sanan21}.
\end{RelatedWork}

\section{While loops}\labelsect{loop}

The definition of a while loop,
\(
  \While b \Do c \Od \defs \nu x \spot \If b \Then (c \Seq x) \Else \Nil \Fi,
\) 
is in terms of a recursion involving a conditional \refdef{while}.%
\footnote{One known subtlety of fixed points is that the degenerate case of this definition  
$\While \true \Do \Nil \Od \refsto (\InfFP x \spot x) = \Abort$.
If either the guard of the loop or its body require at least one step the loop is no longer degenerate.
This is not an issue in the context of refinement because the only specification
that is refined by a degenerate loop is equivalent to $\Abort$.
}
As usual, a fixed point of the form $\InfFP(\lambda x \spot body)$ is abbreviated to $(\InfFP x \spot body)$.
The Hoare logic rule for reasoning about a loop, 
$\While b \Do c \Od$,
for sequential programs uses 
an invariant $p$ that is maintained by the loop body whenever $b$ holds initially~\cite{Hoare69a}.
To show termination a variant expression $w$ is used \cite{Gries81}.
The loop body must strictly decrease $w$ according to a well-founded relation $(\WFrelation{\_}{\_})$
whenever $b$ holds initially,
unless the body establishes the negation of the loop guard.
The relation $(\WFrelation{\_}{\_})$ is assumed to be transitive 
(otherwise just take its transitive closure instead);
its reflexive closure is written $(\WFrelationeq{\_}{\_})$.
The notation, $\Dec{w}$, stands for the relation between states for which $w$ decreases
(i.e.\ $\Comprehension{}{\WFrelation{\Eval{w}{\sigma}}{\Eval{w}{\sigma'}}}{(\sigma,\sigma')}$) and
$\Deceq{w}$ stands for the relation between states for which $w$ decreases or is unchanged
(i.e.\ $\Comprehension{}{\WFrelationeq{\Eval{w}{\sigma}}{\Eval{w}{\sigma'} }}{(\sigma,\sigma')}$).

The law for while loops needs to rule out interference invalidating the loop invariant $p$ or
increasing the variant $w$.
The invariant $p$ and variant $w$ must tolerate interference satisfying
the rely condition $r$ and hence $p$ must be stable under $r$ and $p \dres r \subseteq \Deceq{w}$.

To explain \reflaw{rely-loop-early}, 
which specifies proof obligations sufficient to show
\begin{displaymath}
    \guar{g} \together \rely{r} \together \Pre{p} \Seq \Spec{}{}{\Finrel{q} \rres (p \inter b_f)} 
    ~~\refsto~~ 
    \While b \Do c \Od~ ,
\end{displaymath}
we use the figure below containing a 
control flow graph for a $\While$ loop
that has been augmented by annotations either side of the
dashed arcs indicating that the relations $\Finrel{r}$ and $\Deceq{w}$ are
satisfied by the environment.
Weak correctness is considered first and then termination.
The loop invariant $p$ is assumed to hold at entry.
The invariant is assumed to be stable under $r$ while evaluation of the guard $b$ takes place
and because the guard is assumed to be single reference,
its value is that in one of the states during its evaluation,
call this its \emph{evaluation state}, $\sigma_1$.
\par\noindent
\begin{minipage}{0.55\textwidth}
If $b$ evaluates to false, $p \inter \pnegate{b}$ holds in the evaluation state $\sigma_1$
but although $p$ is stable under $r$, $\pnegate{b}$ may not be.
To handle that, it is assumed there is a set of states $b_f$
that is stable under $r$ and such that $p \inter \pnegate{b} \subseteq b_f$.
Because the loop terminates when $b$ evaluates to false,
the loop establishes the postcondition $p \inter b_f$.
If the guard evaluates to true, $p \inter b$ holds in $b$'s evaluation state $\sigma_1$.
Again $b$ may not be stable under $r$ and so 
a set of states $b_t$ that is stable under $r$ is used,
where $p \inter b \subseteq b_t$.
That set of states is satisfied on entry to the body of the loop
and the body is required to re-establish $p$,
thus re-establishing the invariant for further iterations of the loop.
\end{minipage}%
\begin{minipage}{0.45\textwidth}
\begin{center}
\input{loop}
\end{center}
\end{minipage}\\[0.5ex]
The loop body is also required to establish the postcondition $\Finrel{q}$,
which must tolerate $r$ from $p$.
The reason for using $\Finrel{q}$ (instead of $q$) is to allow for the case where it is the environment 
that reduces the variant and the loop body does nothing (the reflexive case),
and the case in which the environment achieves $q$ or $\Finrel{q}$ 
while the body is executing and the body also achieves $q$.
Of course, if $q$ is reflexive and transitive, $q = \Finrel{q}$.

To show termination a variant expression $w$ is used.
The following version of the while loop rule allows for
the body of the loop to not reduce the variant provided it stably establishes 
the negation of the loop guard.
It makes use of an extra set of states $b_x$ that if satisfied on termination
of the body of the loop ensures that the loop terminates.
Because $b_x$ is stable under $r$, if it holds at the end of the body of the loop,
it still holds when the loop condition is evaluated and 
ensures it evaluates to false.
Each iteration of the loop must either
establish $b_x$
or 
reduce $w$ according to a well-founded relation $(\WFrelation{\_}{\_})$.
The termination argument would not be valid if the environment could increase $w$,
so it is assumed the environment maintains the reflexive closure of the ordering,
i.e.\ $p \dres r \subseteq \Deceq{w}$.

\begin{lawx}[rely-loop-early]
Given sets of states $p$, $b_t$, $b_f$ and $b_x$,
relations $q$ and $r$,
reflexive relation $g$,
a boolean expression $b$ that is single-reference under $r$,
a variant expression $w$
and a relation $(\WFrelation{\_}{\_})$ that is well-founded,
such that 
$p \subseteq \TypeOf{b}{\Boolean}$,
$\Finrel{q} \rres p$ tolerates $r$ from $p$,
$b_t$, $b_f$ and $b_x$ are stable under $p \dres r$, and
\[
   \begin{array}{cccc}
    p \dres r  \subseteq  \Deceq{w} 
    ~~~&~~~p \inter b  \subseteq  b_t 
    ~~~&~~~p \inter \pnegate{b}  \subseteq  b_f 
    ~~~&~~~p \inter b_x  \subseteq  \pnegate{b} 
   \end{array}
\]
then if for all $k$,
\begin{equation}
  \guar{g} \together \rely{r} \together \Pre{b_t \inter p \inter \GeEval{k}{w}} \Seq \Spec{}{}{\Finrel{q} \rres (p\inter (\GtEval{k}{w} \union b_x))} \refsto c
  \labelprop{loop-assumption}
\end{equation}
then,
\(
    \guar{g} \together \rely{r} \together \Pre{p} \Seq \Spec{}{}{\Finrel{q} \rres (p \inter b_f)} 
    ~~\refsto~~ 
    \While b \Do c \Od.
\)
\end{lawx}

\begin{proof}
Because a while loop is defined in terms of recursion and a conditional command,
the proof makes use of \reflaw{well-founded-recursion} and \reflaw{rely-conditional}.
We introduce the following two abbreviations.
\begin{eqnarray*}
  f & \defs & \lambda x \spot \If b \Then (c \Seq x) \Else \Nil \Fi \\
  s & \defs & \guar{g} \together \rely{r} \together \Pre{p} \Seq \Spec{}{}{\Finrel{q} \rres (p \inter b_f)}
\end{eqnarray*}
The law can be restated using $s$ and proven as follows.
\begin{eqnarray}
 && s \refsto \While b \Do c \Od \nonumber \\
 & \Leftrightarrow & \Why{definitions of a while loop \refdef{while} and $f$} \nonumber \\
 && s \refsto \InfFP f \nonumber \\
 &\Leftarrow & \Why{by \reflaw{well-founded-recursion}} \nonumber \\
 && \Pre{b_x} \Seq s \refsto \InfFP f \land \labelprop{case-early} \\
 && \forall k \spot \Pre{\EqEval{k}{w}} \Seq s \refsto \If b \Then (c \Seq \Pre{\GtEval{k}{w} \union b_x} \Seq s) \Else \Nil \Fi \labelprop{case-induction}
\end{eqnarray}
The first condition \refprop{case-early} holds because if $b_x$ holds initially,
from the assumptions $p \inter b_x \subseteq \overline{b}$ 
and hence the conditional must take the null $\Else$ branch.
That allows one to choose $b_t$ to be $\emptyset$
and hence allows any command for the ``then'' part of the conditional
so one can choose $c \Seq \InfFP f$.
The detailed proof of \refprop{case-early} follows, starting with expanding the definition of $s$.
\begin{align*}
&  \guar{g} \together \rely{r} \together \Pre{b_x \inter p} \Seq \Spec{}{}{\Finrel{q} \rres (p \inter b_f)}
 \Refsto*[by \reflaw*{rely-conditional} with $\emptyset$ as its $b_t$ and $b_x$ as its $b_f$;  as $\Finrel{q} \rres (p \inter b_f)$ tolerates $r$ from $b_x \inter p$]
  \guar{g} \together 
    \If b\begin{array}[t]{l}
      \Then~
       (\rely{r} \together \Pre{\emptyset \inter p} \Seq \Spec{}{}{\Finrel{q} \rres(p \inter b_f)}) \\
    \Else~~~
       (\rely{r} \together \Pre{b_x \inter p} \Seq \Spec{}{}{\Finrel{q} \rres (p \inter b_f)}) \Fi
    \end{array}
 \Refsto*[by \reflaw{rely-remove} twice and \reflaw{guar-conditional-distrib}]
    \If b \Then 
      (\guar{g} \together \Pre{\emptyset} \Seq \Spec{}{}{\Finrel{q} \rres(p \inter b_f)}) 
    \Else 
      (\guar{g} \together \Pre{b_x \inter p} \Seq \Spec{}{}{\Finrel{q} \rres (p \inter b_f)}) \Fi
 \Refsto*[precondition $\emptyset$ allows any refinement; \reflem{spec-to-test} as $p \inter b_x \subseteq p \inter b_f$]
  \If b \Then c \Seq \InfFP f \Else \Nil \Fi
 \Refsto*[folding fixed point. i.e. $f(\InfFP f) = \InfFP f$]
  \InfFP f
\end{align*}%
To show the second condition \refprop{case-induction},
for any $k$ consider the following refinement with the definition of $s$ substituted in.
\begin{align*}
&  \Pre{\EqEval{k}{w}} \Seq (\guar{g} \together \rely{r} \together \Pre{p} \Seq \Spec{}{}{\Finrel{q} \rres (p \inter b_f)})
 \Refsto*[by \reflem{assert-distrib} and merging preconditions \refprop{seq-assert-assert}]
  \guar{g} \together \rely{r} \together \Pre{p \inter \EqEval{k}{w}} \Seq \Spec{}{}{\Finrel{q} \rres (p \inter b_f)}
 \Refsto*[weaken the precondition \refprop{assert-weaken} so that it is stable under $r$]
  \guar{g} \together \rely{r} \together \Pre{p \inter \GeEval{k}{w}} \Seq \Spec{}{}{\Finrel{q} \rres (p \inter b_f)}
 \Refsto*[by \reflaw{rely-conditional} as $\Finrel{q} \rres (p \inter b_f)$ tolerates $r$ from $p$,  $\GeEval{k}{w}$ stable]
  \guar{g} \together \If b
    \begin{array}[t]{l}
      \Then~
       (\rely{r} \together \Pre{b_t \inter p \inter \GeEval{k}{w}} \Seq \Spec{}{}{\Finrel{q} \rres (p \inter b_f)}) \\
      \Else~
       (\rely{r} \together \Pre{b_f \inter p \inter \GeEval{k}{w}} \Seq \Spec{}{}{\Finrel{q} \rres (p \inter b_f)}) \Fi
    \end{array}
 \Refsto*[by \reflaw{guar-conditional-distrib} as $g$ is reflexive]
  \If b \Then 
    (\guar{g} \together \rely{r} \together \Pre{b_t \inter p \inter \GeEval{k}{w}} \Seq \Spec{}{}{\Finrel{q} \rres (p \inter b_f)}) \\
&\t1\Else~
    (\guar{g} \together \rely{r} \together \Pre{b_f \inter p \inter \GeEval{k}{w}} \Seq \Spec{}{}{\Finrel{q} \rres (p \inter b_f)}) \Fi
\end{align*}%
To complete the refinement in \refprop{case-induction} we need to show both the following.
\begin{align}
  \guar{g} \together \rely{r} \together \Pre{b_t \inter p \inter \GeEval{k}{w}} \Seq \Spec{}{}{\Finrel{q} \rres (p \inter b_f)} & \refsto c \Seq \Pre{\GtEval{k}{w} \union b_x} \Seq s \labelprop{while-then} \\
  \guar{g} \together \rely{r} \together \Pre{b_f \inter p \inter \GeEval{k}{w}} \Seq \Spec{}{}{\Finrel{q} \rres (p \inter b_f)} & \refsto \Nil \labelprop{while-else}
\end{align}
First, \refprop{while-then} is shown as follows.
\begin{align*}
&  \guar{g} \together \rely{r} \together \Pre{b_t \inter p \inter \GeEval{k}{w}} \Seq \Spec{}{}{\Finrel{q} \rres (p \inter b_f)}
 \Refsto*[by \reflaw{spec-seq-introduce}]
  \guar{g} \together \rely{r} \together \Pre{b_t \inter p \inter \GeEval{k}{w}} \Seq \Spec{}{}{\Finrel{q} \rres (p \inter (\GtEval{k}{w} \union b_x))} \Seq \\\t1
&    \Pre{p \inter (\GtEval{k}{w} \union b_x)} \Seq \Spec{}{}{\Finrel{q} \rres (p \inter b_f)}
 \Refsto*[use the assumption \refprop{loop-assumption} to refine to $c$, and definition of $s$]
  c \Seq \Pre{\GtEval{k}{w} \union b_x} \Seq s
\end{align*}%
Second \refprop{while-else} is refined as follows.
\begin{align*}
&  \guar{g} \together \rely{r} \together \Pre{b_f \inter p \inter \GeEval{k}{w}} \Seq \Spec{}{}{\Finrel{q} \rres (p \inter b_f)}
 \Refsto*[using \reflaw{spec-strengthen-under-pre}; remove the precondition \refprop{assert-remove}]
  \guar{g} \together \rely{r} \together \Spec{}{}{\id{}}
  \Refsto*[by \reflem{spec-to-test} as $\dom{(\id{} \inter \id{})} = \Sigma$ and $\cgd{\Sigma} = \Nil$
  ]
  \Nil
  \qedhere
\end{align*}
\end{proof}

\begin{exax}[while-loop]
The following example uses \reflaw{rely-loop-early} to introduce a loop that 
repeatedly attempts to remove an element $i$ from a set $w$ under interference that 
may remove elements from $w$.
It is assumed the implementation uses a compare-and-swap operation that may fail due to interference.
For termination, it uses the finite set $w$ as the variant expression under the superset ordering,
which is well founded on finite sets.
If the loop body does not succeed in removing $i$ due to interference,
that interference must have removed some element (possibly $i$) from $w$ 
and hence reduced the variant.
The specification (\ref{ex-rem-from-set-spec}) from \refsect{introduction} is repeated here.%
\footnote{Here we use a frame of $w$, rather than having $i' = i$ in the guarantee as in (\ref{ex-rem-from-set-spec}).} 
\begin{align*}
 & \Rguar{w \supseteq w' \land w - w' \subseteq \{i\}} \together
    \Rrely{w \supseteq w' \land i' = i} \together {} \\
 &\t1 \SPre{w \subseteq \Subrange{0}{N-1} \land i \in \Subrange{0}{N-1}} \Seq 
      \RSpec{w}{}{i' \not\in w'} \\
 \refsto~
 & \While i \in w \Do \\
 & \t1 \Rguar{w \supseteq w' \land w - w' \subseteq \{i\}} \together \Rrely{w \supseteq w' \land i' = i} \together {} \\
 & \t2      \SPre{w \subseteq \Subrange{0}{N-1} \land i \in \Subrange{0}{N-1}} \Seq \RSpec{w}{}{w \supset w' \lor i' \not\in w'} \\
 & \Od
\end{align*}
The form of loop introduction rule that includes the negation of the guard as an alternative
is sometimes referred to as an \emph{early-termination} version.
In this case the early termination version is essential.
For the application of \reflaw*{rely-loop-early},
the invariant $p$ is $\Set{w \subseteq \Subrange{0}{N-1} \land i \in \Subrange{0}{N-1}}$,
the loop test $b$ is $\Set{i \in w}$,
$b_t$ is $\Set{\true}$ (as interference may remove $i$ from $w$),
both $b_f$ and $b_x$ are $\Set{i \notin w}$,
the postcondition $q$ is $\Rel{\true}$,
and the variant expression is $w$ under the well-founded superset ordering $\WFrelation{\_}{\_}$ (as $w$ is finite).
One subtlety is that it may be the interference, rather than the body of the loop,
that removes $i$ from $w$ and hence establishes the postcondition;
either way the loop terminates with the desired postcondition.
Another subtlety is that interference may remove $i$ just after $w$ is sampled for the loop guard evaluation
but before the loop body begins, 
and hence the loop body cannot guarantee that $w$ is decreased either by the program or the interference.
For this reason it is essential that the loop body have the (early termination) alternative $\Rel{i' \notin w'}$ in its postcondition.
The condition $\Set{i \in w}$ is single reference under the rely $\Rel{w \supseteq w' \land i' = i}$
because both $i$ and $w$ are single reference under the rely and $i$ is invariant under the rely.
It is straightforward that the invariant implies that the type of $\Set{i \in w}$ is boolean,
and that $b_t$, $b_f$ and $b_x$ are stable under the rely.
Because $q$ is $\Rel{\true}$, 
the proviso that $\Finrel{q} \rres p$ tolerates $r$ from $p$ corresponds to 
$p$ being stable under $r$, which is straightforward.
The final proof obligation \refprop{loop-assumption} corresponds to showing the following for all $k$.
\begin{align*}
  &\Rguar{w \supseteq w' \land w - w' \subseteq \{i\}} \together 
     \Rrely{w \supseteq w' \land i' = i} \together {} \\
  & \t1 \SPre{w \subseteq \Subrange{0}{N-1} \land i \in \Subrange{0}{N-1} \land k \supseteq w} \Seq \\
  &  \t1 \RSpec{w}{}{w' \subseteq \Subrange{0}{N-1} \land i' \in \Subrange{0}{N-1} \land (k \supset w' \lor i' \notin w')} \\
  & \refsto 
     \Rguar{w \supseteq w' \land w - w' \subseteq \{i\}} \together \Rrely{w \supseteq w' \land i' = i} \together {}\\
  & \t1 \SPre{w \subseteq \Subrange{0}{N-1} \land i \in \Subrange{0}{N-1}} \Seq \RSpec{w}{}{w \supset w' \lor i' \not\in w'})
\end{align*}
The refinement holds because 
the frame of $w$ combined with the rely $\Rel{i' = i}$ implies $i$ is unmodified,
and \reflaw{spec-strengthen-with-trading} can be used to complete the proof (see \refexa{loop-body}).
\end{exax}

\begin{lawx}[rely-loop]
Given set of states $p$, $b_t$ and $b_f$,
relations $q$ and $r$,
reflexive relation $g$,
a variant expression $w$
and a transitive relation $(\WFrelation{\_}{\_})$ that is well-founded,
if $b$ is a boolean expression that is single-reference under $r$,
$p \subseteq \TypeOf{b}{\Boolean}$,
$\Finrel{q} \rres p$ tolerates $r$ from $p$,
$b_t$ and $b_f$ are stable under $p \dres r$,
and
\[
   \begin{array}{cccc}
    p \dres r \subseteq \Deceq{w} 
    ~~~&~~~p \inter b \subseteq b_t 
    ~~~&~~~p \inter \pnegate{b} \subseteq b_f
   \end{array}
\]
then if for all $k$,
\begin{equation*}
    \guar{g} \together \rely{r} \together \Pre{b_t \inter p \inter \GeEval{k}{w}} \Seq \Spec{}{}{\Finrel{q} \rres (p\inter \GtEval{k}{w})} \refsto c
\end{equation*}
then,
\(
    \guar{g} \together \rely{r} \together \Pre{p} \Seq \Spec{}{}{(\Deceq{w} \inter \Finrel{q}) \rres (p \inter b_f)} 
    ~~\refsto~~ 
    \While b \Do c \Od.
\)
\end{lawx}

\begin{proof}
The proof follows from \reflaw*{rely-loop-early} using $\emptyset$ for $b_x$ and
$(\Deceq{w} \inter q)$ for $q$.
\end{proof}

\begin{RelatedWork}
The simpler \reflaw{rely-loop} is proved using \reflaw{rely-loop-early},
which handles the ``early termination'' case when the body does not necessarily
reduce the variant but instead a condition that ensures the loop guard is (stably) false is established.
For sequential programs, the early termination variant is usually proved in terms 
of the simpler law by using a more complex variant involving a pair consisting of
the loop guard and a normal variant under a lexicographical order~\cite{Gries81}; 
that variant decreases if the body changes the guard from true to false.
Interestingly, such an approach is not possible in the case of concurrency
because it may be the environment that establishes that the guard is false
rather than the loop body, and in that case the body may not decrease the variant pair.

Our use of a variant expression is in line with showing termination 
for sequential programs in Hoare logic.
It differs from the approach used by Coleman and Jones~\cite{CoJo07},
which uses a well-founded relation $rw$ on the state to show termination
in a manner similar to the way a well-founded relation is used to 
show termination of a $\While$ loop for a sequential program in VDM~\cite{Jones90a}.
In order to cope with interference, Coleman and Jones require 
that the rely condition $r$ implies the reflexive transitive closure of $rw$,
i.e.\ $r \subseteq rw^*$.
However, that condition is too restrictive 
because, if an environment step does not satisfy $rw^+$,
it must not change the state at all.
That issue was addressed in~\cite{HayesJonesColvin14TR} by requiring
the weaker condition $r \subseteq (rw^+ \union \id{X})$,
where $X$ is the set of variables on which the well-foundedness of relation $rw$ depends
(i.e.\ the smallest set of variables, $X$, such that $\id{X} \semi rw \semi \id{X} \subseteq rw$).

In the approach used here, the requirement on the environment 
is that it must not increase the variant expression.
If the environment does not decrease the value of the variant expression,
it must ensure that the variant is unchanged, 
rather than the complete state of the system is unchanged 
as required by Coleman and Jones~\cite{CoJo07}.
It is also subtly more general than the approach used in~\cite{HayesJonesColvin14TR}
because here we require that the variant is not increased by the environment
but allow variables referenced in the variant to change,
e.g.\ if the variant is $(x \bmod N)$, 
the environment may add $N$ to $x$ without changing the value of the variant.
The approach using a variant expression is also easier to comprehend 
compared to that in~\cite{HayesJonesColvin14TR}.

In the approach used by Coleman and Jones~\cite{CoJo07}
the postcondition for the loop body $rw$ is required to be transitive and well founded.
Well foundedness is required to show termination and hence $rw$ cannot be reflexive.
In the version used here, termination is handled using a variant and 
hence the postcondition of the loop body can be the same as the overall specification, i.e.\ $\Finrel{q}$.
Because $q \subseteq \Finrel{q}$, \reflaw{rely-loop-early} can be weakened to a law 
that has a body with a postcondition of $q$.

Wickerson et al.~\cite{DBLP:conf/esop/WickersonDP10,Wickerson10-TR} 
only consider partial correctness, as do Prensa Nieto~\cite{PrensaNieto03}
and San{\'a}n et al.\ \cite{Sanan21}.
Dingel~\cite{Dingel02} makes use of a natural number valued variant
that, like here, cannot be increased by the environment.
Unlike the other approaches, \reflaw{rely-loop-early} allows for the early termination case,
which unlike in the sequential case cannot be proven from 
the non-early termination variant \reflaw{rely-loop}
and hence they cannot handle the refinement in \refexa{while-loop}. 
\end{RelatedWork}

\section{Refinement of removing element from a set}\labelsect{ex-remove}

The laws developed in this paper have been used for the refinement 
of some standard concurrent algorithms in~\cite{HayesJones18}
and the reader is referred there for additional examples,
including a parallel version of the prime number sieve,
which includes an operation to remove an element from a set,
for which we present a refinement to code of the example specification (\ref{ex-rem-from-set-spec}),
which we repeat here.
\begin{align*}
 & \Rguar{w \supseteq w' \land w - w' \subseteq \{i\}} \together
    \Rrely{w \supseteq w' \land i' = i} \together {} \\
 & \t1\SPre{w \subseteq \Subrange{0}{N-1} \land i \in \Subrange{0}{N-1}} \Seq 
      \RSpec{w}{}{i' \not\in w'} \numberthis\label{ex-rem-from-set2}
\end{align*}
If the machine on which the operation is to be implemented provides an atomic instruction
to remove an element from a set (represented as a bitmap) the implementation would be trivial.
Here we assume the machine has an atomic compare-and-swap (CAS) instruction.
The CAS makes use of one shared variable,
$w$, the variable to be updated,
and two local variables,
$pw$, a sample of the previous value of $w$,
and
$nw$, the new value which $w$ is to be updated to.
The CAS is used by sampling $w$ into $pw$, 
calculating the new value $nw$ in terms of $pw$,
and then executing the CAS to atomically update $w$ to $nw$ if $w$ is still equal to $pw$,
otherwise it fails and leaves $w$ unchanged.
The CAS has the following specification, repeated from \refdef{CAS}. 
  \begin{align*}
 CAS \defs\ 
 &    \Frame{w}{\Ratomicrel{
                                               (w = pw \implies w' = nw) \land 
                                               (w \neq pw \implies w' = w)  \numberthis\labeldef{CAS2}
                      }}
\end{align*}
Because the execution of a CAS may fail if interference modifies $w$ 
between the point at which $w$ was sampled (into $pw$) 
and the point at which the CAS reads $w$,
a loop is required to repeat the use of the CAS until $i \not\in w$.
The first refinement step is to introduce a loop
that terminates when $i$ has been removed from $w$
using \reflaw{rely-loop-early} -- see \refexa{while-loop} for details of the application of the law.
\begin{align*}
 (\ref{ex-rem-from-set2}) \refsto~
 & \While i \in w \Do \\
 & \t1 (\Rguar{w \supseteq w' \land w - w' \subseteq \{i\}} \together \Rrely{w \supseteq w' \land i' = i} \together {} \\
 & \t1 \SPre{w \subseteq \Subrange{0}{N-1} \land i \in \Subrange{0}{N-1}} \Seq \RSpec{w}{}{w \supset w' \lor i' \not\in w'}) \numberthis\label{loop-body} \\
 & \Od
\end{align*}
At this point we need additional local variables $pw$ and $nw$.
This paper does not cover local variable introduction laws 
(see \cite{FormaliSE_localisation}).
A local variable is handled here by requiring each local variable name to be fresh,
adding it to the frame, and assuming it is unchanged in the rely condition.%
\footnote{This is equivalent to treating the local variables as global variables
that are not modified by the environment of the program.}
The body of the loop can be decomposed into a sequential composition of three steps:
sampling $w$ into $pw$, calculating $nw$ as $pw - \{i\}$, and applying the CAS.
The guarantee does not contribute to the refinement steps here, 
and hence only the refinement of the remaining components is shown.
\begin{align*}
  & \Rrely{w \supseteq w' \land i' = i \land nw' = nw \land pw' = pw} \together {} \\
  & \t1 \RSpec{nw,pw,w}{}{w \supset w' \lor i' \notin w'} \\
 \refsto~ & \Why{by \reflaw{spec-seq-introduce} twice -- see \refexa*{spec-seq-introduce}} \\
  & \Rrely{w \supseteq w' \land i' = i \land nw' = nw \land pw' = pw} \together {} \\
  & \t1 \RSpec{nw,pw,w}{}{w \supseteq pw' \land pw' \supseteq w'}  \Seq \SPre{pw \supseteq w} \Seq \numberthis\label{read-w} \\
  & \t1 \RSpec{nw,pw,w}{}{nw' = pw - \{i\} \land pw' = pw \land pw' \supseteq w' \land i' = i} \Seq \numberthis\label{assign-nw} \\
  & \t1 \SPre{pw \supseteq w \land nw = pw - \{i\}}; \RSpec{nw,pw,w}{}{pw \supset w' \lor i' \notin w'}  \numberthis\label{CAS-part} \\
 \refsto~ & \Why{by \reflaw{frame-restrict} thrice -- see \refexa{frame-restrict}} \\
  & \Rrely{w \supseteq w' \land i' = i \land nw' = nw \land pw' = pw} \together {} \\
  & \t1 \RSpec{pw}{}{w \supseteq pw' \land pw' \supseteq w'}  \Seq \numberthis\label{xread-w} \\
  & \t1 \SPre{pw \supseteq w} \Seq \RSpec{nw}{}{nw' = pw - \{i\} \land pw' \supseteq w'} \Seq \numberthis\label{xassign-nw} \\
  & \t1 \SPre{pw \supseteq w \land nw = pw - \{i\}}; \RSpec{w}{}{pw \supset w' \lor i' \notin w'} \numberthis\label{xCAS-part} \\
\end{align*}
The postconditions in (\ref{read-w}), (\ref{assign-nw}) and (\ref{CAS-part}) 
were chosen to tolerate interference that may remove elements from $w$,
for example,
in (\ref{read-w}) the value of $w$ is captured in the local variable $pw$ 
but because elements may be removed from $w$ via interference
(\ref{read-w}) can only ensure that $w \supseteq pw' \land pw' \supseteq w'$.

The refinement of (\ref{xread-w})
in the guarantee context from (\ref{loop-body}) 
uses \reflaw{rely-assign-monotonic}.
The guarantee holds trivially as $w$ is not modified.
\begin{align*}
   & \Rguar{w \supseteq w' \land w - w' \subseteq \{i\}} \together 
      \Rrely{w \supseteq w' \land i' = i \land nw' = nw \land pw' = pw} \together {} \\
   &  \t1    \RSpec{pw}{}{w \supseteq pw' \land pw' \supseteq w'} \\
 \refsto~& \Why{by \reflaw{rely-assign-monotonic} -- see \refexa{assign-pw}} \\
    & pw := w
\end{align*}
Specification (\ref{xassign-nw}) involves an update to a local variable ($nw$) 
to an expression involving only local variables ($pw$ and $i$)
and hence its refinement can use a simpler assignment law.
\begin{align*}
  & \Rguar{w \supseteq w' \land w - w' \subseteq \{i\}} \together 
     \Rrely{w \supseteq w' \land i' = i \land nw' = nw \land pw' = pw} \together {} \\
  & \t1  \SPre{pw \supseteq w} \Seq \RSpec{nw}{}{nw' = pw - \{ i \} \land pw' \supseteq w'}  \\
 \refsto~ & \Why{by \reflaw{local-expr-assign} -- see \refexa{assign-nw}} \\
  & nw := pw - \{i\}
\end{align*}
The refinement of (\ref{xCAS-part}) 
introduces an atomic specification command using \reflaw{atomic-spec-introduce} and then 
strengthens its postcondition using \reflaw{atomic-spec-strengthen-post} and
weakens its precondition using \reflaw{atomic-spec-weaken-pre} and weakens its rely 
to convert it to a form corresponding to the definition of a CAS \refdef{CAS2}
--- see \refexa{intro-CAS} for details.
Note that the atomic step of the CAS satisfies the guarantee,
so the guarantee is eliminated as part of the application of \reflaw{atomic-spec-introduce}.
\begin{align*}
  & \Rguar{w \supseteq w' \land w - w' \subseteq \{i\}} \together 
      \Rrely{w \supseteq w' \land i' = i \land nw' = nw \land pw' = pw} \together {} \\
  & \t1  \SPre{pw \supseteq w \land nw = pw - \{ i \}} \Seq \RSpec{w}{}{pw \supseteq w' \lor i' \notin w'}  \\
 \refsto~&
  CAS
\end{align*}
The accumulated code from the refinement gives the implementation of the operation.
\begin{align*}
 & \While i \in w \Do
      pw := w \Seq 
      nw := pw - \{i\} \Seq 
      CAS
    \Od
\end{align*}
Note that if the CAS succeeds, $i$ will no longer be in $w$ and the loop will terminate
but, if the CAS fails, $w \neq pw$ and hence $pw \supset w$ 
because the precondition of the CAS states that $pw \supseteq w$ and
because $w$ can only decrease, 
i.e.\ the variant $w$ must have decreased over the body of the loop under the superset ordering.
The interference may also have removed $i$ from $w$ 
but that is picked up when the loop guard is tested.

\section{Isabelle/HOL mechanisation}\labelsect{Isabelle}

This section discusses the formalisation of our theories in the Isabelle/HOL interactive theorem prover~\cite{IsabelleHOL}. 
The Isabelle theories consist of two main components:
a formalisation of the trace model overviewed in \refsect{model} and detailed in \cite{DaSMfaWSLwC}, and
a formalisation of the concurrent refinement algebra presented in this paper, 
which builds upon earlier work in~\cite{FMJournalAtomicSteps}, which formalised the core rely-guarantee algebra.
The current refinement algebra is built up as a hierarchy of theories based on the axiomisation in \reffig{axioms},
where each of the axioms has been shown to hold in the semantic model.
The refinement laws presented in the paper are proven on the basis of the algebraic theories.%
\footnote{The Isabelle proofs tend to be more detailed than those in this paper; the latter are designed to be more readable.}

At a high-level, the Isabelle theories come as a set of \textit{locales} (proof environments), which have been parameterised by the primitive algebraic operators (sequential composition, weak conjunction, parallel composition, and the lattice operators) and primitive commands (program-step, environment-step, test, and $\Abort$). The primitive algebraic operators and primitive commands are each assumed to satisfy the axioms presented in \reffig{axioms}.
They are used as the basis for defining the derived commands in \reffig{commands}, such as relies, guarantees, expressions, and programming constructs. In the theory, all properties -- including those of the derived commands -- are proved from the axioms of those primitives and lemmas/laws that are built up in a hierarchy of theories.

A consequence of this axiomatic, parameterised theory structure is that any semantic model providing suitable definitions of the primitives can gain access to the full library of theorems by using Isabelle's \textit{instantiate} command and dispatching the required axioms.

A new contribution in this paper is a set of laws for reasoning about expressions under interference (\refsect{expressions}). These were also interesting from a mechanisation point of view.
The syntax of expressions described in \refsect{expressions} is encoded as a standard (inductive) Isabelle datatype~\cite{BlanchetteIsabelleDatatypes} with four cases: constants, variable reference, unary expression and binary expression. The unary expression and binary expression cases are recursive in that they themselves accept sub-expressions (see Figure~\ref{fig:isabelleexpressions}).
To provide as much generality to the theory as possible, 
the unary operator (of the Isabelle function type ${'v} \Rightarrow {'v}$) 
and the binary operator (of type ${'v} \Rightarrow {'v} \Rightarrow {'v}$)
within expressions are both parameters to the datatype. 
Moreover, the theory is polymorphic in the type of values in the expression 
(allowing the choice of value type to depend on the program refinement), 
and in the representation of the program state space 
(which becomes another locale parameter, in the form of a variable getter- and setter- function pair). 
There is a minimum requirement that the value universe contains values for \textsf{true} and \textsf{false} 
if using conditional or loop commands. 
This latter requirement is achieved by requiring the value type to implement an Isabelle type class 
that provides values for \textsf{true} and \textsf{false}. 

\lstdefinelanguage{isabelle}
{
  morekeywords={
    datatype
  },
  basicstyle={\itshape},
  keywordstyle={\normalfont\bfseries},
  escapeinside={(*}{*)},
  sensitive=true, 
  morecomment=[l]{//}, 
  morecomment=[s]{/*}{*/} 
}

\begin{figure}
\begin{lstlisting}[language=isabelle,columns=fullflexible]
datatype ('s, 'v) expr = 
    Constant "'v" (*$|$*)
    Variable "'s (*$\Rightarrow$*)'v" (*$|$*)
    UnaryOp "'v (*$\Rightarrow$*)'v" "('s,'v) expr" (*$|$*)
    BinaryOp "'v (*$\Rightarrow$*)'v (*$\Rightarrow$*)'v" "('s,'v) expr" "('s,'v) expr"
\end{lstlisting}
\caption{Isabelle datatype definition for expressions, 
where \textit{'s} denotes the type of the state space and
\textit{'v} denotes the type of values in the expression.}\label{fig:isabelleexpressions}
\end{figure}

The approach of encoding expressions used in this work, wherein the abstract syntax of expressions is explicit, is commonly called \textit{deep embedding}~\cite{ZeydaAxiomaticValueModel}. It contrasts with \textit{shallow embedding}, popular in some other program algebra mechanisations~\cite{ZaydaUTP-AFP}, where one omits to model the abstract syntax of expressions explicitly, and instead uses an in-built type of the theorem prover to represent an expression. For example, an expression is modelled as a function from a state to a value, where the theorem prover's in-built function type is used. This shallow embedding usually makes it easier to use the expressions in a program refinement, because all the methods and theorems about functions can be used to reason about expressions.
However, as appealing as this solution is, it is not satisfactory for our purposes, because the shallow embedding approach does not allow one to analyse the expression substructure. Specifically, we cannot define our  expressions as an isomorphism for any function from the program state to a value, because it makes it impossible to unambiguously decompose an expression into the four expression cases: 
whereas the expressions $x+x$ and $2*x$ are equivalent if evaluated in a single state $\sigma$,
i.e.\ $\Eval{(x+x)}{\sigma} = \Eval{(2*x)}{\sigma}$,
in the context of interference from concurrent threads, 
the expressions $x + x$ and $x * 2$ are not equivalent:
under concurrent evaluation, $x +x$ may evaluate to an odd number under interference, 
while $x * 2$ is always even.
In our work, expressions obtain their semantics through the evaluation command $\Test{e}_k$, 
which turns an expression into a command. 
In Isabelle, this takes the form of a recursive function on the expression datatype.

\section{Conclusions}\labelsect{conclusions}

Our overall goal is to provide mechanised support for deriving concurrent programs from specifications
via a set of refinement laws,
with the laws being proven with respect to a simple core theory.
The rely/guarantee approach of Jones~\cite{Jones83b} forms the basis for our approach,
but we generalise the approach as well as provide a formal foundation 
that allows one to prove new refinement laws.
Our approach is based on a core concurrent refinement algebra~\cite{FMJournalAtomicSteps}
with a trace-based semantics~\cite{DaSMfaWSLwC}.
The core theory, semantics and refinement laws have all been developed as 
Isabelle/HOL theories (\refsect{Isabelle}).

Precondition assertions (\refsect{language}),
guarantees (\refsect{guarantees}), 
relies (\refsect{relies}), 
and
partial and total specifications (\refsect{specifications}) 
are each treated as separate commands in a wide-spectrum language.
The commands are defined in terms of our core language primitives
allowing straightforward proofs of laws for these constructs in terms of the core theory.
Our refinement calculus approach differs from that of Xu et al.\ \cite{XuRoeverHe97},
Prensa Nieto~\cite{PrensaNieto03} and San{\'a}n et al.\ \cite{Sanan21}
whose approaches use Jones-style five-tuples similar to Hoare logic.
The latter two also disallow nested parallel compositions --- they allow a (multi-way) parallel at the top level only.
Our approach also differs from that of Dingel~\cite{Dingel02} who uses a monolithic, four-component
(pre, rely, guarantee, and post condition) specification command.
Treating the concepts as separate commands allows simpler laws about the individual commands
to be developed in isolation.
The commands are combined using our base language operators,
including its novel weak conjunction operator.
That allows more complex laws that involve multiple constructs to be developed and
proven within the theory.
We support postcondition specifications (\refsect{specifications}),
which encode Jones-style postconditions within our theory
and for which we have developed a comprehensive theory supporting both partial and total correctness.
In addition, we have defined atomic specifications (\refsect{atomic-spec}),
which mimics the style of specification used by Dingel~\cite{Dingel02}
for specifying abstract operations on concurrent data structures that can be implemented as a single atomic step
with stuttering allowed before and after.
In practical program refinements, most steps involve refining just a pre-post specification or
a combination of a rely command with a pre-post specification.
Noting that guarantees distribute over programming constructs 
(e.g.\ sequential and parallel composition, conditionals, and loops),
guarantees only need to be considered for refinements to assignments or atomic specifications.
This makes program derivations simpler as one does not need to carry around the complete quintuple of Jones
or quadruple of Dingle.

The laws in this paper are ``semantic'' in the sense that tests and assertions use sets of states,
and the relations used in relies, guarantees and specifications are in terms of sets of pairs of states.
Hence the laws can be considered generic with respect to the language (syntax)
used to express sets of states (e.g.\ characteristic predicates on values the program variables) and
relations (e.g.\ predicates on the before and after values of program variables).
That allows the theory to be applied to standard state-based theories like
B~\cite{Abrial96}, VDM~\cite{Jones90a} and Z~\cite{Hayes93,Woodcock96}.
Only a handful of the laws are sensitive to the representation of the program state $\Sigma$,
and hence it is reasonably straightforward to adapt the laws to different state representations.
Such a change would affect the update in an assignment, 
and the semantics of accessing variables within expressions in 
assignments and as guards in conditionals and loops.

Whereas Xu et al.\ \cite{XuRoeverHe97}, Prensa Nieto~\cite{PrensaNieto03}, Dingel~\cite{Dingel02}, San{\'a}n et al.\ \cite{Sanan21} and Schellhorn et al.~\cite{SchellhornTEPR14}) assume expression evaluation is atomic,
in our approach expressions are not assumed to be atomic
and are defined in terms of our core language primitives (\refsect{expressions}).
An interesting challenge is handling expression and guard evaluation in the context
of interference, as is required to develop non-blocking implementations
in which the values of the variables in guards or assignments may be changed 
by interference from other threads.
Expression evaluation also leads to anomalies, such as the possibility of 
the expression $x=x$ evaluating to false if $x$ is modified between accesses to $x$.
The approach taken here generalises that taken by others 
\cite{ColemanVSTTE08,Wickerson10-TR,CoJo07,HayesBurnsDongolJones12}
who assume an expression only contains a single variable that is unmodified by the interference
and that variable is only referenced once in the expression.
That condition ensures that the evaluation of an expression under interference
corresponds to evaluating it in one of the states during its execution.
Our approach makes use of a weaker requirement that the expression is single reference under the rely condition
(\refdefi*{single-reference-under-rely}) that also guarantees that property.
Because we have defined expressions in terms of our core language primitives,
we are able to prove the key lemmas about single-reference expressions
and then use those lemmas to prove general laws for constructs containing expressions,
including assignments (\refsect{assignments}), 
conditionals (\refsect{conditional}) and 
while loops (\refsect{loop}).

Like Schellhorn et al.~\cite{SchellhornTEPR14}, 
we have included recursion in our language and use it to define while loops.
Our laws for recursion and while loops are more general in that they provide for 
early termination of recursions and loops.
The generality of the refinement laws makes them more useful in practice
(see the {\color{blue}related work} sections throughout this paper).

Brookes~\cite{Brookes-full-abstraction} and Dingel~\cite{Dingel02} make use of a trace semantics
that treats commands as being semantically equivalent if their sets of traces are
equivalent modulo finite stuttering and mumbling (see \refsect{specifications}).
The approach taken here is subtly different.
Our specification command is defined so that
it implicitly allows for finite stuttering and mumbling:
it is closed under finite stuttering and mumbling. 
Further, our encoding of programming language constructs (code), such as assignments and conditionals,
is defined in such a way that if $c$ and $d$ are code, 
and $c$ is semantically equivalent to $d$ modulo finite stuttering, 
then $c$ and $d$ are refinement equivalent.
For example, equivalences such as 
$\If \true \Then c \Else d \Fi = c$ for $c$ and $d$ code,
can be handled in the algebra.
Our approach handles finite stuttering and mumbling (in a different way)
while allowing refinement equivalence to be handled as equality,
which allows finite stuttering and mumbling to be handled in the algebra,
rather than the trace semantics.

In developing our refinement laws,
we have endeavoured to make the laws as general as possible.
The proof obligations for each law have been derived as part of the proof process for the law,
so that they are just what is needed to allow the proof to go through.
Many of our laws are more general than laws found in the 
related work on rely/guarantee concurrency.
Our more general laws allow one to tackle refinements 
that are not possible using other approaches. 
In practice, when applying our laws, many of our proof obligations are straightforward to prove;
the difficult proof obligations tend to correspond to the ``interesting'' parts of the refinement,
where it needs to explicitly cope with non-trivial interference.
The laws in this paper are sufficient to develop practical concurrent programs
but the underlying theory makes it straightforward to
\begin{itemize}
\item
develop new laws for existing constructs or combinations of constructs,
\item
add new data types, such as arrays,
and
\item
extend the language with new constructs, such as 
multiway parallel,%
\footnote{Our Isabelle theories include this.}
a $\kw{switch}$ command, 
a $\kw{for}$ command,
or a simultaneous (or parallel) assignment~\cite{CPL63},
and associated laws.
\end{itemize}
The building blocks and layers of theory we have used allow for simpler proofs 
than those based, for example, on an operational semantics for the
programming language~\cite{CoJo07}.

This paper is based on our concurrent refinement algebra developed in~\cite{FMJournalAtomicSteps},
which includes the proof of the parallel introduction law summarised in \refsect{parallel}.
The focus of the current paper is on laws for refining a specification in the context of interference, 
i.e.\ refining each of the threads in a parallel composition in isolation.
The tricky part is handling interference on shared variables.
The laws developed here have been used for the refinement 
of some standard concurrent algorithms in~\cite{HayesJones18}
and the reader is referred there for additional examples.

\subsection{Future work}\labelsect{future}

We are actively pursuing adding 
generalised invariants \cite{ReynoldsCraft81,LamportSchneiderConstraints85,TaIitRC,TaIitRC-SCP,TaIitRC-OtRC}, 
evolution guarantees \cite{Jones91m,ColletteJones00a},
and local variable blocks \cite{FormaliSE_localisation} to the language,
along with the necessary rely/guarantee laws to handle them.
That work shares the basic theory used in this paper 
but extends it with additional primitive operators
for handling variable localisation~\cite{FormaliSE_localisation,DaSMfaWSLwC,MPC19CylindricAlgebra}.
Local variables also allow one to develop theories for procedure parameter passing mechanisms
such as value and reference parameters,
and both generalised invariants and localisation are useful tools to support data refinement.

Some concurrent algorithms (such as spin lock)
do not give a guarantee of termination (under interference that is also performing locks)
but do guarantee termination in the absence of interference.
A partial version of an atomic specification command is more appropriate for specifying such algorithms.
Future work also includes developing a while loop rule for refining from a partial specification
that allows the loop to not terminate under interference 
but guarantees termination if the environment satisfies a temporal logic property.

Concurrent threads may need to wait for access to a resource or 
on a condition (e.g.~a buffer is non-empty).
Handling resources and termination of operations that may wait are further extensions
we are actively pursuing; initial ideas for incorporating these may be found in~\cite{REFINE2018}.
The approach in the current paper assumes a sequentially consistent memory model and
hence additional ``fencing'' is needed for use on a multi-processor with a weak memory model.
Additional work is needed to include the appropriate fencing to restore the desired behaviour.
The use of a concept of a resource allows one to link control/locking variables
with the data they control/lock~\cite{REFINE2018}, thus allowing the generation of appropriate fencing.

Conceptually, the abstract state space used within this paper could also incorporate a heap 
and use logics for reasoning about heaps,
such as separation logic~\cite{Reynolds02,BrookesSepLogic07} and 
relational separation logic~\cite{Yang-relational-separation-logic},
but detailed investigation of such instantiations is left for future work.

The ``possible values'' notation~\cite{JonesPierce10,PVEaCfC} provides a richer notation for expressing postconditions. 
The possible values postcondition $x' \in \posvals{e}$ states that the final value of $x$
is one of the possible values of $e$ in one of the states during the execution of the command.
Developing the theory to handle more expressive postconditions with possible values
is left as future work because representing a postcondition as a binary relation
is not expressive enough to handle possible values.

\section*{Acknowledgements}

Thanks are due to
Callum Bannister,
Robert Colvin,
Diego Machado Dias,
Julian Fell,
Tom Manderson,
Joshua Morris,
Andrius Velykis,
Kirsten Winter,
and our anonymous reviewers
for feedback on ideas presented in this paper
and/or contributions to the supporting Isabelle/HOL theories.
Special thanks go to Cliff Jones for his continual feedback and encouragement 
during the course of developing this research.

\bibliographystyle{alpha}
\bibliography{ms}

\newpage\printindex

\end{document}

%% file: rely-guar.tex
\begin{picture}(0,0)%
\includegraphics{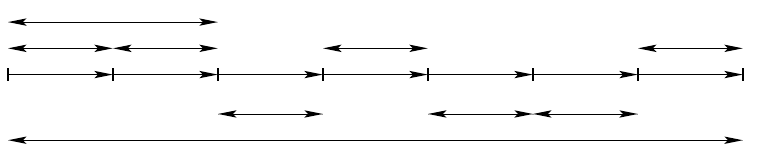}%
\end{picture}%
\setlength{\unitlength}{2763sp}%
\begingroup\makeatletter\ifx\SetFigFont\undefined%
\gdef\SetFigFont#1#2#3#4#5{%
  \reset@font\fontsize{#1}{#2pt}%
  \fontfamily{#3}\fontseries{#4}\fontshape{#5}%
  \selectfont}%
\fi\endgroup%
\begin{picture}(8655,1887)(1111,-3814)
\put(9751,-2761){\makebox(0,0)[lb]{\smash{{\SetFigFont{8}{9.6}{\rmdefault}{\mddefault}{\updefault}{\color[rgb]{0,0,0}$\checkmark$}%
}}}}
\put(1801,-2686){\makebox(0,0)[b]{\smash{{\SetFigFont{8}{9.6}{\rmdefault}{\mddefault}{\updefault}{\color[rgb]{0,0,0}$\estepd$}%
}}}}
\put(1801,-2386){\makebox(0,0)[b]{\smash{{\SetFigFont{8}{9.6}{\rmdefault}{\mddefault}{\updefault}{\color[rgb]{0,0,1}$r$}%
}}}}
\put(3001,-2686){\makebox(0,0)[b]{\smash{{\SetFigFont{8}{9.6}{\rmdefault}{\mddefault}{\updefault}{\color[rgb]{0,0,0}$\estepd$}%
}}}}
\put(4201,-2686){\makebox(0,0)[b]{\smash{{\SetFigFont{8}{9.6}{\rmdefault}{\mddefault}{\updefault}{\color[rgb]{0,0,0}$\pstepd$}%
}}}}
\put(5476,-2686){\makebox(0,0)[b]{\smash{{\SetFigFont{8}{9.6}{\rmdefault}{\mddefault}{\updefault}{\color[rgb]{0,0,0}$\estepd$}%
}}}}
\put(6601,-2686){\makebox(0,0)[b]{\smash{{\SetFigFont{8}{9.6}{\rmdefault}{\mddefault}{\updefault}{\color[rgb]{0,0,0}$\pstepd$}%
}}}}
\put(7801,-2686){\makebox(0,0)[b]{\smash{{\SetFigFont{8}{9.6}{\rmdefault}{\mddefault}{\updefault}{\color[rgb]{0,0,0}$\pstepd$}%
}}}}
\put(9001,-2686){\makebox(0,0)[b]{\smash{{\SetFigFont{8}{9.6}{\rmdefault}{\mddefault}{\updefault}{\color[rgb]{0,0,0}$\estepd$}%
}}}}
\put(3001,-2386){\makebox(0,0)[b]{\smash{{\SetFigFont{8}{9.6}{\rmdefault}{\mddefault}{\updefault}{\color[rgb]{0,0,1}$r$}%
}}}}
\put(5401,-2386){\makebox(0,0)[b]{\smash{{\SetFigFont{8}{9.6}{\rmdefault}{\mddefault}{\updefault}{\color[rgb]{0,0,1}$r$}%
}}}}
\put(9001,-2386){\makebox(0,0)[b]{\smash{{\SetFigFont{8}{9.6}{\rmdefault}{\mddefault}{\updefault}{\color[rgb]{0,0,1}$r$}%
}}}}
\put(1126,-2836){\makebox(0,0)[rb]{\smash{{\SetFigFont{8}{9.6}{\rmdefault}{\mddefault}{\updefault}{\color[rgb]{0,0,1}$p$}%
}}}}
\put(5476,-3736){\makebox(0,0)[b]{\smash{{\SetFigFont{8}{9.6}{\rmdefault}{\mddefault}{\updefault}{\color[rgb]{1,0,0}$q$}%
}}}}
\put(4201,-3436){\makebox(0,0)[b]{\smash{{\SetFigFont{8}{9.6}{\rmdefault}{\mddefault}{\updefault}{\color[rgb]{1,0,0}$g$}%
}}}}
\put(7801,-3436){\makebox(0,0)[b]{\smash{{\SetFigFont{8}{9.6}{\rmdefault}{\mddefault}{\updefault}{\color[rgb]{1,0,0}$g$}%
}}}}
\put(6601,-3436){\makebox(0,0)[b]{\smash{{\SetFigFont{8}{9.6}{\rmdefault}{\mddefault}{\updefault}{\color[rgb]{1,0,0}$g$}%
}}}}
\put(2401,-2086){\makebox(0,0)[b]{\smash{{\SetFigFont{8}{9.6}{\rmdefault}{\mddefault}{\updefault}{\color[rgb]{0,0,1}$r$}%
}}}}
\put(1201,-3061){\makebox(0,0)[b]{\smash{{\SetFigFont{8}{9.6}{\rmdefault}{\mddefault}{\updefault}{\color[rgb]{0,0,0}$\sigma_0$}%
}}}}
\put(2401,-3061){\makebox(0,0)[b]{\smash{{\SetFigFont{8}{9.6}{\rmdefault}{\mddefault}{\updefault}{\color[rgb]{0,0,0}$\sigma_1$}%
}}}}
\put(3601,-3061){\makebox(0,0)[b]{\smash{{\SetFigFont{8}{9.6}{\rmdefault}{\mddefault}{\updefault}{\color[rgb]{0,0,0}$\sigma_2$}%
}}}}
\put(4801,-3061){\makebox(0,0)[b]{\smash{{\SetFigFont{8}{9.6}{\rmdefault}{\mddefault}{\updefault}{\color[rgb]{0,0,0}$\sigma_3$}%
}}}}
\put(6001,-3061){\makebox(0,0)[b]{\smash{{\SetFigFont{8}{9.6}{\rmdefault}{\mddefault}{\updefault}{\color[rgb]{0,0,0}$\sigma_4$}%
}}}}
\put(7201,-3061){\makebox(0,0)[b]{\smash{{\SetFigFont{8}{9.6}{\rmdefault}{\mddefault}{\updefault}{\color[rgb]{0,0,0}$\sigma_5$}%
}}}}
\put(8401,-3061){\makebox(0,0)[b]{\smash{{\SetFigFont{8}{9.6}{\rmdefault}{\mddefault}{\updefault}{\color[rgb]{0,0,0}$\sigma_6$}%
}}}}
\put(9601,-3061){\makebox(0,0)[b]{\smash{{\SetFigFont{8}{9.6}{\rmdefault}{\mddefault}{\updefault}{\color[rgb]{0,0,0}$\sigma_7$}%
}}}}
\end{picture}%

%% file: assignment.tex
\begin{picture}(0,0)%
\includegraphics{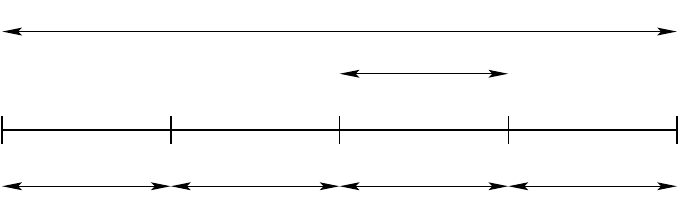}%
\end{picture}%
\setlength{\unitlength}{2960sp}%
\begingroup\makeatletter\ifx\SetFigFont\undefined%
\gdef\SetFigFont#1#2#3#4#5{%
  \reset@font\fontsize{#1}{#2pt}%
  \fontfamily{#3}\fontseries{#4}\fontshape{#5}%
  \selectfont}%
\fi\endgroup%
\begin{picture}(7244,2286)(1179,-3664)
\put(5701,-2611){\makebox(0,0)[b]{\smash{{\SetFigFont{9}{10.8}{\rmdefault}{\mddefault}{\updefault}{\color[rgb]{0,0,0}$\Opt{}$}%
}}}}
\put(1201,-2461){\makebox(0,0)[b]{\smash{{\SetFigFont{9}{10.8}{\rmdefault}{\mddefault}{\updefault}{\color[rgb]{0,0,0}$\sigma$}%
}}}}
\put(3001,-2461){\makebox(0,0)[b]{\smash{{\SetFigFont{9}{10.8}{\rmdefault}{\mddefault}{\updefault}{\color[rgb]{0,0,0}$\sigma_1$}%
}}}}
\put(4801,-2461){\makebox(0,0)[b]{\smash{{\SetFigFont{9}{10.8}{\rmdefault}{\mddefault}{\updefault}{\color[rgb]{0,0,0}$\sigma_2$}%
}}}}
\put(6601,-2461){\makebox(0,0)[b]{\smash{{\SetFigFont{9}{10.8}{\rmdefault}{\mddefault}{\updefault}{\color[rgb]{0,0,0}$\sigma_3$}%
}}}}
\put(8401,-2461){\makebox(0,0)[b]{\smash{{\SetFigFont{9}{10.8}{\rmdefault}{\mddefault}{\updefault}{\color[rgb]{0,0,0}$\sigma'$}%
}}}}
\put(1201,-3136){\makebox(0,0)[b]{\smash{{\SetFigFont{9}{10.8}{\rmdefault}{\mddefault}{\updefault}$p$}}}}
\put(3001,-3136){\makebox(0,0)[b]{\smash{{\SetFigFont{9}{10.8}{\rmdefault}{\mddefault}{\updefault}{\color[rgb]{0,0,0}$eq(k,e)$}%
}}}}
\put(3901,-3586){\makebox(0,0)[b]{\smash{{\SetFigFont{9}{10.8}{\rmdefault}{\mddefault}{\updefault}{\color[rgb]{0,0,0}$\Finrel{r}$}%
}}}}
\put(5701,-3586){\makebox(0,0)[b]{\smash{{\SetFigFont{9}{10.8}{\rmdefault}{\mddefault}{\updefault}{\color[rgb]{0,0,0}$\id{\overline{x}}$}%
}}}}
\put(6601,-3136){\makebox(0,0)[b]{\smash{{\SetFigFont{9}{10.8}{\rmdefault}{\mddefault}{\updefault}{\color[rgb]{0,0,0}$\EqEval{k}{x}$}%
}}}}
\put(7501,-3586){\makebox(0,0)[b]{\smash{{\SetFigFont{9}{10.8}{\rmdefault}{\mddefault}{\updefault}{\color[rgb]{0,0,0}$\Finrel{r}$}%
}}}}
\put(4801,-1561){\makebox(0,0)[b]{\smash{{\SetFigFont{9}{10.8}{\rmdefault}{\mddefault}{\updefault}{\color[rgb]{0,0,0}$q$}%
}}}}
\put(5701,-2011){\makebox(0,0)[b]{\smash{{\SetFigFont{9}{10.8}{\rmdefault}{\mddefault}{\updefault}{\color[rgb]{0,0,0}$g$}%
}}}}
\put(2101,-3586){\makebox(0,0)[b]{\smash{{\SetFigFont{9}{10.8}{\rmdefault}{\mddefault}{\updefault}{\color[rgb]{0,0,0}$\Finrel{r}$}%
}}}}
\put(4801,-3136){\makebox(0,0)[b]{\smash{{\SetFigFont{9}{10.8}{\rmdefault}{\mddefault}{\updefault}{\color[rgb]{0,0,0}$\Finrel{r} \limg \EqEval{k}{e} \rimg$}%
}}}}
\put(2101,-2611){\makebox(0,0)[b]{\smash{{\SetFigFont{9}{10.8}{\rmdefault}{\mddefault}{\updefault}{\color[rgb]{0,0,0}$\Idle$}%
}}}}
\put(7501,-2611){\makebox(0,0)[b]{\smash{{\SetFigFont{9}{10.8}{\rmdefault}{\mddefault}{\updefault}{\color[rgb]{0,0,0}$\Idle$}%
}}}}
\put(3901,-2611){\makebox(0,0)[b]{\smash{{\SetFigFont{9}{10.8}{\rmdefault}{\mddefault}{\updefault}{\color[rgb]{0,0,0}$\Idle$}%
}}}}
\end{picture}%

%% file: assignment2.tex
\begin{picture}(0,0)%
\includegraphics{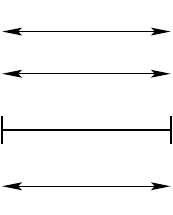}%
\end{picture}%
\setlength{\unitlength}{2960sp}%
\begingroup\makeatletter\ifx\SetFigFont\undefined%
\gdef\SetFigFont#1#2#3#4#5{%
  \reset@font\fontsize{#1}{#2pt}%
  \fontfamily{#3}\fontseries{#4}\fontshape{#5}%
  \selectfont}%
\fi\endgroup%
\begin{picture}(1844,2286)(4779,-3664)
\put(5701,-1561){\makebox(0,0)[b]{\smash{{\SetFigFont{9}{10.8}{\rmdefault}{\mddefault}{\updefault}{\color[rgb]{0,0,0}$q$}%
}}}}
\put(4801,-2461){\makebox(0,0)[b]{\smash{{\SetFigFont{9}{10.8}{\rmdefault}{\mddefault}{\updefault}{\color[rgb]{0,0,0}$\sigma_2$}%
}}}}
\put(6601,-2461){\makebox(0,0)[b]{\smash{{\SetFigFont{9}{10.8}{\rmdefault}{\mddefault}{\updefault}{\color[rgb]{0,0,0}$\sigma_3$}%
}}}}
\put(5701,-2011){\makebox(0,0)[b]{\smash{{\SetFigFont{9}{10.8}{\rmdefault}{\mddefault}{\updefault}{\color[rgb]{0,0,0}$g$}%
}}}}
\put(5701,-3586){\makebox(0,0)[b]{\smash{{\SetFigFont{9}{10.8}{\rmdefault}{\mddefault}{\updefault}{\color[rgb]{0,0,0}$\id{\overline{x}}$}%
}}}}
\put(4801,-3136){\makebox(0,0)[b]{\smash{{\SetFigFont{9}{10.8}{\rmdefault}{\mddefault}{\updefault}{\color[rgb]{0,0,0}$p_1\,k$}%
}}}}
\put(6601,-3136){\makebox(0,0)[b]{\smash{{\SetFigFont{9}{10.8}{\rmdefault}{\mddefault}{\updefault}{\color[rgb]{0,0,0}$\EqEval{k}{x}$}%
}}}}
\put(5701,-2611){\makebox(0,0)[b]{\smash{{\SetFigFont{9}{10.8}{\rmdefault}{\mddefault}{\updefault}{\color[rgb]{0,0,0}$\Opt{}$}%
}}}}
\end{picture}%

%% file: if.tex
\begin{picture}(0,0)%
\includegraphics{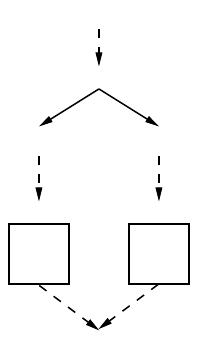}%
\end{picture}%
\setlength{\unitlength}{3158sp}%
\begingroup\makeatletter\ifx\SetFigFont\undefined%
\gdef\SetFigFont#1#2#3#4#5{%
  \reset@font\fontsize{#1}{#2pt}%
  \fontfamily{#3}\fontseries{#4}\fontshape{#5}%
  \selectfont}%
\fi\endgroup%
\begin{picture}(1980,3502)(5611,-4793)
\put(6601,-2386){\makebox(0,0)[b]{\smash{{\SetFigFont{10}{12.0}{\rmdefault}{\mddefault}{\updefault}{\color[rgb]{0,0,0}$\sigma_1$}%
}}}}
\put(7051,-2386){\makebox(0,0)[lb]{\smash{{\SetFigFont{10}{12.0}{\sfdefault}{\mddefault}{\updefault}{\color[rgb]{0,0,0}$\pnegate{b}$}%
}}}}
\put(6601,-2086){\makebox(0,0)[b]{\smash{{\SetFigFont{10}{12.0}{\sfdefault}{\mddefault}{\updefault}{\color[rgb]{0,0,0}$p$}%
}}}}
\put(6526,-1786){\makebox(0,0)[rb]{\smash{{\SetFigFont{10}{12.0}{\sfdefault}{\mddefault}{\updefault}{\color[rgb]{0,0,0}$\Finrel{r}$}%
}}}}
\put(6601,-1486){\makebox(0,0)[b]{\smash{{\SetFigFont{10}{12.0}{\sfdefault}{\mddefault}{\updefault}{\color[rgb]{0,0,0}$p$}%
}}}}
\put(6226,-2386){\makebox(0,0)[rb]{\smash{{\SetFigFont{10}{12.0}{\sfdefault}{\mddefault}{\updefault}{\color[rgb]{0,0,0}$b$}%
}}}}
\put(7126,-3061){\makebox(0,0)[rb]{\smash{{\SetFigFont{10}{12.0}{\sfdefault}{\mddefault}{\updefault}{\color[rgb]{0,0,0}$\Finrel{r}$}%
}}}}
\put(7201,-3736){\makebox(0,0)[b]{\smash{{\SetFigFont{10}{12.0}{\sfdefault}{\mddefault}{\updefault}{\color[rgb]{0,0,0}else}%
}}}}
\put(7201,-3436){\makebox(0,0)[b]{\smash{{\SetFigFont{10}{12.0}{\sfdefault}{\mddefault}{\updefault}{\color[rgb]{0,0,0}$p \inter b_f$}%
}}}}
\put(7201,-2761){\makebox(0,0)[b]{\smash{{\SetFigFont{10}{12.0}{\sfdefault}{\mddefault}{\updefault}{\color[rgb]{0,0,0}$p \inter \pnegate{b}$}%
}}}}
\put(5926,-3061){\makebox(0,0)[rb]{\smash{{\SetFigFont{10}{12.0}{\sfdefault}{\mddefault}{\updefault}{\color[rgb]{0,0,0}$\Finrel{r}$}%
}}}}
\put(6001,-3436){\makebox(0,0)[b]{\smash{{\SetFigFont{10}{12.0}{\sfdefault}{\mddefault}{\updefault}{\color[rgb]{0,0,0}$p \inter b_t$}%
}}}}
\put(6001,-2761){\makebox(0,0)[b]{\smash{{\SetFigFont{10}{12.0}{\sfdefault}{\mddefault}{\updefault}{\color[rgb]{0,0,0}$p \inter b$}%
}}}}
\put(6151,-4411){\makebox(0,0)[rb]{\smash{{\SetFigFont{10}{12.0}{\sfdefault}{\mddefault}{\updefault}{\color[rgb]{0,0,0}$\Finrel{r}$}%
}}}}
\put(7051,-4411){\makebox(0,0)[lb]{\smash{{\SetFigFont{10}{12.0}{\sfdefault}{\mddefault}{\updefault}{\color[rgb]{0,0,0}$\Finrel{r}$}%
}}}}
\put(6601,-4711){\makebox(0,0)[b]{\smash{{\SetFigFont{10}{12.0}{\sfdefault}{\mddefault}{\updefault}{\color[rgb]{0,0,0}$q$}%
}}}}
\put(6001,-3961){\makebox(0,0)[b]{\smash{{\SetFigFont{10}{12.0}{\sfdefault}{\mddefault}{\updefault}{\color[rgb]{0,0,0}$q$}%
}}}}
\put(7201,-3961){\makebox(0,0)[b]{\smash{{\SetFigFont{10}{12.0}{\sfdefault}{\mddefault}{\updefault}{\color[rgb]{0,0,0}$q$}%
}}}}
\put(5626,-3886){\makebox(0,0)[rb]{\smash{{\SetFigFont{10}{12.0}{\sfdefault}{\mddefault}{\updefault}{\color[rgb]{0,0,0}$\Finrel{r}$}%
}}}}
\put(7576,-3886){\makebox(0,0)[lb]{\smash{{\SetFigFont{10}{12.0}{\sfdefault}{\mddefault}{\updefault}{\color[rgb]{0,0,0}$\Finrel{r}$}%
}}}}
\put(6001,-3736){\makebox(0,0)[b]{\smash{{\SetFigFont{10}{12.0}{\sfdefault}{\mddefault}{\updefault}{\color[rgb]{0,0,0}then}%
}}}}
\end{picture}%

%% file: loop.tex
\begin{picture}(0,0)%
\includegraphics{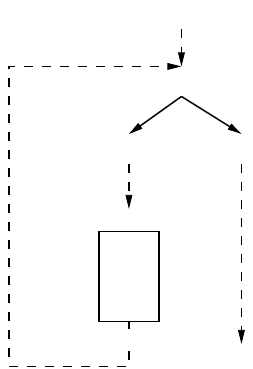}%
\end{picture}%
\setlength{\unitlength}{3158sp}%
\begingroup\makeatletter\ifx\SetFigFont\undefined%
\gdef\SetFigFont#1#2#3#4#5{%
  \reset@font\fontsize{#1}{#2pt}%
  \fontfamily{#3}\fontseries{#4}\fontshape{#5}%
  \selectfont}%
\fi\endgroup%
\begin{picture}(2505,3667)(4786,-4883)
\put(6676,-1711){\makebox(0,0)[lb]{\smash{{\SetFigFont{10}{12.0}{\rmdefault}{\mddefault}{\updefault}{\color[rgb]{0,0,0}$\Deceq{w}$}%
}}}}
\put(6601,-2086){\makebox(0,0)[b]{\smash{{\SetFigFont{10}{12.0}{\sfdefault}{\mddefault}{\updefault}{\color[rgb]{0,0,0}$p$}%
}}}}
\put(7126,-3361){\makebox(0,0)[rb]{\smash{{\SetFigFont{10}{12.0}{\sfdefault}{\mddefault}{\updefault}{\color[rgb]{0,0,0}$\Finrel{r}$}%
}}}}
\put(7276,-3361){\makebox(0,0)[lb]{\smash{{\SetFigFont{10}{12.0}{\sfdefault}{\mddefault}{\updefault}{\color[rgb]{0,0,0}$\Deceq{w}$}%
}}}}
\put(6601,-1411){\makebox(0,0)[b]{\smash{{\SetFigFont{10}{12.0}{\sfdefault}{\mddefault}{\updefault}{\color[rgb]{0,0,0}$p$}%
}}}}
\put(6301,-2311){\makebox(0,0)[rb]{\smash{{\SetFigFont{10}{12.0}{\rmdefault}{\mddefault}{\updefault}{\color[rgb]{0,0,0}$b$}%
}}}}
\put(6901,-2311){\makebox(0,0)[lb]{\smash{{\SetFigFont{10}{12.0}{\rmdefault}{\mddefault}{\updefault}{\color[rgb]{0,0,0}$\pnegate{b}$}%
}}}}
\put(6526,-1711){\makebox(0,0)[rb]{\smash{{\SetFigFont{10}{12.0}{\sfdefault}{\mddefault}{\updefault}{\color[rgb]{0,0,0}$\Finrel{r}$}%
}}}}
\put(6001,-3061){\makebox(0,0)[rb]{\smash{{\SetFigFont{10}{12.0}{\sfdefault}{\mddefault}{\updefault}{\color[rgb]{0,0,0}$\Finrel{r}$}%
}}}}
\put(6151,-3061){\makebox(0,0)[lb]{\smash{{\SetFigFont{10}{12.0}{\sfdefault}{\mddefault}{\updefault}{\color[rgb]{0,0,0}$\Deceq{w}$}%
}}}}
\put(7201,-2761){\makebox(0,0)[b]{\smash{{\SetFigFont{10}{12.0}{\sfdefault}{\mddefault}{\updefault}{\color[rgb]{0,0,0}$p \inter \pnegate{b}$}%
}}}}
\put(7276,-4786){\makebox(0,0)[b]{\smash{{\SetFigFont{10}{12.0}{\sfdefault}{\mddefault}{\updefault}{\color[rgb]{0,0,0}$p \inter b_f$}%
}}}}
\put(5701,-4036){\makebox(0,0)[rb]{\smash{{\SetFigFont{10}{12.0}{\sfdefault}{\mddefault}{\updefault}{\color[rgb]{0,0,0}$\Finrel{r}$}%
}}}}
\put(6451,-4036){\makebox(0,0)[lb]{\smash{{\SetFigFont{10}{12.0}{\sfdefault}{\mddefault}{\updefault}{\color[rgb]{0,0,0}$\Deceq{w}$}%
}}}}
\put(6076,-3811){\makebox(0,0)[b]{\smash{{\SetFigFont{10}{12.0}{\sfdefault}{\mddefault}{\updefault}{\color[rgb]{0,0,0}body}%
}}}}
\put(6076,-4186){\makebox(0,0)[b]{\smash{{\SetFigFont{10}{12.0}{\rmdefault}{\mddefault}{\updefault}{\color[rgb]{0,0,0}$\Finrel{q}$}%
}}}}
\put(6076,-2761){\makebox(0,0)[b]{\smash{{\SetFigFont{10}{12.0}{\sfdefault}{\mddefault}{\updefault}{\color[rgb]{0,0,0}$p \inter b$}%
}}}}
\put(4801,-2536){\makebox(0,0)[rb]{\smash{{\SetFigFont{10}{12.0}{\sfdefault}{\mddefault}{\updefault}{\color[rgb]{0,0,0}$\Finrel{r}$}%
}}}}
\put(4951,-2536){\makebox(0,0)[lb]{\smash{{\SetFigFont{10}{12.0}{\sfdefault}{\mddefault}{\updefault}{\color[rgb]{0,0,0}$\Deceq{w}$}%
}}}}
\put(6076,-4636){\makebox(0,0)[b]{\smash{{\SetFigFont{10}{12.0}{\sfdefault}{\mddefault}{\updefault}{\color[rgb]{0,0,0}$p \inter (gt\,k\,w) \union b_x)$}%
}}}}
\put(6151,-3436){\makebox(0,0)[b]{\smash{{\SetFigFont{10}{12.0}{\sfdefault}{\mddefault}{\updefault}{\color[rgb]{0,0,0}$p \inter b_t \inter (ge\,k\,w)$}%
}}}}
\put(6601,-2386){\makebox(0,0)[b]{\smash{{\SetFigFont{10}{12.0}{\rmdefault}{\mddefault}{\updefault}{\color[rgb]{0,0,0}$\sigma_1$}%
}}}}
\end{picture}%